\documentclass{jfm}
\usepackage{graphicx}
\usepackage{epstopdf, epsfig}
\usepackage[normalem]{ulem}
\usepackage{tikz}
\usepackage{amsmath}
\usepackage{bm}
\usepackage{color}
\usepackage{float}
\usepackage{wrapfig, subcaption, setspace, booktabs}

\newcommand{\be}{\begin{equation}}
\newcommand{\ee}{\end{equation}}

\shorttitle{Local streamline geometry characterization}
\shortauthor{Rishita Das and Sharath S. Girimaji}

\title{Characterization of velocity-gradient dynamics in incompressible turbulence using local streamline geometry}

\author{Rishita Das\aff{1}
  \corresp{\email{rishitadas@tamu.edu}}
  \and Sharath S. Girimaji\aff{2}}

\affiliation{\aff{1}Department of Aerospace Engineering, Texas A\&M University,
College Station, TX 77843, USA
\aff{2}Department of Ocean Engineering, Texas A\&M University,
College Station, TX 77843, USA}

\begin{document}

\maketitle
\begin{abstract}


This study develops a comprehensive description of local streamline geometry and uses the resulting shape features to characterize velocity gradient ($A_{ij}=\partial u_i/\partial x_j$) dynamics.
The local streamline {\em geometric shape parameters} and {\em scale-factor (size)} are extracted from $A_{ij}$ by extending the linearized critical point analysis.  In the present analysis, 
$A_{ij}$  is factorized into its magnitude ($A \equiv \sqrt{A_{ij}A_{ij}}$) and normalized tensor $b_{ij} \equiv A_{ij}/A$.
The geometric shape is shown to be determined exclusively by four $b_{ij}$ parameters --  second invariant, $q$; third invariant, $r$; intermediate strain-rate eigenvalue, $a_2$; and, angle between vorticity and intermediate strain-rate eigenvector, $\omega_2$. Velocity gradient magnitude $A$ plays a role only in determining the scale of the local streamline structure. 
Direct numerical simulation data of forced isotropic turbulence ($Re_\lambda \sim 200 - 600$) is used to establish streamline shape and scale distribution and, then to characterize velocity-gradient dynamics. 
Conditional mean trajectories (CMTs) in $q$-$r$ space reveal important non-local features of pressure and viscous dynamics which are not evident from the $A_{ij}$-invariants. 
Two distinct types of $q$-$r$ CMTs demarcated by a separatrix are identified. The inner trajectories are dominated by inertia-pressure interactions and the viscous effects play a significant role only in the outer trajectories. 
Dynamical system characterization of inertial, pressure  and viscous effects in the $q$-$r$ phase space is developed. Additionally, it is shown that the residence time of $q$-$r$ CMTs through different topologies correlate well with the corresponding population fractions. These findings not only lead to improved understanding of non-local dynamics, but also provide an important foundation for developing Lagrangian velocity-gradient models.

\end{abstract}

\section{\label{sec:intro}Introduction}

The structure of velocity-gradient tensor (VGT, $A_{ij}=\partial u_i/\partial x_j$) and its evolution in a turbulent flow provide valuable insight into key turbulence processes including non-local pressure and viscous effects. The last few decades have witnessed many important advances toward understanding internal structure of  $A_{ij}$  \citep{ashurst1987alignment,kerr1987histograms} and describing local streamline topology in terms of $A_{ij}$ invariants \citep{chong1990general}. 
The topological classification of local streamline structure has enabled further advances in (i) identifying key universal features of local streamline structure \citep{soria1994study,blackburn1996topology,chacin1996study,chacin2000dynamics,elsinga2010universal}, and (ii)  characterization of important velocity gradient processes conditioned upon topology \citep{martin1998dynamics,ooi1999study,elsinga2010evolution,
atkinson2012lagrangian,bechlars2017evolution}.
Other studies on the structure of $A_{ij}$ have led to improved understanding of internal alignment properties, characteristic length scales and non-normality in different topologies \citep{hamlington2008direct,chevillard2008modeling,danish2018multiscale,keylock2018schur}.
The improved comprehension of turbulence dynamics derived from these studies has been employed to develop Lagrangian VGT evolution models \citep{cantwell1992exact,girimaji1990diffusion,girimaji1995modified,
martin1998dynamics2,chertkov1999lagrangian,jeong2003velocity,chevillard2006lagrangian,
chevillard2008modeling,johnson2016closure}. 
While much progress has been made, our comprehension of VGT dynamics, specifically the non-local pressure and viscous processes, remains incomplete and the closure models need further improvement.

The goal of this study is to develop a more complete description of local streamline geometry that can be used for enhanced characterization of velocity gradient dynamics. An important feature of the approach is that it combines $A_{ij}$ internal structural features \citep{ashurst1987alignment} with topological classification \citep{chong1990general} to render a more complete geometric basis for conditioning non-local velocity gradient processes.
At the very outset, it is important to formally distinguish between topology and geometry.
The geometry of an object is constituted by its shape and size 
\citep{yale1968geometry,smart1998modern}.
Shape and size are quantified in terms of shape-parameters and scale-factor, respectively.
Geometric shape is defined as the structural characteristics of the object, that is invariant to translation, rotation and reflection and is independent of size.
Topology, on the other hand, is a class of geometric shapes that have similar connectivity and can be transformed from one to the other by continuous deformation \citep{kinsey1993topology,blackett2014elementary}. 
Clearly, a description of streamline geometry requires additional details to those derived from a topological classification. 


The objectives of the present study are three-fold. First, we derive a description of local streamline shape-parameters by extending critical point analysis \citep{perry1975critical,perry1987description}. Then, we employ direct numerical simulation (DNS) data to determine the statistical distribution of  local shape-parameters and scale-factor of streamlines in turbulent flow fields. Finally, we use DNS data to develop a dynamical system characterization of non-local pressure and viscous processes conditioned on key shape parameters.

The remaining sections of this study are arranged as follows. Section \ref{sec:eqns} develops the framework for the description of local streamline shape.
Section \ref{sec:data} provides a brief description of the DNS data used for analysis.
The probability distributions of streamline shape parameters and the mean scale-factor for different geometric-shapes are presented in section \ref{sec:Res1}.
Section \ref{sec:Res2} investigates the conditional mean trajectories (CMTs) and characterizes the role of inertial, pressure and viscous processes in the evolution of streamline shape.
Finally, the key findings of the study are summarized in section \ref{sec:conc}.


\section{\label{sec:eqns}Complete characterization of local streamline geometry}

We first reiterate the fundamental distinction between topological and geometric 
description of an object in the context of the present work.
These concepts are then used to establish that the $A_{ij}$-invariants cannot
uniquely describe the local streamline geometry or shape. Then, we develop a complete description of local streamline geometry in terms of
shape parameters and scale factor.

\subsection{\label{ssec:eqns1} Geometry and topology}

The \textit{geometry} of any object can be described with two principal attributes - shape and size. 
\textit{Geometric shape} is the structural characterization of an object that is independent of size and invariant under translation, rotation, reflection and any other similarity transformation \citep{yale1968geometry,smart1998modern}. Shape
describes internal structural arrangement and is parameterized by the unique combination of angles between edges and ratios of lengths of edges.
\textit{Scale-factor} is a measure of the size of such a geometric shape, which can be scaled through simple enlargement or shrinking without altering the angles and ratios of distances. It is also known as stretching factor  \citep{yale1968geometry} or ratio of similarity \citep{smart1998modern}.
In general the number of shape-parameters required to describe an object depends upon the complexity of its geometry and its dimensionality. 
In the present context, the shrinking and enlargement are the same in all directions (isotropic) and therefore only one scale-factor is required to specify the size. 
The geometries of two objects are {\em similar} if all the shape parameters are identical. If two objects have identical shape parameters and scale factors, they are called {\em congruent.}


\textit{Topology} describes a set of geometric shapes that exhibit the following attributes  \citep{kinsey1993topology,smart1998modern,blackett2014elementary}:
(i)  different shapes of the set can be transformed from one to the other by continuous deformations or homeomorphisms, such as stretching, compression, torsion and shearing; and (ii)
 all shapes of the set have the same connectivity (e.g. singly or doubly connected).
Thus, topology does not completely define the specific geometric shape of an object, but identifies a set of shapes with certain common features. 
We will next examine  topology and geometry of local streamline structures.

\subsection{\label{ssec:eqns2} Streamline structure}


Fluid particle evolution equation forms the basis of the streamline structure analysis. The position of a fluid particle in a velocity field evolves according to
\begin{equation}
     \frac{dx_i}{dt} = u_{i}
\label{eq:ui1}
\end{equation}
The deformation of infinitesimal material line and area elements \citep{orszag1970comments,monin2013statistical,girimaji1990material} can be inferred from this equation.
This Lagrangian description can also be used to describe the shape of streaklines and streamlines in a flow. 
Using critical point analysis and first order Taylor series expansion of the velocity field \citep{perry1975critical,perry1987description,perry1994topology}, the particle trajectory is governed by   
\begin{equation}
     \frac{dx_i}{dt} = A_{ij}x_j
\label{eq:ui2}
\end{equation}
Subject to simplification of a steady flow and spatially uniform velocity gradient field in the immediate vicinity of the free-slip critical point, the solution trajectories from the above equation represent the local instantaneous streamlines. 
The streamline structure can be described in different levels of details such as 
(i) topological classification, or (ii) full geometry description.


\subsubsection{\label{ssec:eqns2a} Topological classification of streamlines}

\begin{figure}
\centering
\includegraphics[width=0.85\textwidth]{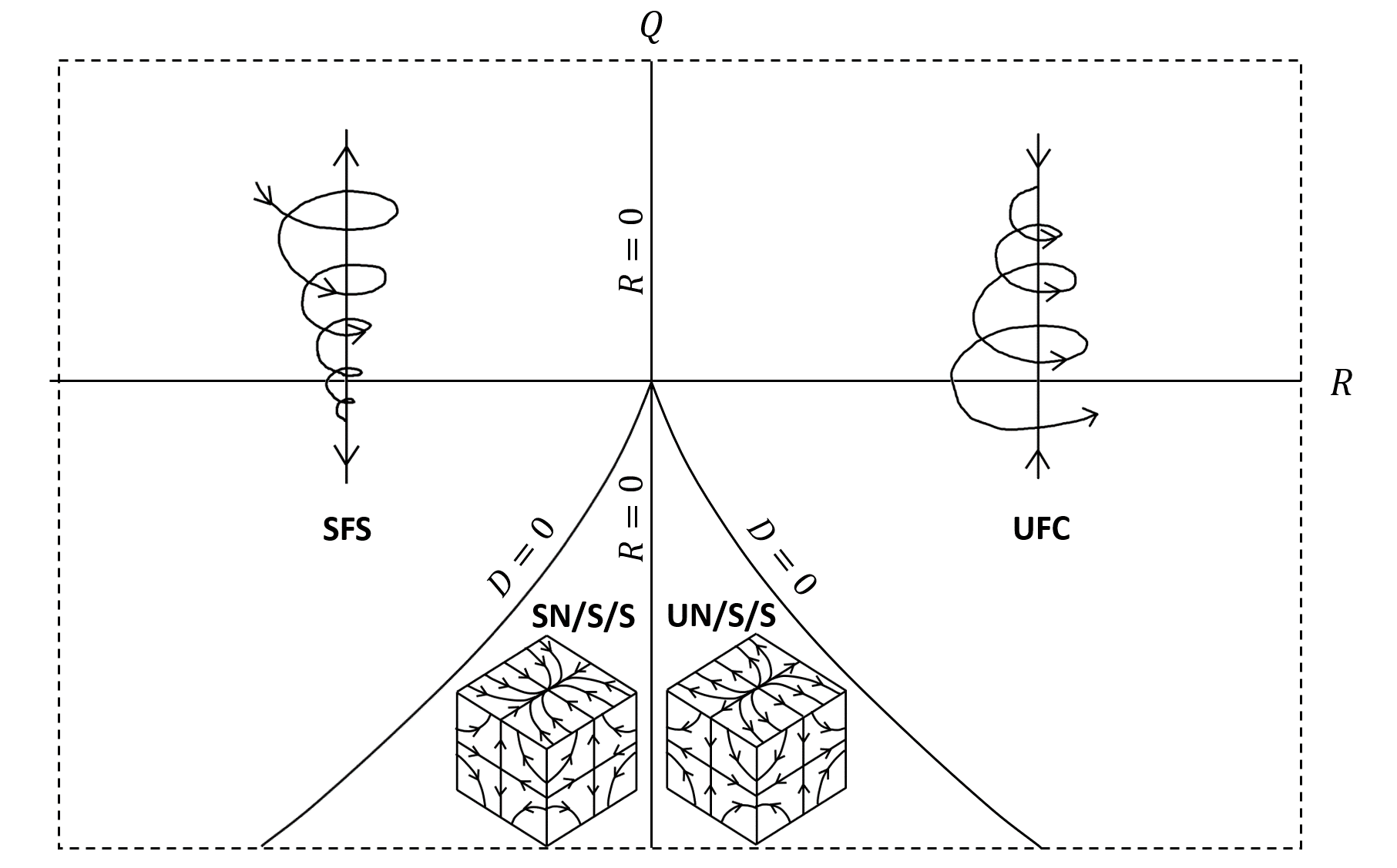}
\caption{Classification of local three-dimensional streamlines into non-degenerate topologies in $Q$-$R$ plane \citep{chong1990general} for incompressible turbulence. The curved solid lines are discriminant $D=0$ lines.}
\label{fig:QR_topo}
\end{figure}

The topological classification of local streamline structure in incompressible flows can be achieved with only the VGT invariants \citep{chong1990general}, 
\begin{equation}
    Q = -\frac{1}{2}A_{ij}A_{ji} \;\;  \text{and} \;\;   R = -\frac{1}{3}A_{ij}A_{jk}A_{ki} 
\end{equation}
The $R=0$ and the discriminant $D = Q^3 + ({27}/{4})R^2 = 0$ lines divide the $Q$-$R$ plane into four regions, each representing a topology, as depicted in figure \ref{fig:QR_topo}. 
These four non-degenerate three-dimensional topologies are stable focus stretching (SFS), unstable focus compression (UFC), unstable-node/saddle/saddle (UN/S/S) and stable-node/saddle/saddle (SN/S/S).
This classification groups together streamline-shapes that are ``topologically equivalent'' or geometrically homeomorphic.
In other words, different shapes related to each other by affine transformations are categorized as the same topology. 
A representative streamline shape of each topology, based on a canonical form of VGT, is shown in figure \ref{fig:QR_topo}.

The topological description in $Q$-$R$ plane does not uniquely describe the geometric shape of streamlines. This is best illustrated by the following example of an elliptic (closed) two-dimensional streamline flow. 
A VGT of the form \citep{blaisdell1996simulation},
\begin{eqnarray}\centering
    \bm{A}= \left[
    \begin{array}{ccc}
    0 & -\gamma-e\\
    \gamma-e & 0
    \end{array}\right]		\;\;\;\;\text{where} \;\;	0 < |e| < |\gamma|
\label{eq:Aij1}
\end{eqnarray}
describes local elliptic streamlines in the $x_1$-$x_2$ plane with major axis along $x_1$ direction.
It is easy to recognize that in terms of VG magnitude ($A \equiv \sqrt{A_{ij}A_{ij}}$) and invariants, the parameters $\gamma$ and $e$ are:
\begin{equation}
    \gamma = \sqrt{\frac{1}{2}+\frac{Q}{A^2}} \;\;, \;\; e = \sqrt{\frac{1}{2}-\frac{Q}{A^2}} 
\end{equation}
It may be recalled here that $R=0$ for all two-dimensional streamlines. 
The shape of an ellipse is defined by the aspect ratio $E$, which is the ratio of major to minor axes of the ellipse. It can be shown that the aspect ratio of the elliptic streamlines is \citep{blaisdell1996simulation}:
\begin{equation}
    E = \sqrt{\frac{\gamma+e}{\gamma-e}} = \sqrt{\frac{\sqrt{\frac{1}{2}+\frac{Q}{A^2}} + \sqrt{\frac{1}{2}-\frac{Q}{A^2}}}{\sqrt{\frac{1}{2}+\frac{Q}{A^2}} - \sqrt{\frac{1}{2}-\frac{Q}{A^2}}}}
\label{eq:AR}
\end{equation}
It is evident from equation (\ref{eq:AR}) that for the same value of $Q$, different values of VG magnitude $A$ results in different aspect ratios ($E$), i.e. different shapes, of the local streamlines. 
For example, for the same value of $Q$, $A^2=2Q$ represents circular streamlines while $A^2>>Q$ represents an infinitely elongated ellipse.
Clearly, $Q$ does not uniquely describe the aspect ratio.
On the other hand the variable
\begin{equation}
     q \equiv \frac{Q}{A^2}
\end{equation}
uniquely determines the aspect ratio of the ellipse, i.e.
\begin{equation}
    E(q) = \sqrt{\frac{\sqrt{\frac{1}{2}+q} + \sqrt{\frac{1}{2}-q}}{\sqrt{\frac{1}{2}+q} - \sqrt{\frac{1}{2}-q}}}
\label{eq:Eq}
\end{equation}
Therefore, in this two-dimensional case, $q$ is the only shape-parameter and it uniquely characterizes the streamline shape.

The velocity components at any location ($x_1,x_2$) on a streamline can be obtained using equations (\ref{eq:ui2}) and (\ref{eq:Aij1}) as follows,
\begin{equation}
    \begin{split}
       & u_1 = A_{1k}x_k = - \frac{A}{\sqrt{2}} \bigg ( \sqrt{\frac{1}{2} + q} + \sqrt{\frac{1}{2} -q} \bigg) x_2  \\ 
       & u_2 = A_{2k}x_k = \frac{A}{\sqrt{2}} \bigg ( \sqrt{\frac{1}{2} + q} - \sqrt{\frac{1}{2} -q} \bigg) x_1 
    \end{split}
\label{eq:ui3}
\end{equation}
Clearly, each of the velocity components scale by a factor of $A$. Therefore, $A$ only influences the speed along the streamline, the shape of which is defined by $q$. 
From equation (\ref{eq:ui3}), the major and minor axis lengths of any given streamline are
\begin{equation}
\begin{split}
     & L_{maj} (A,q) \propto \frac{1}{A} \times {\bigg ( \sqrt{\frac{1}{2}+q}-\sqrt{\frac{1}{2}-q} \bigg ) }^{-1} \\ 
      & L_{min} (A,q) \propto  \frac{1}{A} \times {\bigg (\sqrt{\frac{1}{2}+q}+\sqrt{\frac{1}{2}-q} \bigg)}^{-1} \;  
    \label{eq:F2}
\end{split}
\end{equation}
If there are two geometrically similar ellipses of same aspect ratio $E(q)$ and different VG magnitudes - $A_1$ and $A_2$, then one can be scaled to the other by the ratio,
\begin{equation}
    \frac{L_{maj1} (A_1,q)}{L_{maj2} (A_2,q)} = \frac{L_{min1} (A_1,q)}{L_{min2} (A_2,q)} = \frac{A_2}{A_1}
\end{equation}  
Therefore, the {scale-factor} of an elliptic streamline is inversely proportional to the VG magnitude $A$. 

This example illustrates that the shape of the ellipse is not uniquely defined by $Q$ as it combines the streamline shape and scale information. 
Similarly in three-dimensional flow, the invariants $Q$ and $R$ can not uniquely define the streamline shape or scale. 
Any point in the $Q$-$R$ plane can represent multiple streamline shapes, which are homeomorphic but not necessarily \textit{similar}.


\subsubsection{\label{ssec:eqns2b}Geometric description of streamlines}

From equation (\ref{eq:ui2}), it is evident that all elements of $A_{ij}$ must be known to fully describe local streamline geometry. The locally linearized velocity vector field is given by, 
\begin{equation}
    u_i = A_{ij}x_j
\label{eq:uA}
\end{equation}
Since a streamline is always tangential to the velocity, the equation of streamline can be obtained as follows:
\begin{equation}
    d\vec{s} \times \vec{u} = 0 \;\;\;\; \text{where} \;\; d\vec{s} = dx_1 \hat{i} + dx_2 \hat{j} + dx_3 \hat{k} 
\end{equation}
Here, $d\vec{s}$ is the infinitesimal arc-length vector along the streamline. Therefore,
\begin{equation}
    (u_3dx_2 - u_2 dx_3) \hat{i} + (u_1dx_3 - u_3dx_1) \hat{j} + (u_2dx_1 - u_1dx_2) \hat{k} = 0
\end{equation}
Now, setting each vector component to zero, we obtain the following differential equations,
\begin{equation}
    \frac{dx_2}{dx_3} = \frac{u_2}{u_3} = \frac{A_{2j}x_j}{A_{3k}x_k} \;\;, \;\;\frac{dx_3}{dx_1} = \frac{u_3}{u_1} = \frac{A_{3j}x_j}{A_{1k}x_k} \;\;,\;\;\frac{dx_1}{dx_2} = \frac{u_1}{u_2} = \frac{A_{1j}x_j}{A_{2k}x_k}
\label{eq:slope1}
\end{equation}
The above differential equations can be integrated to obtain the equations describing streamlines.

Our goal is to derive a geometric description that separates shape and size features. 
We start by factorizing the VGT $\bm{A}$ into its magnitude $A$ and normalized VGT or VG structure tensor $\bm{b}$ \citep{das2019reynolds} such that
\begin{equation}
    A_{ij} = A{b_{ij}} 
\label{eq:3a}
\end{equation}
We now seek the streamline structure corresponding to the $b_{ij}$-field. We define a local velocity field given by
\begin{equation}
    u^*_i \equiv b_{ij} x_j
\label{eq:ub}
\end{equation}
The corresponding streamlines are given by, 
\begin{equation}
     d\vec{s^*} \times \vec{u^*} = 0 \;\;\;\; \text{where} \;\; d\vec{s^*} = dx^*_1 \hat{i} + dx^*_2 \hat{j} + dx^*_3 \hat{k}
\end{equation}
Thus, the $b_{ij}$-streamlines are defined by the following equations,
\begin{equation}
    \frac{dx^*_2}{dx^*_3} = \frac{u^*_2}{u^*_3} = \frac{b_{2j}x_j}{b_{3k}x_k} \;\;, \;\;\frac{dx^*_3}{dx^*_1} = \frac{u^*_3}{u^*_1} = \frac{b_{3j}x_j}{b_{1k}x_k}\;\;,\;\;\frac{dx^*_1}{dx^*_2} = \frac{u^*_1}{u^*_2} = \frac{b_{1j}x_j}{b_{2k}x_k}
\label{eq:slope2}
\end{equation} 
Using equations (\ref{eq:slope1}), (\ref{eq:3a}) and (\ref{eq:slope2}), we can write the following identities involving $A_{ij}$- and $b_{ij}$- streamlines.
\begin{equation}
    \frac{dx^*_2}{dx^*_3} \equiv \frac{dx_2}{dx_3} \;\;,\;\; \frac{dx^*_3}{dx^*_1} \equiv \frac{dx_3}{dx_1} \;\;,\;\; \frac{dx^*_1}{dx^*_2} \equiv \frac{dx_1}{dx_2}
\end{equation}  
Thus, all the internal structure information including ratios of distances and alignments of $A_{ij}$- and $b_{ij}$- streamlines are identical. Therefore, the shape-parameters of the two streamlines are the same. 
It is now clear from equation (\ref{eq:slope2}) that all the $b_{ij}$-elements are required in order to determine the streamline shape.

From equations (\ref{eq:uA}) and (\ref{eq:ub}), at any location ($\vec{x}=\vec{x}_0$) of a streamline, it can be shown that
\begin{equation}
     \vec{u} = \bm{A}\vec{x}_0 = A \; \bm{b} \vec{x}_0 = A \vec{u^*} \implies  \vec{u}= A \vec{u^*} 
\end{equation}
Therefore, it is clear that the velocities of $A_{ij}$- and $b_{ij}$- streamlines only differ in magnitude by a factor of $A$. 
As shown for the case of two-dimensional elliptic streamlines, this implies that the streamline scale-factor is inversely proportional to $A$.
From the analysis so far, we conclude the following:
\begin{enumerate}
    \item The shape features of the streamline geometry are entirely contained in $b_{ij}$.
    \item All independent elements of $b_{ij}$ are required to completely describe the streamline geometric shape.
    \item The VG magnitude $A$ determines the streamline scale-factor and is inversely proportional to the scale-factor.
\end{enumerate}

Now we examine the properties of normalized VGT ($\bm{b}$) in order to understand the shape characteristics of local streamline geometry.
The second and third invariants of $\bm{b}$,
\begin{equation}
    q=-\frac{1}{2}b_{ij} b_{ji} = \frac{Q}{A^2} \;\;\;\; \text{and} \;\;\; r = -\frac{1}{3}b_{ij}b_{jk}b_{ki} = \frac{R}{A^3}
\label{eq:3}
\end{equation}
classify streamline topology as before, and have the added advantage of separating shape from scale.
The other advantage is that $b_{ij}$-elements are bounded as follows:
\begin{equation}
-\sqrt{\frac{2}{3}} \leq b_{ij} \leq \sqrt{\frac{2}{3}} \;\; \forall \;\; i=j \;\;\; \text{and} \;\;\; -1 \leq  b_{ij} \leq 1  \;\; \forall \;\; i \neq j
\end{equation}

We now seek the smallest set of independent $b_{ij}$-elements or parameters required to characterize the normalized VGT and hence the local streamline shape. 
The tensor $\bm{b}$ can be decomposed into its symmetric and anti-symmetric counterparts as follows
\begin{equation}
    \bm{b} = \bm{s} + \bm{w}\;, \;\;\; \text{where} \;\;  s_{ij}=(b_{ij}+b_{ji})/2 \;\; \text{and} \;\; w_{ij}=(b_{ij}-b_{ji})/2
\label{eq:4}
\end{equation}
Here, $s_{ij}$ is normalized strain-rate tensor and $w_{ij}$ is normalized rotation-rate tensor. 
Since the orientation of streamlines with respect to the laboratory frame of reference is immaterial for shape description, the tensor components can be considered in any coordinate system of choice to simplify geometric shape description.
Therefore, for present purposes, $\bm{b}$ can be expressed in the principal frame of $\bm{s}$ without any loss of generality:
\begin{eqnarray}
    \bm{b} =
    \left[
    \begin{array}{ccc}
    a_1 & 0 & 0\\
    0 & a_2 & 0\\
    0 & 0 & a_3
    \end{array}\right]
    + 
    \left[
    \begin{array}{ccc}
    0 & -\omega_3 & \omega_2\\
    \omega_3 & 0 & -\omega_1\\
    -\omega_2 & \omega_1 & 0
    \end{array}\right]
	\;\;\; \text{where} \;\; a_1 \geq a_2 \geq a_3
\label{eq:5}
\end{eqnarray}
Here, $a_i$ are the normalized strain-rates, i.e. eigenvalues of tensor $\bm{s}$.
For incompressible flow, $a_1 (> 0)$ is the most expansive strain-rate, $a_3 (< 0)$ is the most compressive strain-rate and $a_2$ can be positive, negative or zero. 
The corresponding eigenvectors - $\vec{E}_{a_1}$, $\vec{E}_{a_2}$ and $\vec{E}_{a_3}$ - are mutually orthogonal and constitute the principal directions of the symmetric tensor $\bm{s}$.
Further, $\omega_i$ are the components of normalized vorticity vector ($\vec{\omega}$) along the strain-rate eigen-directions. 
Therefore, these six parameters - $a_1$, $a_2$, $a_3$, $\omega_1$, $\omega_2$ and $\omega_3$ - completely define the normalized VGT and thence the geometric-shape of the local streamlines.

We now seek to further reduce the six parameters into the smallest independent set.
First, we apply the incompressibility condition:
\begin{equation}
    a_1 + a_2 + a_3 = 0 \;\; \implies a_3 = -(a_1+a_2)
    \label{eq:6}
\end{equation}
Then, due to normalization of VGT, we have,
\begin{equation}
         b_{ij}b_{ij} = s_{ij}s_{ij}+ w_{ij}w_{ij} = a_1^2 + a_2^2 + a_3^2 + 2(\omega_1^2 + \omega_2^2 + \omega_3^2) = 1
    \label{eq:7}
\end{equation}
These constraints reduce the system to a total of four functionally independent parameters, which completely specify $\bm{b}$ in the principal frame of $\bm{s}$ and therefore determine the exact local streamline shape.
It is expeditious to choose the frame-independent invariants, $q$ and $r$, as two of the four parameters.

Following the work of \cite{ashurst1987alignment}, which demonstrates a preferential alignment of vorticity vector with the intermediate strain-rate eigenvector in a turbulent flow field,
we choose the other two parameters to be the intermediate strain-rate eigenvalue $a_2$ and the vorticity component along the intermediate strain-rate eigen-direction $|\omega_2|$. 
From equation (\ref{eq:3}), we have
\begin{equation}
    -\frac{1}{2}b_{ij} b_{ji} = \frac{1}{2}(w_{ij}w_{ij}-s_{ij}s_{ij}) = q
    \label{eq:8}
\end{equation}
Note that a positive $q$ implies rotation-dominated flow while a negative $q$ represents strain-dominated flow. 
Using equations (\ref{eq:7}) and (\ref{eq:8}),
\begin{equation}
    s_{ij}s_{ij} = a_1^2 + a_2^2 + a_3^2 = \frac{1}{2}-q
    \label{eq:9a}
\end{equation}
Equations (\ref{eq:6}) and (\ref{eq:9a}) can be used to show that:
\begin{equation}
    a_1 = \frac{1}{2}(-a_2+\sqrt{1-3a_2^2-2q}) \;\;\;\; \text{and} \;\;\;\; a_3 = \frac{1}{2}(-a_2-\sqrt{1-3a_2^2-2q})
    \label{eq:10}
\end{equation}
Therefore, $a_1$ and $a_3$ may be calculated from the four parameter set. Next, the third invariant in equation (\ref{eq:3}) can be expanded as follows
\begin{equation}
    \begin{split}
    & r = -\frac{1}{3}b_{ij}b_{jk}b_{ki} = -\frac{1}{3}(s_{ij}s_{jk}s_{ki} + 3s_{ij}w_{jk}w_{ki}) \\
    \implies \;\;\;\;\;\;\;\; & r = - a_1a_2a_3 - a_1\omega_1^2-a_2\omega_2^2-a_3\omega_3^2
    \end{split}
    \label{eq:11}
\end{equation}
From equations (\ref{eq:7}) and (\ref{eq:8}), we also obtain,
\begin{equation}
    w_{ij}w_{ij}= 2(\omega_1^2 + \omega_2^2 + \omega_3^2) = \frac{1}{2} + q
    \label{eq:9b}
\end{equation}
Solving equations (\ref{eq:11}) and (\ref{eq:9b}) leads to:
\begin{equation}
    |\omega_1| = \frac{1}{2\sqrt{2}}\sqrt{\frac{- 8a_2^3 - 8r  + a_2(3-2q-12\omega_2^2) + (1+2q-4\omega_2^2)\sqrt{1-3a_2^2-2q}}{\sqrt{1-3a_2^2-2q}}}
    \label{eq:12a}
\end{equation}
\begin{equation}
    |\omega_3| = \frac{1}{2\sqrt{2}}\sqrt{\frac{8a_2^3 + 8r  - a_2(3-2q-12\omega_2^2) + (1+2q-4\omega_2^2)\sqrt{1-3a_2^2-2q}}{\sqrt{1-3a_2^2-2q}}}
    \label{eq:12b}
\end{equation}
Equations (\ref{eq:10}), (\ref{eq:12a}) and (\ref{eq:12b}) exhibit that $a_1$, $a_3$, $|\omega_1|$ and $|\omega_3|$ can be completely and uniquely determined in terms of $q$, $r$, $a_2$ and $|\omega_2|$, thus completely specifying the tensor $\bm{b}$.
Therefore, each combination of $q$, $r$, $a_2$ and $|\omega_2|$ represents a unique geometric shape.
These four quantities are now designated as the shape-parameters. 


\begin{figure}
\centering
\includegraphics[width=0.6\textwidth]{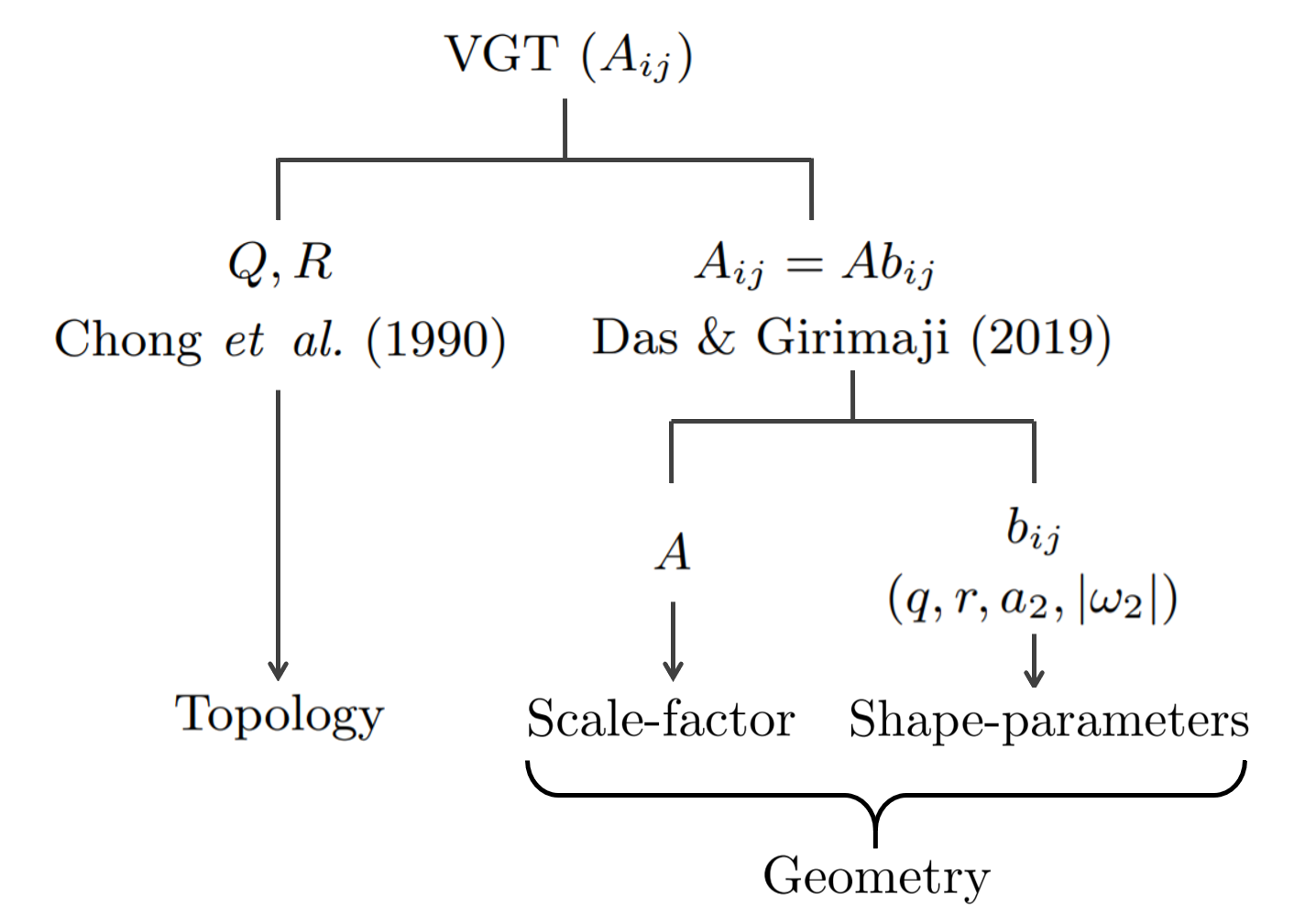}
\caption{Summary of the key points of different frameworks for studying local flow streamline structure}
\label{fig:flowchart}
\end{figure}

The streamline-shape and its scale-factor together constitute the complete geometry of the local streamlines.
Figure \ref{fig:flowchart} summarizes the important distinctions between topological and geometric descriptions of streamlines.


\subsubsection{\label{ssec:eqns2c}Kinematic bounds of shape-parameters}

We now seek to establish the bounds of the shape-parameters. 
Let us first determine the bounds of the invariant parameter $q$. From equations (\ref{eq:9a}) and (\ref{eq:9b}), we can write
\begin{equation}
    s_{ij}s_{ij} = \frac{1}{2}-q \geq 0 \;\;\;\; \text{and} \;\;\;\; w_{ij}w_{ij} = \frac{1}{2}+q \geq 0
    \label{eq:16a}
\end{equation}
leading to
\begin{equation}
    -\frac{1}{2} \leq q \leq \frac{1}{2}
    \label{eq:16}
\end{equation}
Applying the conditions $a_1 \geq a_2 \geq a_3$ in equation (\ref{eq:10}) and solving for $a_2$ in the resulting inequality equations, we obtain
\begin{eqnarray}
    -\sqrt{\frac{1-2q}{12}} \leq a_2 \leq \sqrt{\frac{1-2q}{12}}
    \label{eq:13}
\end{eqnarray}
The most expansive strain-rate ($a_1$) is non-negative by definition and attains its maximum value when $a_2$ is minimum. Thus,  
\begin{eqnarray}
    0 \leq a_1 \leq \sqrt{\frac{1-2q}{3}}
    \label{eq:13b}
\end{eqnarray}
Similarly, the most compressive strain-rate ($a_3$) has the following bounds,
\begin{eqnarray}
    -\sqrt{\frac{1-2q}{3}} \leq a_3 \leq 0
    \label{eq:13c}
\end{eqnarray}
From equation (\ref{eq:9b}) it is seen that all vorticity components have identical bounds:
\begin{eqnarray}
   |\omega_i|^2 \leq {\frac{q}{2}+\frac{1}{4}} \;\;\; \implies \;\;\;  -\sqrt{\frac{q}{2}+\frac{1}{4}} \leq \omega_i \leq \sqrt{\frac{q}{2}+\frac{1}{4}} \;\;\;\;\;\; \forall \;\;\;\; i={1,2,3}
    \label{eq:15}
\end{eqnarray}
Determining the bounds of third invariant $r$ is quite involved and the steps are not displayed here.
Substituting the upper bounds of $\omega_1$ and $\omega_3$ (equation \ref{eq:15}) into equations (\ref{eq:12a}) and (\ref{eq:12b})  and applying the bounds of $a_2$ and $|\omega_2|$ (equations \ref{eq:13} and \ref{eq:15}) leads to the following inequality:
\begin{equation}
    -\frac{1+q}{3} \bigg( \frac{1-2q}{3} \bigg)^{1/2} \leq r \leq \frac{1+q}{3} \bigg( \frac{1-2q}{3} \bigg)^{1/2}
    \label{eq:17a}
\end{equation}
This represents the kinematic bounds of $r$ for a given value of $q$ and therefore the boundary of the realizable region of $q$-$r$ plane. The extreme values of $r$ occur at $q=0$. Therefore, the absolute bounds of $r$ (also derived by \cite{wang2014kinematic} in a different context) are
\begin{eqnarray}
    -\frac{\sqrt{3}}{9} \leq r \leq \frac{\sqrt{3}}{9}
    \label{eq:17}
\end{eqnarray}
Equations (\ref{eq:16}), (\ref{eq:13}), (\ref{eq:15}) and (\ref{eq:17a}) define the kinematic bounds of the shape-parameters - $q$, $a_2$, $|\omega_2|$ and $r$.


\subsubsection{\label{ssec:eqns2d}Characterization of geometric-shape in $q$-$r$ plane}

\begin{figure}
\centering
\includegraphics[width=0.9\textwidth]{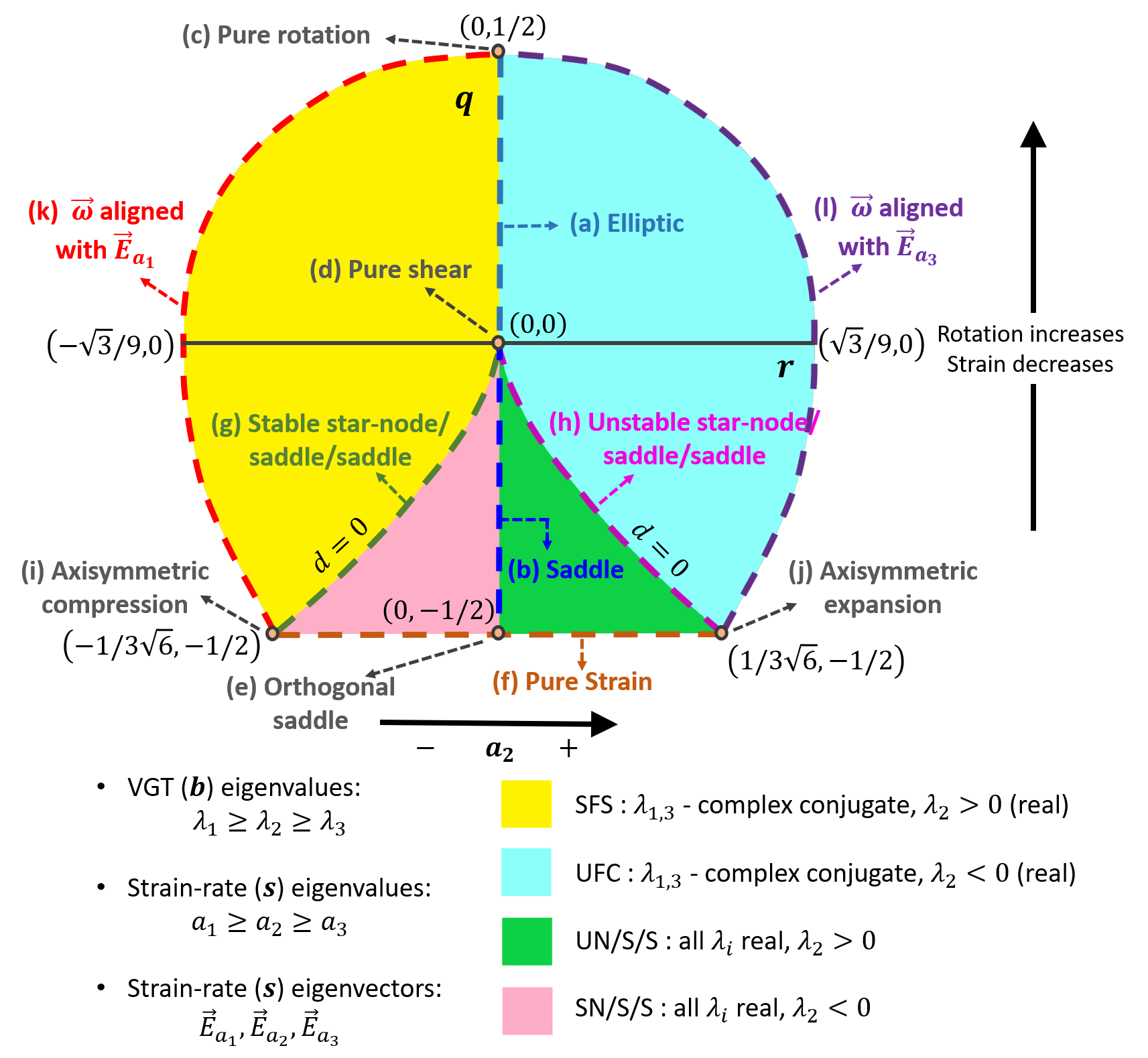}
\caption{Description of streamline shape represented by the $q$-$r$ plane}
\label{fig:qrmap}
\end{figure}

\begin{table}  
  \begin{center}
\setlength{\tabcolsep}{5pt}
  \begin{tabular}{lcc}
      Streamline shape & Location in ($q$,$r$) plane & Description \\[3pt]
      \hline
      \vspace{0.2cm}
      $(a)$ Elliptic: & $q>0, r=0$ & $\lambda_{1,3}=\pm i\lambda,\lambda_2=0$ 
      \\
      \vspace{0.2cm}
      $(b)$ Saddle: & $q<0, r=0$ & $ \lambda_{1,3}= \pm \lambda, \lambda_2=0$ \\
      \vspace{0.2cm}
      $(c)$ Pure rotation: & $q=\frac{1}{2},r=0$ & $\lambda_{1,3}=\pm i\lambda, \lambda_2=0, a_{i}=0$\\
      \vspace{0.2cm}
      $(d)$ Pure shear: & $q=0,r=0$ & All $\lambda_i=0$ \\
      \vspace{0.2cm}
      $(e)$ Orthogonal saddle: & $q=-\frac{1}{2},r=0$ & $\lambda_{1,3}=a_{1,3}$, $\vec{\omega}=0$ \\
      \vspace{0.2cm}
      $(f)$ Pure strain: & $q=-\frac{1}{2}$ & $\lambda_i=a_i$, $\vec{\omega}=0$ \\ 
      \vspace{0.2cm}
      $(g)$ \begin{tabular}{@{}l@{}}Stable star-node/ \\ saddle/saddle:\end{tabular} & left $d=0$ line & $\lambda_2=\lambda_3=-\frac{\lambda_1}{2}$\\
      \vspace{0.2cm}
      $(h)$ \begin{tabular}{@{}l@{}}Unstable star-node/ \\ saddle/saddle:\end{tabular} & right $d=0$ line & $\lambda_1=\lambda_2=-\frac{\lambda_3}{2}$\\ 
      \vspace{0.2cm}
      $(i)$ Axisymmetric compression: & $q=-\frac{1}{2}, r= -1/3\sqrt{6}$ & $\lambda_2=\lambda_3=-\frac{\lambda_1}{2}$, $\vec{\omega}=0$\\
      \vspace{0.2cm}
      $(j)$ Axisymmetric expansion: & $q=-\frac{1}{2}, r= 1/3\sqrt{6}$ & $\lambda_1=\lambda_2=-\frac{\lambda_3}{2}$, $\vec{\omega}=0$\\ 
      \vspace{0.2cm}
      $(k)$ \begin{tabular}{@{}l@{}}Orthogonal stretching \\ of stable spiral:\end{tabular} & $r=-\frac{1+q}{3} \big( \frac{1-2q}{3} \big)^{\frac{1}{2}}$ & $\vec{\omega}$ aligned with $\vec{E}_{a_1}$ if $a_2=a_3=-\frac{a_1}{2}$\\
      \vspace{0.2cm}
      $(l)$ \begin{tabular}{@{}l@{}}Orthogonal compression \\ of unstable spiral:\end{tabular}  & $r=\frac{1+q}{3} \big( \frac{1-2q}{3} \big)^{\frac{1}{2}}$ & $\vec{\omega}$ aligned with $\vec{E}_{a_3}$ if $a_1=a_2=-\frac{a_3}{2}$\\

  \end{tabular}
  \caption{Streamline shapes along marked points/lines in $q$-$r$ plane of figure \ref{fig:qrmap}}
  \label{tab:qrmap}
  \end{center}
\end{table}

First, we classify the topologies in $q$-$r$ space in a manner similar to that in $Q$-$R$ space \citep{chong1990general}. 
The $r=0$ and discriminant $d = q^3 + ({27}/{4})r^2 =0$ lines divide the plane into four non-degenerate topology types. 
Readers are referred to \cite{ooi1999study} (Fig. 1) for general streamline shapes belonging to the different topologies.
The difference here is that $q$-$r$ provides a mathematically bounded phase plane and each ($q,r$) combination represents a unique streamline shape.
For the sake of completeness, we reiterate this classification in the context of the present framework. 

In focal streamlines ($d>0$), $\bm{b}$ has one real ($\lambda_2$) and two complex conjugate ($\lambda_{1,3}$) eigenvalues:
\begin{enumerate}
  \item Stable focus stretching or SFS streamlines ($r<0$) spiral towards a stable focus while stretching out of the focal plane. 
  \item Unstable focus compression or UFC streamlines ($r>0$) spiral away from the center while being compressed into the focal plane.
\end{enumerate}

In nodal streamlines ($d<0$), $\bm{b}$ has three distinct real eigenvalues ($\lambda_1 > \lambda_2 > \lambda_3$) and three solution planes (not necessarily orthogonal):
\begin{enumerate}
  \setcounter{enumi}{2}
  \item Stable node/saddle/saddle (SN/S/S) streamlines consist of a stable node ($\lambda_2 < 0$ or $r<0$) in one plane and saddle nodes in two planes. 
  \item Unstable node/saddle/saddle (UN/S/S) streamlines consist of an unstable node ($\lambda_2 > 0$ or $r>0$) in one plane and saddle nodes in two planes.
\end{enumerate}


The focus is now on the complete description of various streamline shapes associated with different locations in the $q$-$r$ plane.
We examine the streamline shapes at several lines and points of geometric significance in the $q$-$r$ plane as marked in figure \ref{fig:qrmap}.  
These corresponding streamline shape descriptions are listed in table \ref{tab:qrmap}. 
The details of the various shapes are further discussed below: 

\textit{Two-dimensional streamlines}: The $r=0$ line represents two-dimensional planar flow streamlines. At every point along this line the intermediate eigenvalue of $\bm{b}$ is zero ($\lambda_2=0$) but the intermediate strain-rate eigenvalue ($a_2$) is not necessarily zero. In the event that the vorticity vector ($\vec{\omega}$) is perfectly aligned with the intermediate strain-rate eigenvector ($\vec{E}_{a_2}$), $a_2$ is zero. The specific shapes at different $q$ values along this line are as follows:

\begin{enumerate}

\item \textit{Elliptic}: 
Upper half of the $r=0$ line represents closed elliptic streamlines or centers (\cite{kaplan1958ordinary} Fig. 11-10) with the aspect ratio of the ellipse dependent on the $q$-value as given in equation (\ref{eq:Eq}). 
Here, $\bm{b}$ has one real and two purely imaginary eigenvalues ($\lambda_2, \lambda_{1,3}=\pm i\lambda_{i}$).   

\item \textit{Pure-rotation}: The top-most point in the plane, i.e. ($q=1/2$, $r=0$), represents circular streamlines undergoing pure or solid-body rotation. A pure rotation flow is elliptic flow with aspect ratio $1$. Here, the strain-rate eigenvalues are zero, i.e. $a_1=a_2=a_3=0$, and the normalized VGT $\bm{b}$ is only composed of vorticity/rotation rate tensor ($\bm{w}$).

\item \textit{Pure-shear}: The entire $q=0$ line represents streamlines with equal contributions of strain and rotation ($s_{ij}s_{ij}=w_{ij}w_{ij}$). The $r=0$ point on this line is of particular significance since at this point all the eigenvalues of $\bm{b}$ are zero and therefore it represents only shear deformation of the local fluid element. 
The eigenvalues of $\bm{s}$ and $\bm{w}$ are individually non-zero due to the contribution of shear to each of these tensors.
Thus, the origin of the $q$-$r$ plane represents pure-shear parallel streamlines (\cite{kaplan1958ordinary} Fig. 11-11).

\item \textit{Saddle}: Any point on the lower half ($q<0$) of the $r=0$ line has two real equal and opposite eigenvalues ($\lambda_{1,3}=\pm \lambda$) of $\bm{b}$. 
These points represent open hyperbolic streamlines, constituted by a saddle point with compression in one eigen-direction and expansion in the other (\cite{kaplan1958ordinary} Fig. 11-4). These eigenvectors of $\bm{b}$ are in general oblique and become progressively orthogonal as $q$ approaches its lower limit.

\item \textit{Orthogonal saddle}: At the bottom-most point ($q=-1/2$, $r=0$) on this line, there is no vorticity and the two real eigenvectors of $\bm{b}$ are perpendicular to each other.
This results in a two-dimensional orthogonal saddle with the compressing streamlines perpendicular to the expanding streamlines.  

\end{enumerate}

\textit{Pure-strain}: The entire bottom-most line ($q=-1/2$) represents three-dimensional nodal streamlines with three orthogonal solution planes ($\bm{b}$-eigenvectors) of compression and expansion. This is due to the fact that along this line streamlines have zero vorticity ($\vec{\omega}=0$ and $w_{ij}=0$) and thus, $\bm{b}$ ($= \bm{s}$) is a symmetric tensor. The $q=-1/2$ line, therefore, represents pure-strain streamlines.

\textit{Star-node/saddle/saddle}: The zero discriminant lines demarcating the focal streamlines from nodal streamlines also have a specific streamline shape. At any point on the left $d=0$ line, the tensor $\bm{b}$ has one positive real eigenvalue, $\lambda_1$, and two equal negative real eigenvalues, $\lambda_2=\lambda_3=-\lambda_1/2$. 
This results in a stable symmetrical node or star-node, i.e. straight streamlines directed towards the critical point (see \cite{kaplan1958ordinary} Fig. 11-9a), in one of the three eigenvector planes and saddles in the other two eigenvector planes. 
Similarly, on the right $d=0$ line, $\bm{b}$ has two equal positive eigenvalues, $\lambda_1=\lambda_2=-\lambda_3/2$, thus representing an unstable symmetrical node or star-node (straight streamlines directed away from the critical point) in one of the solution planes and saddles in the other two.

\textit{Axisymmetric compression/expansion}: A special case of the above mentioned shape occurs at the points of intersection of the $d=0$ lines with the pure-strain line, i.e. at the corner points ($q=-1/2,r= \pm 1/3\sqrt{6}$) of the $q$-$r$ plane. Due to the orthogonality of the eigenvector planes, the bottom-left corner of the plane represents axisymmetric compression accompanied by twice as stronger expansion perpendicular to it, forming an elongated tube-like streamline structure.
And the bottom-right corner represents axisymmetric expansion with twice as stronger compression perpendicular to it, forming a flatter disk-like streamline geometry.

\textit{Orthogonal focal stretching/compression}: Next we emphasize on the significance of the left and right boundaries of the $q$-$r$ plane as given by equation (\ref{eq:17a}). When $r$ is at its lower bound for a given $q$ (left boundary), $\vec{\omega}$ is perfectly aligned with the most expansive strain-rate eigenvector ($\vec{E}_{a_1}$) provided $a_1$ is equal to its upper kinematic limit for that $q$ value (or $a_2$ is equal to its lower kinematic limit, implying $a_2=a_3=-a_1/2$). 
In other words, when:
\begin{equation}
     \omega_1 = \omega = \sqrt{\frac{1}{4}+\frac{q}{2}} \;,\; \omega_2=\omega_3=0 \;\;\; \text{and} \;\;\; a_2 = -\sqrt{\frac{1-2q}{12}}
\end{equation}
solving equation (\ref{eq:12a}) yields,
\begin{equation}
     r = -\frac{1+q}{3} \bigg( \frac{1-2q}{3} \bigg)^{1/2} 
\end{equation}
which is the left boundary of the $q$-$r$ plane.
This result is important as it will be shown in section \ref{sec:Res1c} that in a turbulent flow field, $a_1$ achieves its maximum value along the left boundary and therefore $\vec{\omega}$ is indeed most likely aligned with $\vec{E}_{a_1}$ along the left boundary of the plane. This line therefore represents SFS streamlines stretching in a direction perpendicular to the plane of rotation/spiralling.
Similarly, the right boundary of the $q$-$r$ plane represents perfect alignment of $\vec{\omega}$ with the most compressive strain-rate eigenvector ($\vec{E}_{a_3}$) provided $a_3$ is at its most negative limit for a given $q$ value (i.e. $a_1=a_2=-a_3/2$). This line represents UFC streamlines with compression perpendicular to the focal plane. Again, the DNS data substantiates this result.

It is evident that shape-defining geometric properties, such as strain-rate eigenvalues and alignment of vorticity with strain-rate eigendirections, vary in a specific manner across the $q$-$r$ plane. 
There is scope for further characterizing this variation within the non-degenerate topologies of the $q$-$r$ plane, which will be pursued in future work. 



\subsection{\label{ssec:eqns3}Evolution of shape and scale}

We now present the governing equations of the shape parameters in a turbulent flow field. 
From Navier-Stokes equation, one can derive the following governing equations for the elements of normalized VGT components \citep{das2019reynolds},
\begin{equation}\label{eq:b1}
\frac{d{b}_{ij}}{dt'} = -{b}_{ik} {b}_{kj} + {h}_{ij}+{\tau}_{ij} + \frac{1}{3} {b}_{mk} {b}_{km} {\delta}_{ij} + {b}_{ij}  ({b}_{mk}{b}_{kn}  - {h}_{mn} - {\tau}_{mn}){b}_{mn}
\end{equation}
where $dt' \triangleq Adt $ is the normalized time increment. Here, the normalized anisotropic pressure Hessian and viscous diffusion terms, given by,
\begin{equation} \label{eq:b2}
h_{ij}=\frac{H_{ij}}{A^2}  =  \frac{1}{A^2} \bigg (- \frac{\partial^2 p}{\partial x_i \partial x_j}  + \frac{\partial^2 p}{\partial x_k \partial x_k} \frac{\delta_{ij}}{3} \bigg ) \;\; \mbox {and} \;\; \tau_{ij}=\frac{T_{ij}}{A^2}=  \frac{\nu}{A^2} \frac{\partial^2 {A}_{ij}}{\partial x_k \partial x_k}
\end{equation}
represent the non-local physics. 
Further manipulations of equation (\ref{eq:b1}) leads to the following evolution equations \citep{das2019reynolds} for the invariant parameters, $q$ and $r$, 
\begin{equation}\label{eq:b3}
\frac{dq}{dt'} = \underbrace{-3r + 2q b_{ij}b_{ik}b_{kj}}_{I} \underbrace{-h_{ij} (b_{ji} + 2q b_{ij})}_{\mathcal{P}} \underbrace{- \tau_{ij} (b_{ji} + 2qb_{ij})}_{V} 
\end{equation}
\begin{equation}\label{eq:b4}
\frac{dr}{dt'} =  \underbrace{{2}q^2 + 3rb_{ij}b_{ik}b_{kj}}_{I}  \underbrace{- \frac{4}{3}q^2 - h_{ij}(b_{ki}b_{jk} + 3rb_{ij})}_{\mathcal{P}}   \underbrace{ - \tau_{ij}(b_{ki}b_{jk} + 3rb_{ij})}_{V}
\end{equation}
Terms representing the role of different physical processes, namely inertial ($I$), pressure ($\mathcal{P}$) and viscous ($V$) contributions, toward the evolution of $q$ and $r$ are marked in the equation above.
Note that the pressure term in $r$-equation consists of an isotropic part ($= - \frac{4}{3}q^2 $) and an anisotropic part ($ = - h_{mn}(b_{im}b_{ni} + 3rb_{mn}) $). 
However, the pressure contribution in $q$-equation does not involve an isotropic component.
The inertial and isotropic pressure terms constitute the local contribution to the evolution of streamline shape, while the anisotropic pressure and viscous terms represent the non-local effects. The local terms form a closed dynamical system of equations called restricted Euler equations \citep{vieillefosse1984internal}. The non-local terms are unclosed in the system of equations and need to be modeled. Characterization of these terms is one of the principal objectives of this study.

The evolution equations for frame-dependent parameters, $a_2$ and $\omega_2$, are more complicated since they depend on the evolution of eigenvectors of strain-rate tensor. 
Readers are referred to the works of \cite{dresselhaus1992kinematics} and \cite{nomura1998structure} for these governing equations. 

The scale-factor of a streamline geometry varies inversely with $A$ and the evolution equation for $A$ is given below \citep{das2019reynolds},
\begin{equation}\label{eq:A2}
\frac{d }{dt'}(log A) = \frac{1}{2A^3}\frac{dA^2}{dt} = -{b}_{ij}{b}_{ik}{b}_{kj} + {b}_{ij}{h}_{ij} +{b}_{ij}{\tau}_{ij}
\end{equation}
This differential equation, representing the evolution of scale-factor, is clearly a function of shape-parameters and the unclosed pressure and viscous terms. 
The governing equations for shape-parameters ($b_{ij}$, equation \ref{eq:b1}), on the other hand, do not depend on scale-factor. 
Once the shape evolution equations are closed, the scale-factor equation does not require any additional modeling.

\section{\label{sec:data}Numerical simulation data}

Direct numerical simulation (DNS) datasets of incompressible forced homogeneous isotropic turbulence (HIT) have been used in this study to investigate streamline geometry and velocity gradient processes.
All the datasets have been obtained from Donzis research group at Texas A$\&$M University and have been used previously to study intermittency, anomalous exponents, Reynolds number scaling and non-linear depletion \citep{donzis2008dissipation,donzis2010short,donzis2012some,gibbon2014regimes}.
The simulations employ stochastic forcing at large scales to maintain statistical-stationarity in a periodic box of dimensions $2 \pi \times 2 \pi \times 2 \pi$. 
Three different flow cases of Taylor Reynolds numbers, $Re_\lambda=225$, $385$ and $588$ are used for the present analysis. Here, the Taylor Reynolds number is defined as 
\begin{equation} 
	Re_\lambda \equiv u'\lambda/\nu \;\;\;\;\; \text{where} \;\;  \lambda=(15\nu (u')^2/\epsilon)^{1/2}  
\end{equation}
is the Taylor Microscale and $u'$ is root-mean-square (rms) velocity, $\nu$ is kinematic viscosity, and $\epsilon = 2\nu \langle S_{ij}S_{ij} \rangle $  is dissipation rate. 
The DNS datasets have been computed on the following grids: $512^3$ (for $Re_\lambda=225$), $1024^3$ (for $Re_\lambda=385$) and $2048^3$ (for $Re_\lambda=588$).
The resolution levels of the DNS datasets are: $k_{max} \eta = 1.34$ for $Re_\lambda=225$, $1.41$ for $Re_\lambda=385$ and $1.39$ for $Re_\lambda=588$. Here $k_{max}$ is the highest wave number that can be resolved and $\eta$ is the Kolmogorov length scale.
The spatial derivatives are computed using Fourier transforms.

The results are presented in two parts. First, the probability density function of shape parameters - $q,r,a_2,|\omega_2|$ - are examined, followed by the conditional mean scale-factor in the shape-parameter space and the representation of geometric shape in the $q$-$r$ plane. Then, the VGT evolution and the contributions of different physical processes are examined using conditional trajectories in the $q$-$r$ plane.

\section{\label{sec:Res1}Statistical characterization of local streamline geometry}

Four independent shape-parameters - $q$, $r$, $a_2$ and $|\omega_2|$ - determine the local streamline-shape as shown in the previous section.
Since the magnitude of vorticity ($\omega$) is a function of $q$ (equation \ref{eq:9b}), 
we propose a new independent parameter in place of $\omega_2$:
\begin{equation}
    cos \beta = \frac{\omega_2}{\omega} = {\omega_2}\bigg /{\sqrt{\frac{1}{4}+\frac{q}{2}}} 
    \label{eq:cosb}
\end{equation}
Here, $\beta$ is the angle between vorticity and intermediate eigenvector of strain-rate. 

Due to the complexity of illustrating the probability distribution of a four-dimensional state space ($q$, $r$, $a_2$, $|cos \beta|$), joint pdfs of these parameters are presented in two-dimensional phase planes to exhibit the shape characteristics. 
Then, we examine the conditional mean scale-factor as a function of geometric-shape. 
Finally, the conditional averaged $a_2$ and $|cos \beta|$ are investigated in the frame-invariant $q$-$r$ plane. 

\subsection{\label{sec:Res1a} Probability distribution of geometric shape parameters}

\begin{figure}
\centering

\begin{tikzpicture}
\node[above right] (img) at (0,0) {\includegraphics[width=0.48\textwidth,trim={3.5cm 3cm 1cm 5cm},clip]{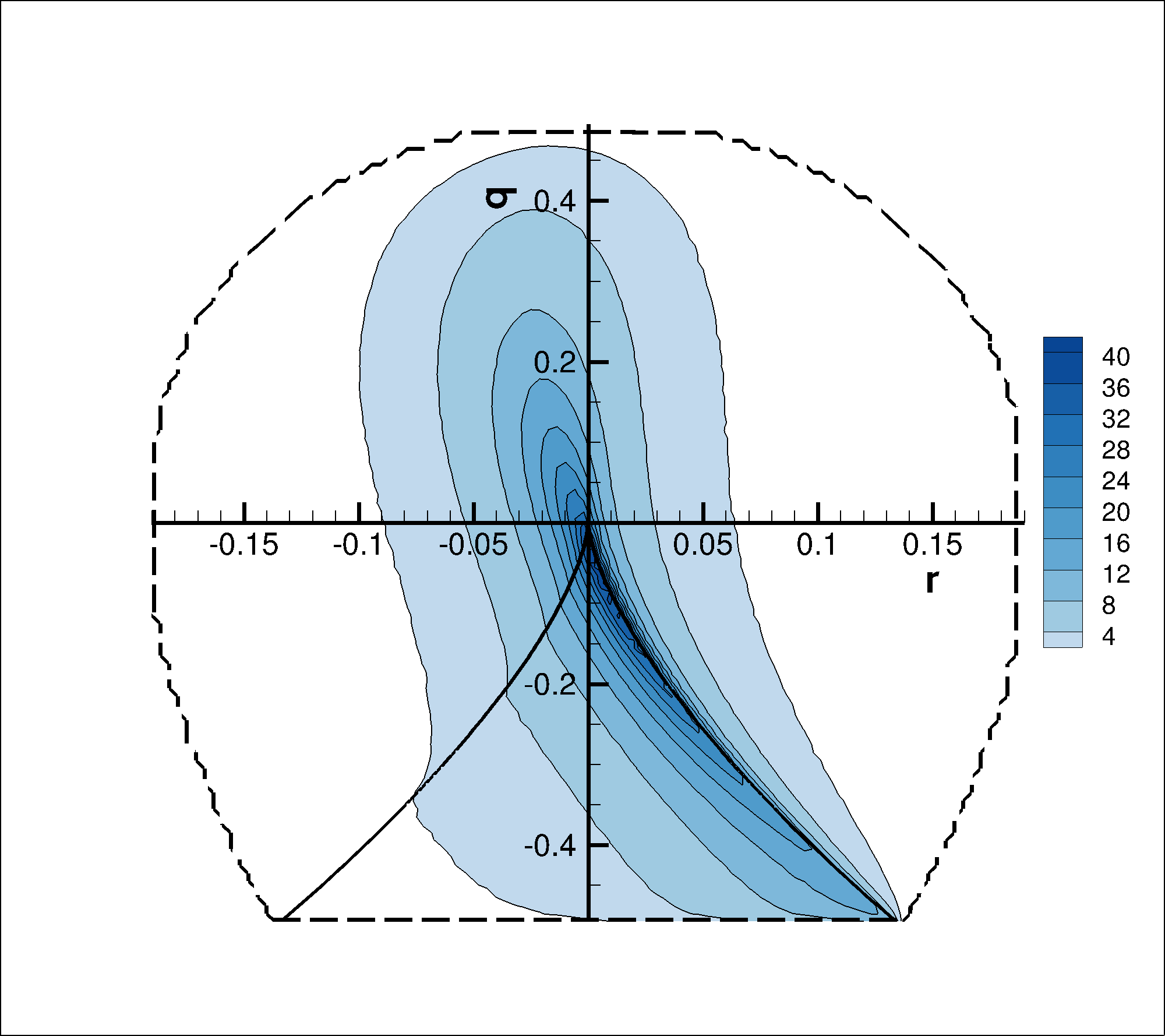}};
\node at (10pt,145pt) {(\textit{a})};
\end{tikzpicture}
\begin{tikzpicture}
\node[above right] (img) at (0,0) {\includegraphics[width=0.48\textwidth,trim={3.5cm 3cm 1cm 5cm},clip]{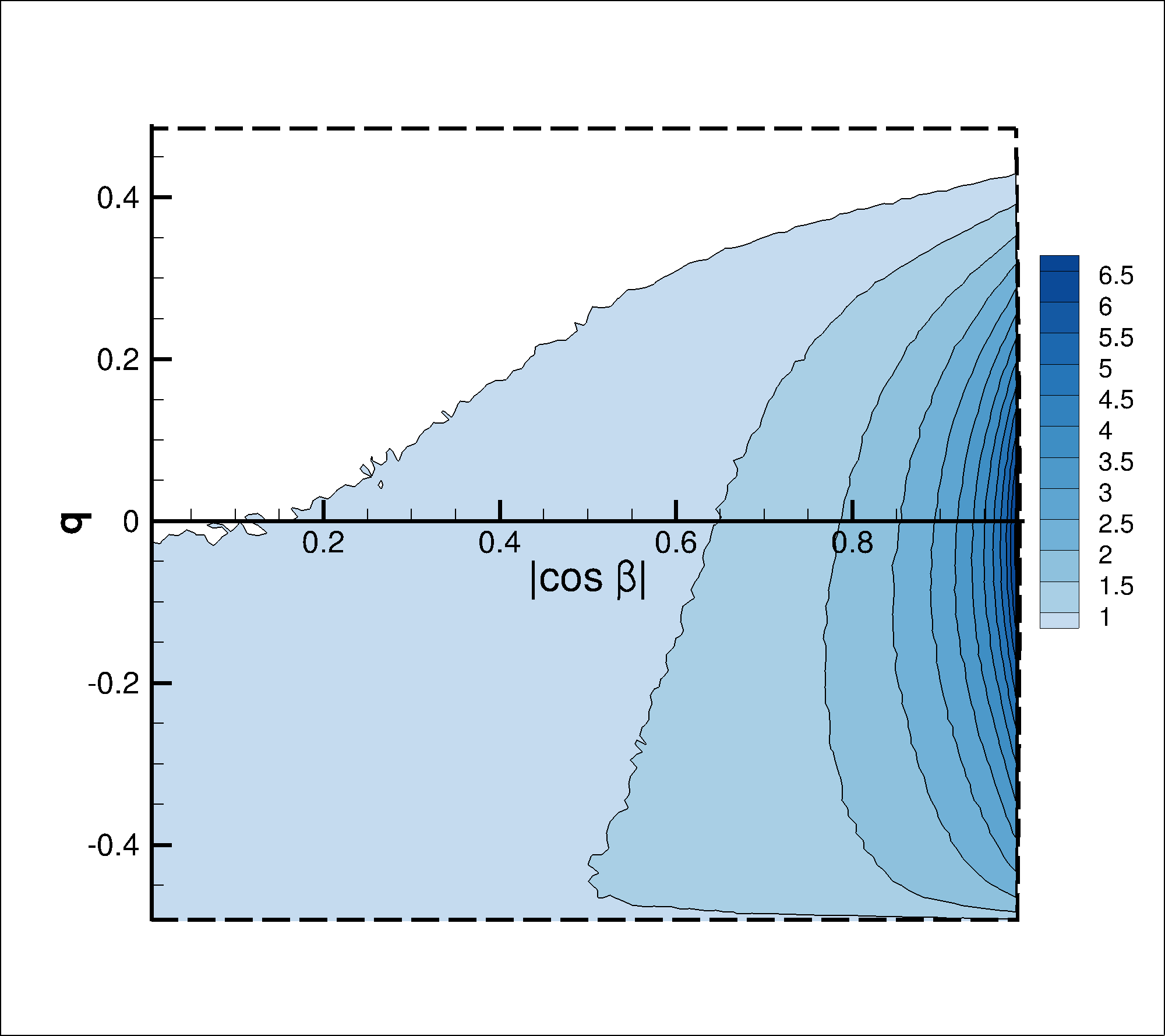}};
\node at (10pt,145pt) {(\textit{b})};
\end{tikzpicture}

\begin{tikzpicture}
\node[above right] (img) at (0,0) {\includegraphics[width=0.48\textwidth,trim={3.5cm 3cm 1cm 5cm},clip]{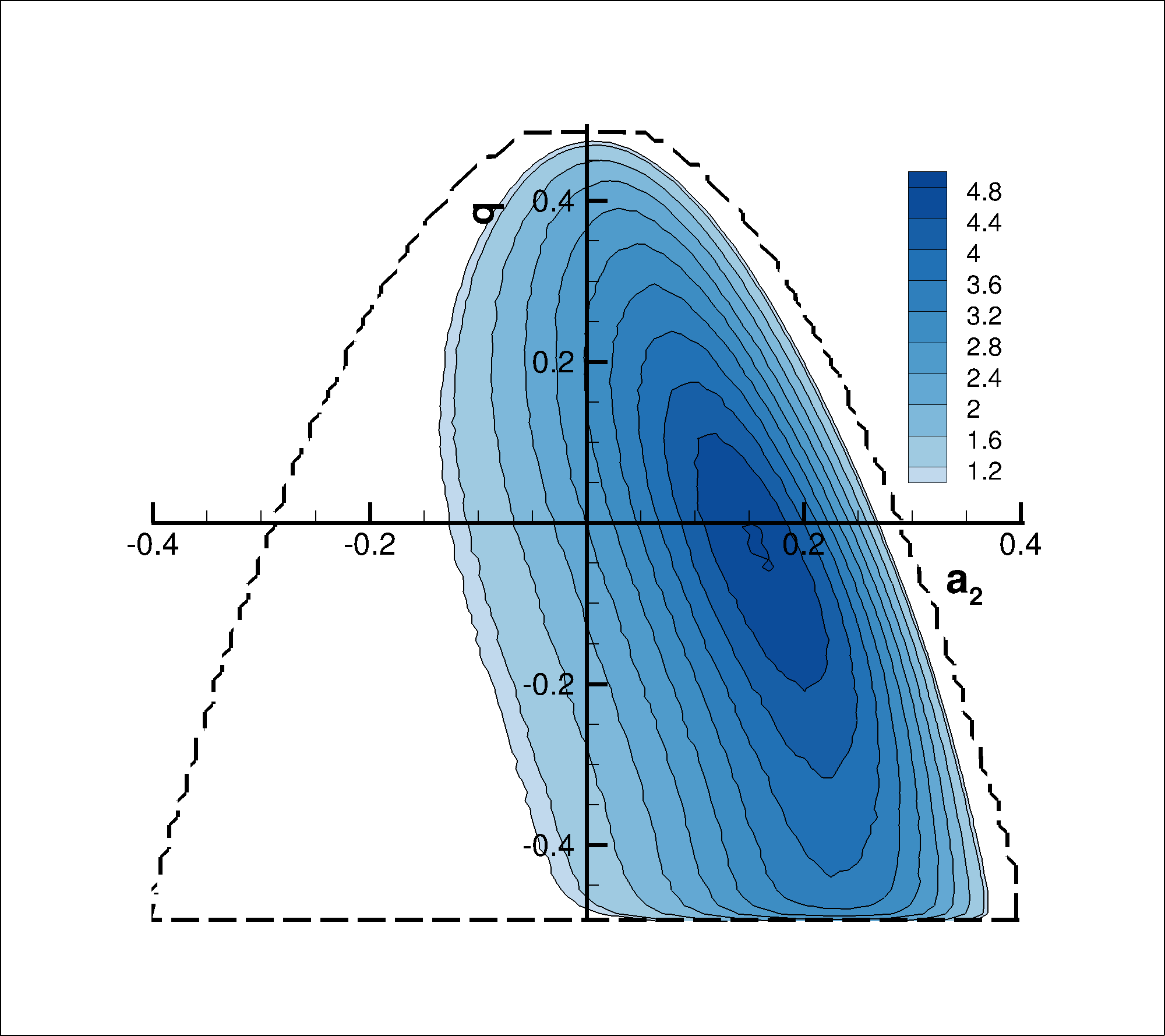}};
\node at (10pt,145pt) {(\textit{c})};
\end{tikzpicture}
\begin{tikzpicture}
\node[above right] (img) at (0,0) {\includegraphics[width=0.48\textwidth,trim={3.5cm 3cm 1cm 5cm},clip]{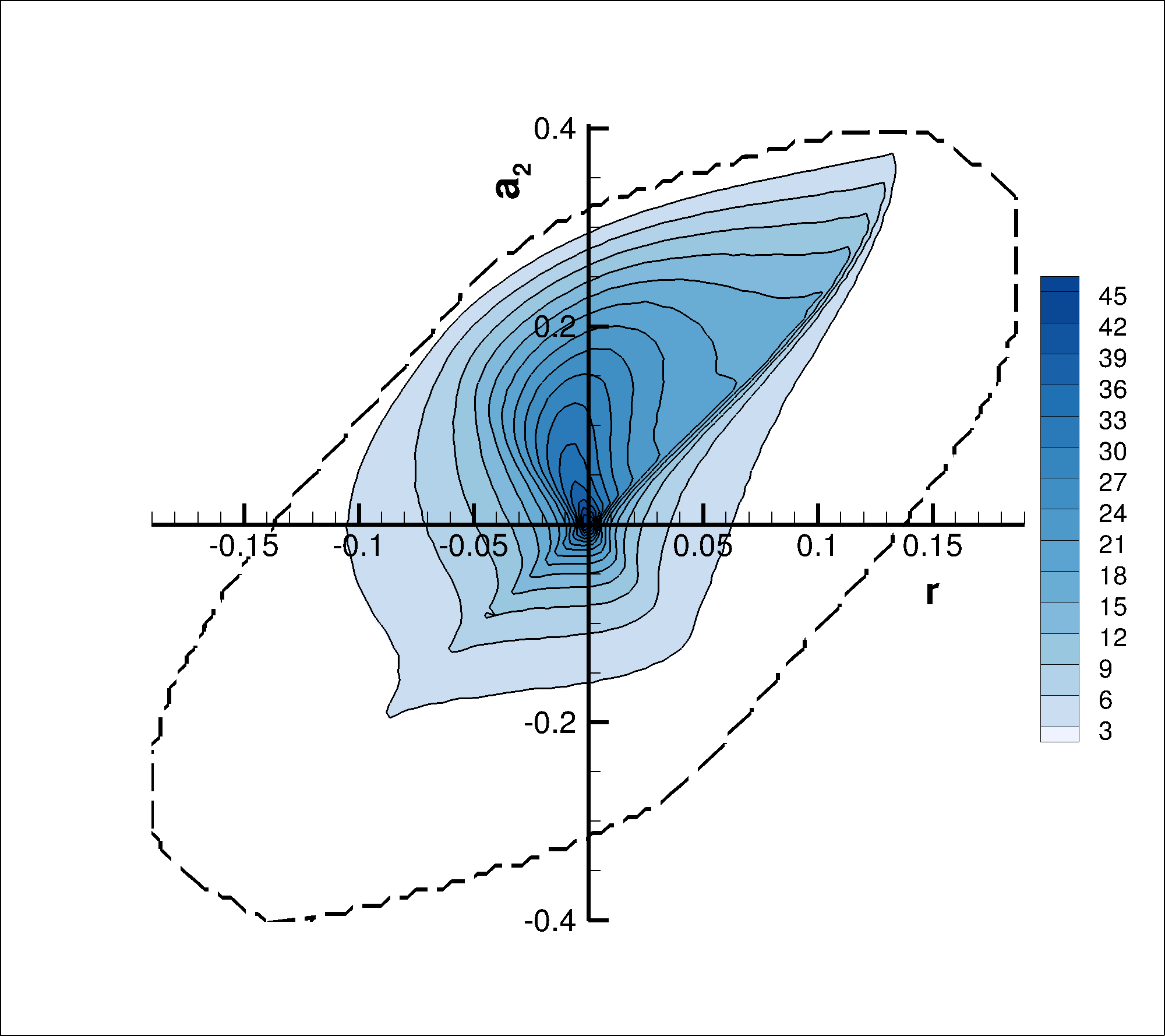}};
\node at (10pt,145pt) {(\textit{d})};
\end{tikzpicture}

\begin{tikzpicture}
\node[above right] (img) at (0,0) {\includegraphics[width=0.48\textwidth,trim={3.5cm 3cm 1cm 5cm},clip]{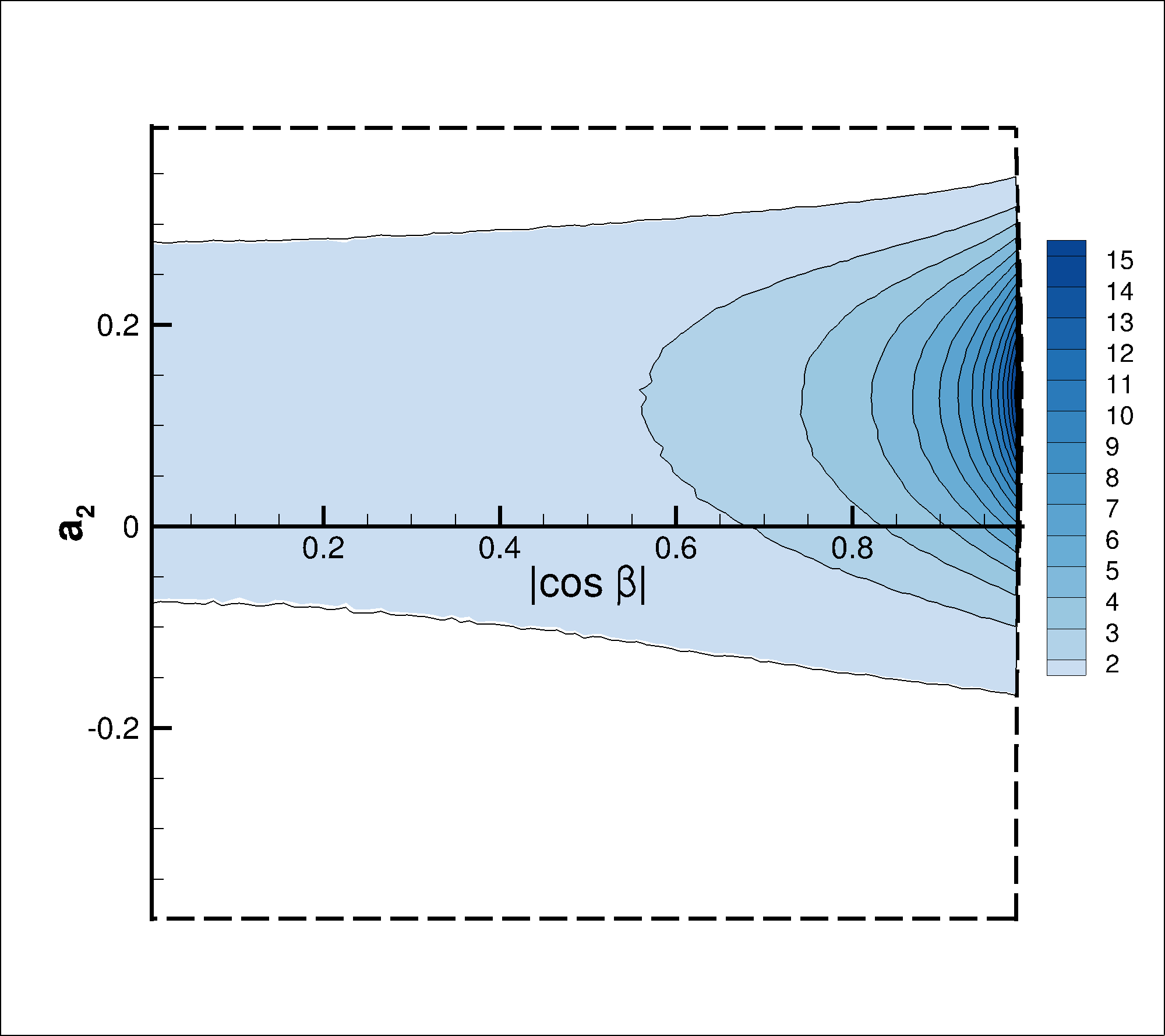}};
\node at (10pt,145pt) {(\textit{e})};
\end{tikzpicture}
\begin{tikzpicture}
\node[above right] (img) at (0,0) {\includegraphics[width=0.48\textwidth,trim={3.5cm 2.5cm 1cm 5cm},clip]{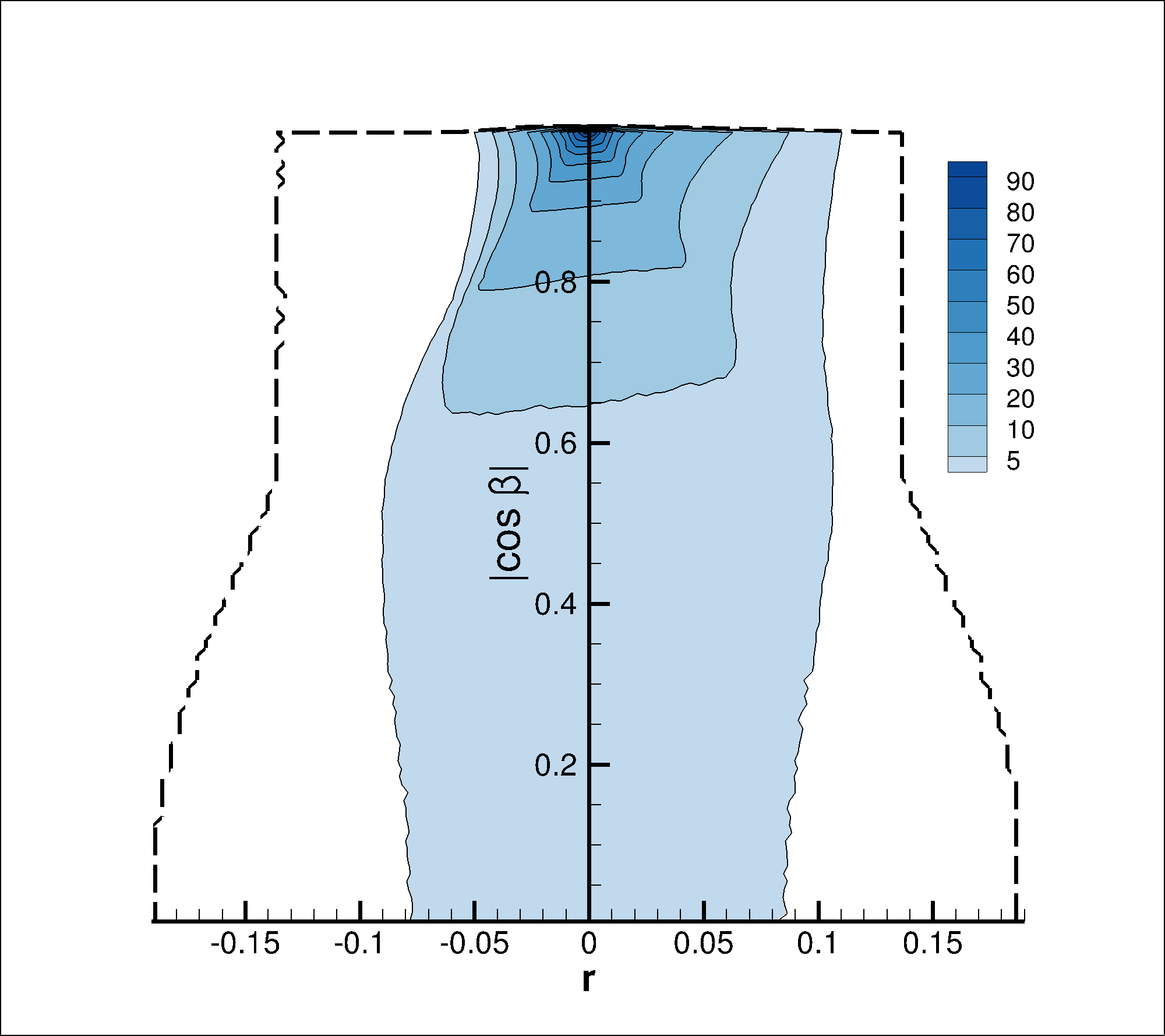}};
\node at (10pt,145pt) {(\textit{f})};
\end{tikzpicture}

\caption{Joint pdf of (a) $q$-$r$ (b) $q$-$|cos \beta|$ (c) $q$-$a_2$ (d) $a_2$-$r$ (e) $a_2$-$|cos \beta|$ (f) $|cos \beta|$-$r$ for $Re_\lambda=225$. Dashed line marks the boundary of the realizable region of the phase plane. The colored contours show $90\%$ of the field. The remaining $10 \%$ of the field occupy the region denoted by white color within the realizable phase space.}
\label{fig:phase6}
\end{figure}
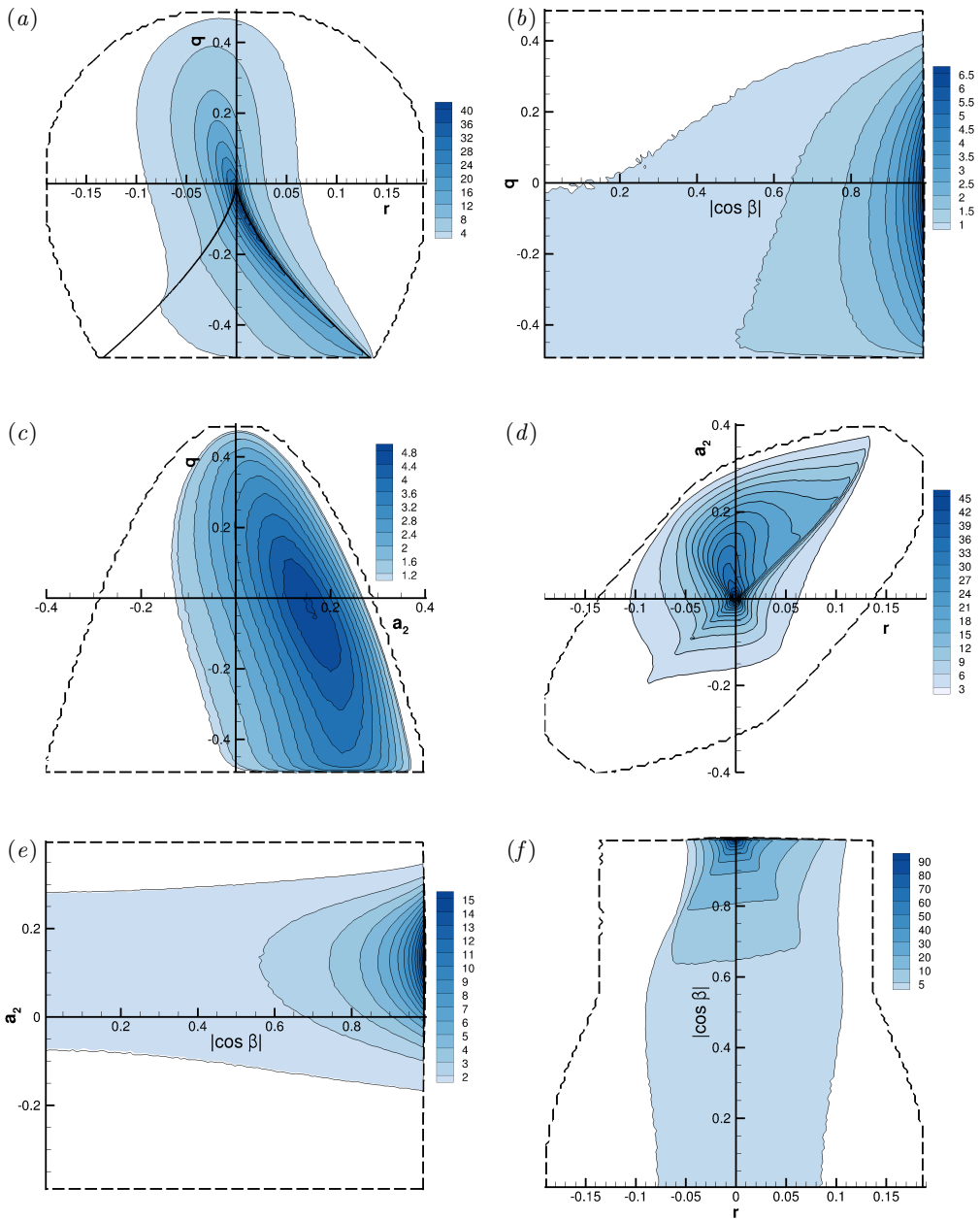

The various shape-parameter joint pdfs are plotted in figure \ref{fig:phase6} for the turbulence flow field of $Re_\lambda=225$. 
We now examine each joint pdf in detail. In what follows, we refer to the region within kinematic bounds of the phase space as the realizable space. 

\begin{enumerate}

\item $\bm{q,r}$ \textbf{pdf} (figure \ref{fig:phase6}a): 
The $q$-$r$ joint pdf resembles the teardrop shape of $Q$-$R$ but with a thicker tail, as previously shown by \cite{das2019reynolds}. 
The distribution exhibits high density along the right $d=0$ or Vieillefosse line \citep{vieillefosse1984internal,bikkani2007role}. Such behavior is also observed in the $Q$-$R$ plane \citep{soria1994study,blackburn1996topology,chong1998turbulence}. 
Joint pdf values as high as $40$ occur along the right discriminant line and decreases monotonically on either side.
Nearly $90\%$ of the total distribution occupy less than half of the total realizable area of the $q$-$r$ plane. 

\item $\bm{q,|cos \beta|}$ \textbf{pdf} (figure \ref{fig:phase6}b): The realizable region of $q$-$|cos \beta|$ phase plane is given by the kinematic bounds: $|cos \beta| \in [0,1]$ and $q \in [-1/2,1/2]$.
The upper and lower boundaries represent pure rotation and pure strain respectively. The right boundary ($|cos \beta|=1$) represents perfect alignment of vorticity with the intermediate strain-rate eigenvector while $|cos \beta|=0$ represents orthogonality between the two.
The joint pdf demonstrates that the highest probability of occurrence is when the vorticity vector is perfectly aligned with the intermediate strain-rate eigenvector, consistent with the findings of \cite{ashurst1987alignment} and other alignment studies in literature. 
The joint pdf further shows that in rotation-dominated streamlines, vorticity is likely to better align with the intermediate strain-rate eigenvector.

\item $\bm{q,a_2}$ \textbf{pdf} (figure \ref{fig:phase6}c): The $q$-$a_2$ plane is bounded within a parabolic realizable domain (equation \ref{eq:13}). 
If $a_2<0$, then expansion ($a_1$) is the strongest strain-rate while $a_2>0$ implies the strongest strain-rate is compression ($a_3$). 
The $q$-$a_2$ probability distribution is more dispersed than other phase spaces, covering a large part of the total realizable area.
It is clear that the intermediate strain-rate eigenvalue ($a_2$) is highly likely to be positive, in agreement with the results of \cite{kerr1987histograms} and \cite{ashurst1987alignment}. 
The highest probability is around $a_2 \approx 0.15$ and $q \approx 0$. 
Furthermore, the intermediate strain-rate is likely to be more expansive in the strain-dominated streamlines than in the rotation-dominated streamlines.

\item $\bm{a_2,r}$ \textbf{pdf} (figure \ref{fig:phase6}d): 
This figure suggests that a turbulent flow field is highly likely to have unstable topologies, i.e. diverging streamlines directed away from the critical point ($r>0$), with expansive intermediate strain-rate ($a_2>0$).
The peak value of the pdf occurs at $a_2 =0$ and $r=0$, which further indicates that nearly planar or two-dimensional local geometries are highly probable.
The pdf contour value is high along the $a_2= 2.17 r$ line. 

\item $\bm{a_2,|cos \beta|}$ \textbf{pdf} (figure \ref{fig:phase6}e): The $a_2$-$|cos \beta|$ plane has a rectangular realizable region with the kinematic bounds of
$-1/\sqrt{6} \leq a_2 \leq 1/\sqrt{6}$ (from equation \ref{eq:13}) and $0 \leq |cos \beta| \leq 1$. 
This joint pdf reinforces the previous observations that the local streamlines in turbulence are most likely to have positive intermediate strain-rate while vorticity is most aligned with the intermediate strain-rate eigen-direction.

\item $\bm{|cos \beta|,r}$ \textbf{pdf} (figure \ref{fig:phase6}f): 
The top and bottom boundaries of this phase plane represent alignment and orthogonality of vorticity with intermediate strain-rate eigen vector.
The left and right boundaries are obtained numerically from the DNS data set.
The probability distribution is highly concentrated around $|cos \beta|=1,r=0$ with a very high pdf value of $90$ - maximum among all the phase spaces. 
This reaffirms that vorticity is most aligned with intermediate strain-rate eigen direction in nearly planar streamlines.
The pdf value reduces nearly symmetrically with increase in $r$ magnitude on either side of the $r=0$ line.

\end{enumerate}


The joint distributions of shape-parameters at $Re_\lambda=385,588$ (not presented separately) are nearly identical to that of $Re_\lambda=225$. 
Therefore, not only $q$-$r$, but all the other shape-parameters exhibit ``universal" distributions in high Reynolds number turbulence. 

\subsection{\label{sec:Res1b}Scale-factor conditioned on shape-parameters}

\begin{figure}
\centering
\begin{tikzpicture}
\node[above right] (img) at (0,0) {\includegraphics[width=0.48\textwidth,trim={3.5cm 3cm 1cm 5cm},clip]{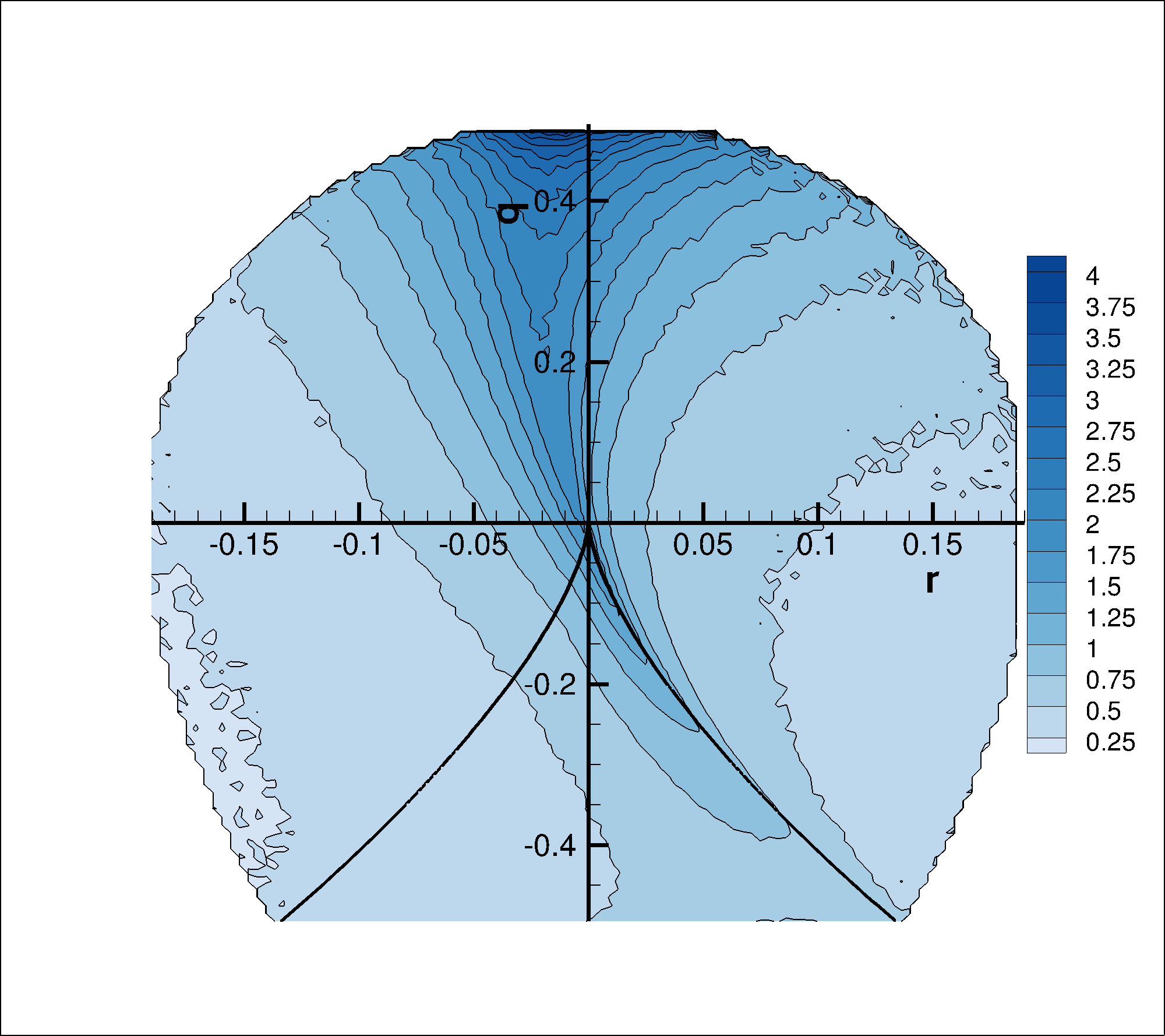}};
\node at (10pt,145pt) {(\textit{a})};
\end{tikzpicture}
\begin{tikzpicture}
\node[above right] (img) at (0,0) {\includegraphics[width=0.48\textwidth,trim={3.5cm 3cm 1cm 5cm},clip]{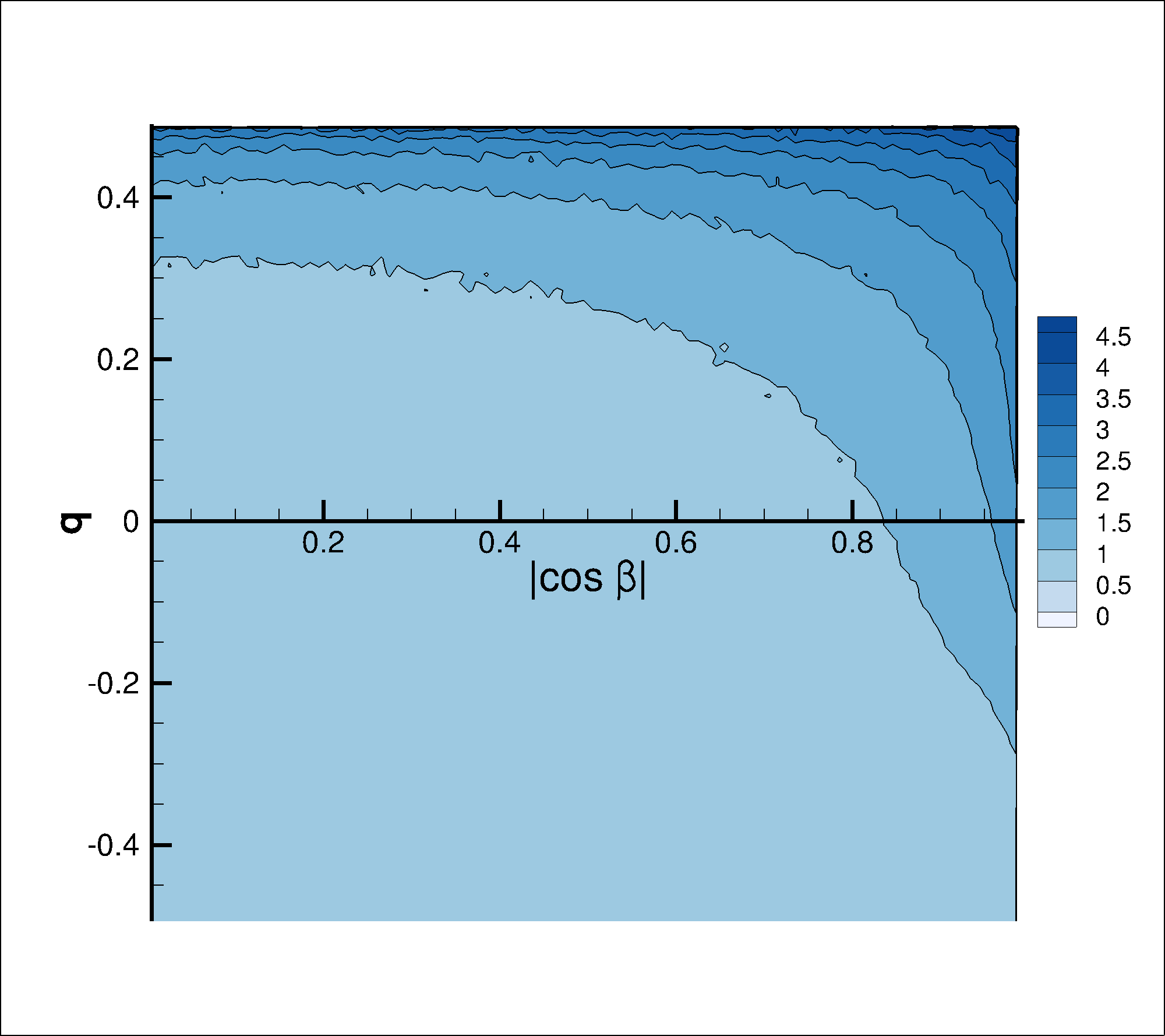}};
\node at (10pt,145pt) {(\textit{b})};
\end{tikzpicture}

\begin{tikzpicture}
\node[above right] (img) at (0,0) {\includegraphics[width=0.48\textwidth,trim={3.5cm 3cm 1cm 5cm},clip]{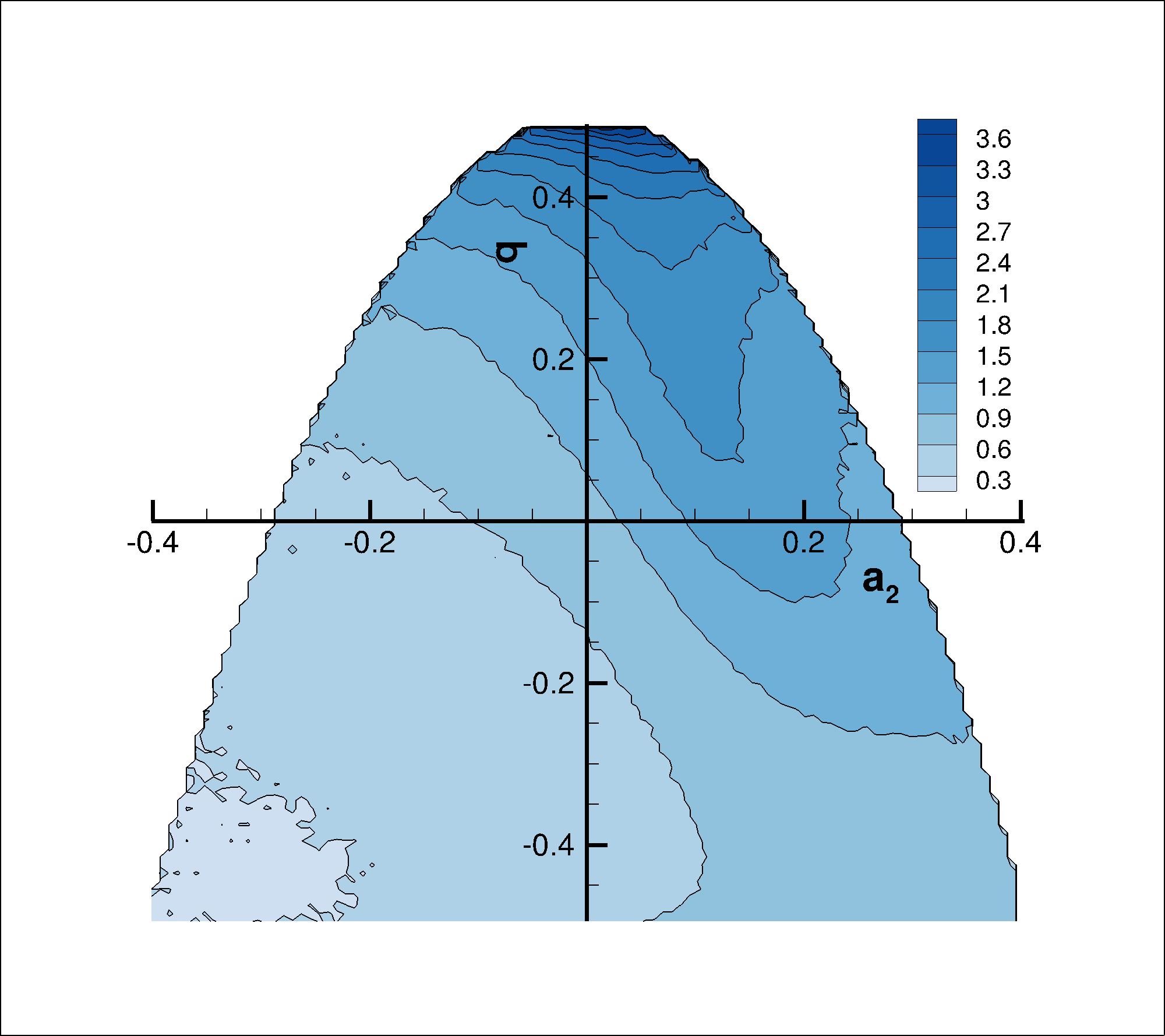}};
\node at (10pt,145pt) {(\textit{c})};
\end{tikzpicture}
\begin{tikzpicture}
\node[above right] (img) at (0,0) {\includegraphics[width=0.48\textwidth,trim={3.5cm 3cm 1cm 5cm},clip]{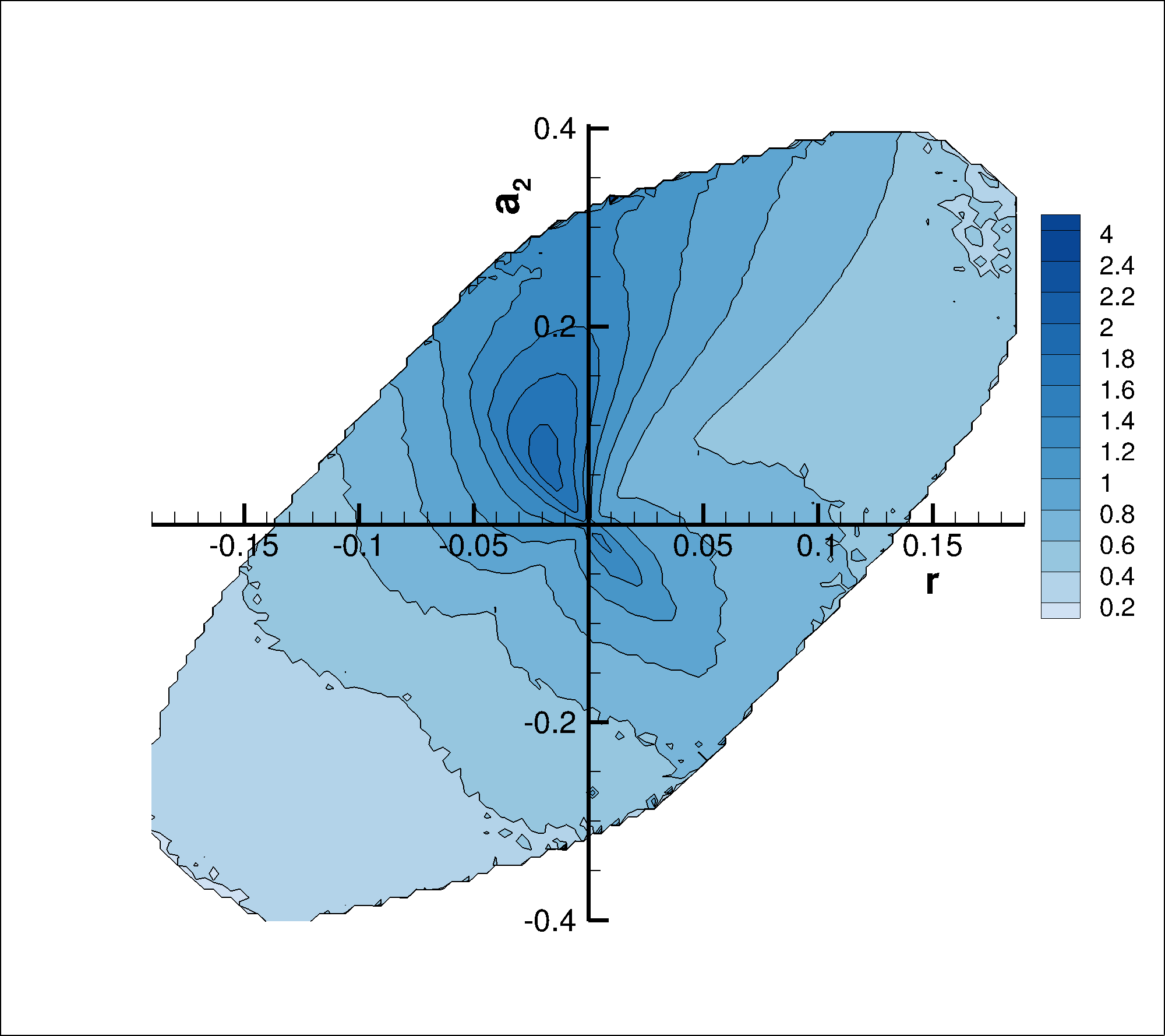}};
\node at (10pt,145pt) {(\textit{d})};
\end{tikzpicture}

\begin{tikzpicture}
\node[above right] (img) at (0,0) {\includegraphics[width=0.48\textwidth,trim={3.5cm 3cm 1cm 5cm},clip]{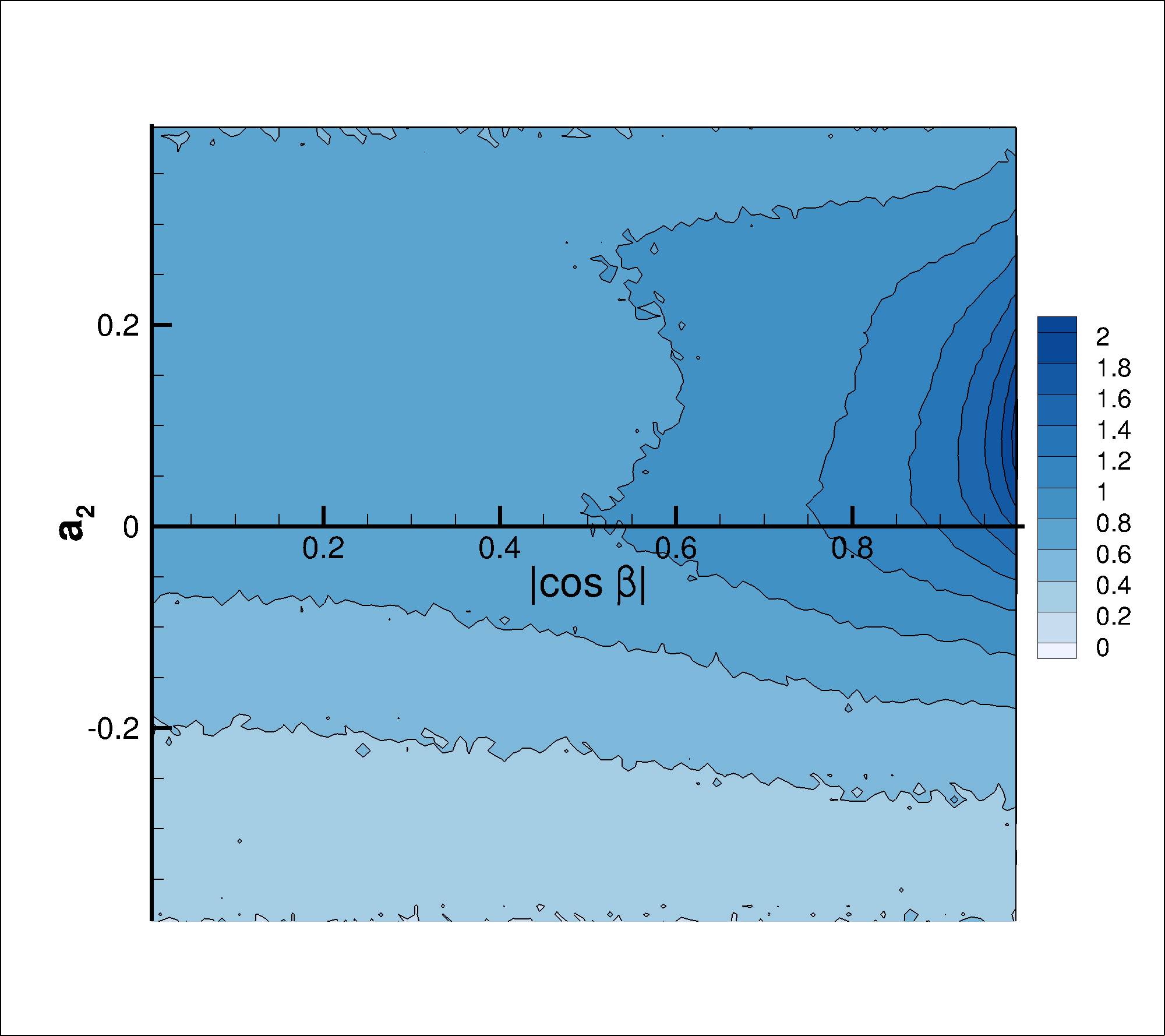}};
\node at (10pt,145pt) {(\textit{e})};
\end{tikzpicture}
\begin{tikzpicture}
\node[above right] (img) at (0,0) {\includegraphics[width=0.48\textwidth,trim={3.5cm 2.5cm 1cm 5cm},clip]{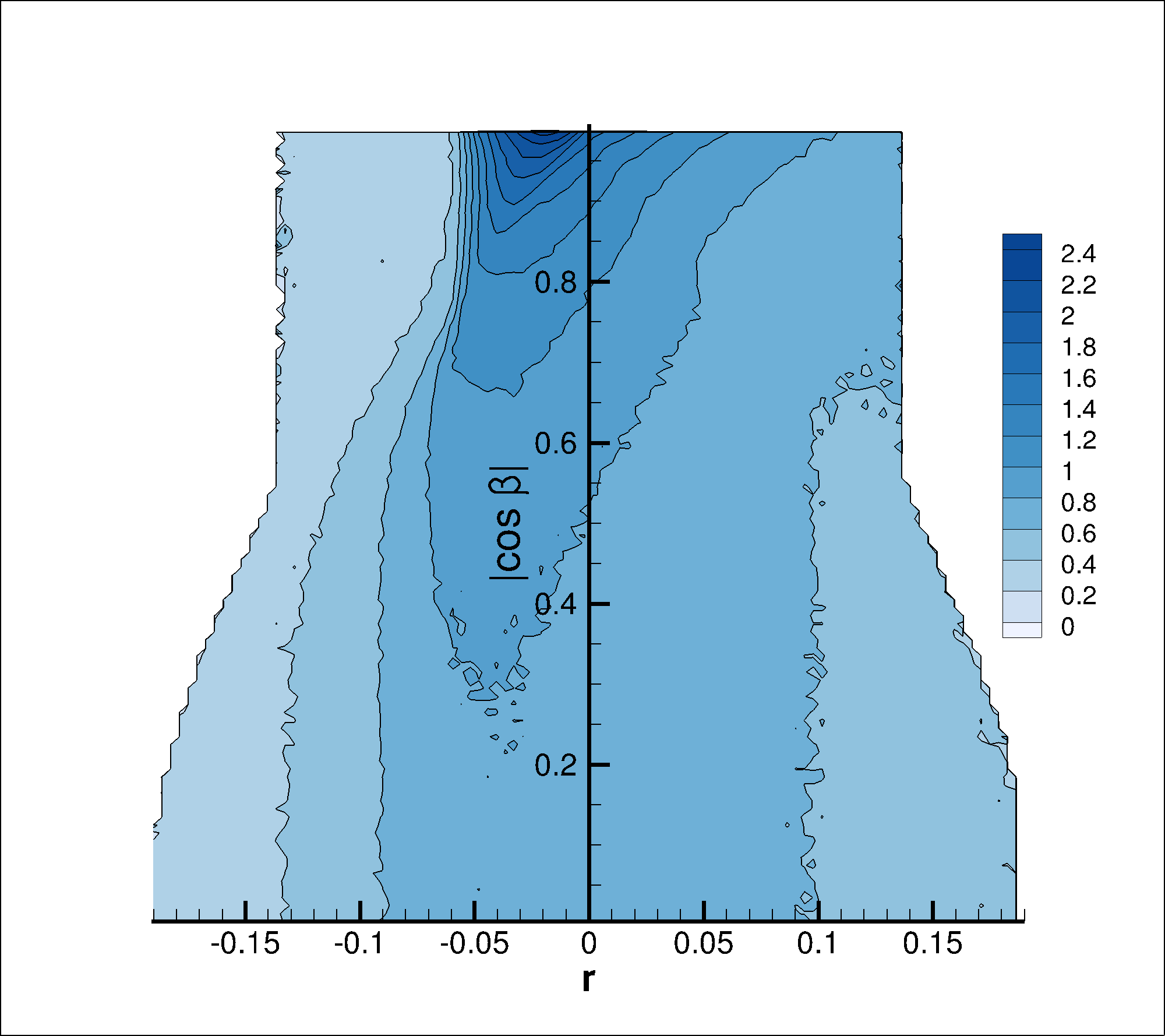}};
\node at (10pt,145pt) {(\textit{f})};
\end{tikzpicture}

\caption{Conditional mean $\langle A^2|x,y \rangle / \langle A^2 \rangle$ in (a) $q$-$r$ ($x=r$,$y=q$) (b) $q$-$|cos \beta|$ (c) $q$-$a_2$ (d) $a_2$-$r$ (e) $a_2$-$|cos \beta|$ (f) $|cos \beta|$-$r$ planes for $Re_\lambda=225$.} 
\label{fig:A2_phase6}
\end{figure}
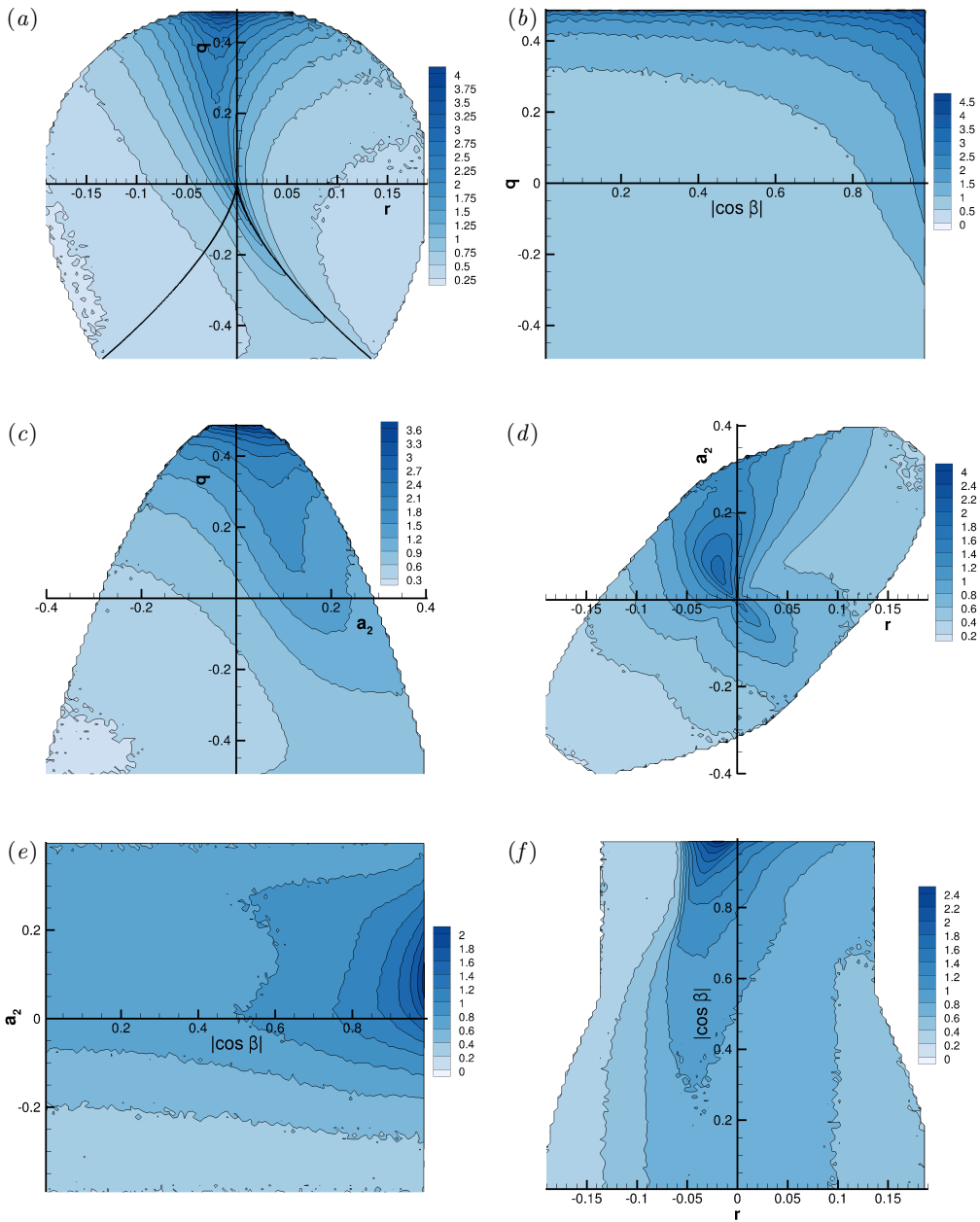

The scale-factor of a given streamline shape is defined by the Frobenius norm of VGT, $A=\sqrt{A_{ij}A_{ij}}$, as shown in section \ref{ssec:eqns2b}.
Higher magnitude $A$ implies a smaller length-scale of the streamlines.
Average VG magnitude squared $\langle A^2 \rangle$ increases monotonically with Reynolds number \citep{yeung2018effects,buaria2019extreme}.
Thus, scale-factor strongly depends on Reynolds number.
The distribution of streamline geometric shape, on the other hand, is nearly invariant with Reynolds number (section \ref{sec:Res1a}).
We now examine the dependence of scale-factor on shape. Figure \ref{fig:A2_phase6} shows the conditional mean distribution of $A^2$, normalized by global mean $\langle A^2 \rangle$, in the phase planes of shape-parameters. The primary observations from these figures are summarized below. 
\begin{enumerate}

\item Figure \ref{fig:A2_phase6} (a-c) indicates that the highest conditional mean VG magnitude or smallest scale-factor values occur when $q \approx 1/2$ (pure-rotation streamlines) and decreases progressively as $q$ decreases.

\item Figure \ref{fig:A2_phase6} (c-e) suggests that $A^2$ tends to be high when intermediate strain-rate eigenvalue is positive and in the range, $a_2 \in (0,0.2)$. 
 
\item VG magnitude is the highest when $r$ is in the slightly negative range as shown in figure \ref{fig:A2_phase6} (a,d,f). This represent stable converging streamlines directed towards the critical point.

\item Figure \ref{fig:A2_phase6} (b,e,f) illustrates that conditional mean $A^2$ is the highest when $|cos \beta| \approx 1$, i.e. vorticity is aligned with the intermediate strain-rate eigen direction. Furthermore, the VG magnitude tends to decrease monotonically with $|cos \beta|$ provided $q>0$, $a_2>0$ and $r$ is slightly negative. 

\item The conditional average in $q$-$r$ plane (figure \ref{fig:A2_phase6} a) further shows that in the rotation-dominated streamlines ($q>0$), $A^2$ is high along a curved line slightly to the left of the $r=0$ line, representing nearly planar SFS streamlines. While in the strain-dominated streamline shapes ($q<0$), $A^2$ is high along the right discriminant line, with a slight preference towards the sheet-like UN/S/S streamline shapes.

\item If $a_2<0$, then the VG magnitude increases monotonically with $a_2$ irrespective of vorticity-strainrate alignment (see figure \ref{fig:A2_phase6} e). 

\end{enumerate}


\subsection{\label{sec:Res1c}Projection of geometric-shape on $q$-$r$ plane}


\begin{figure}
\centering
\begin{tikzpicture}
\node[above right] (img) at (0,0) {\includegraphics[width=0.48\textwidth,trim={3.5cm 3cm 1cm 5cm},clip]{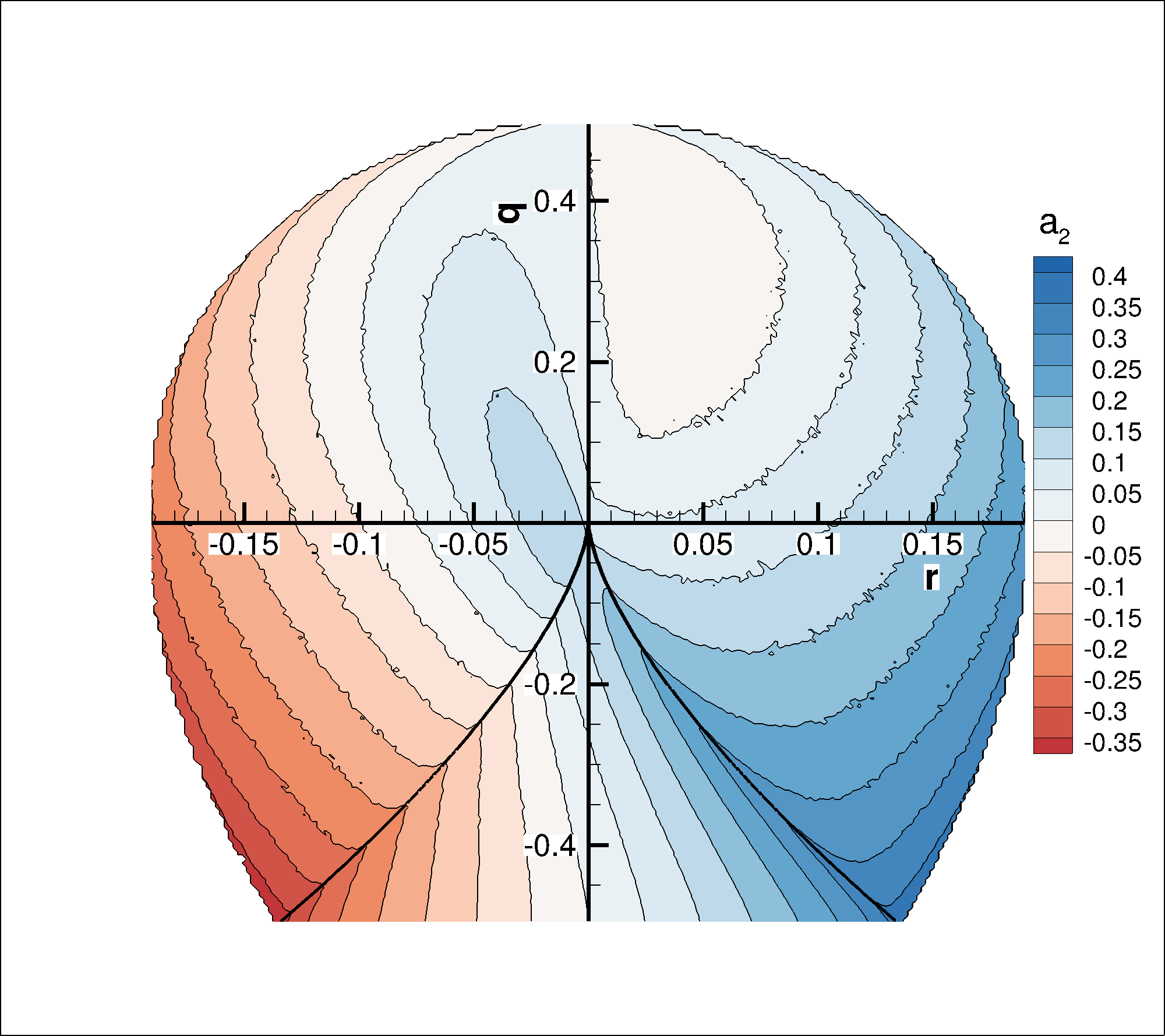}};
\node at (10pt,145pt) {(\textit{a})};
\end{tikzpicture}
\begin{tikzpicture}
\node[above right] (img) at (0,0) {\includegraphics[width=0.48\textwidth,trim={3.5cm 3cm 1cm 5cm},clip]{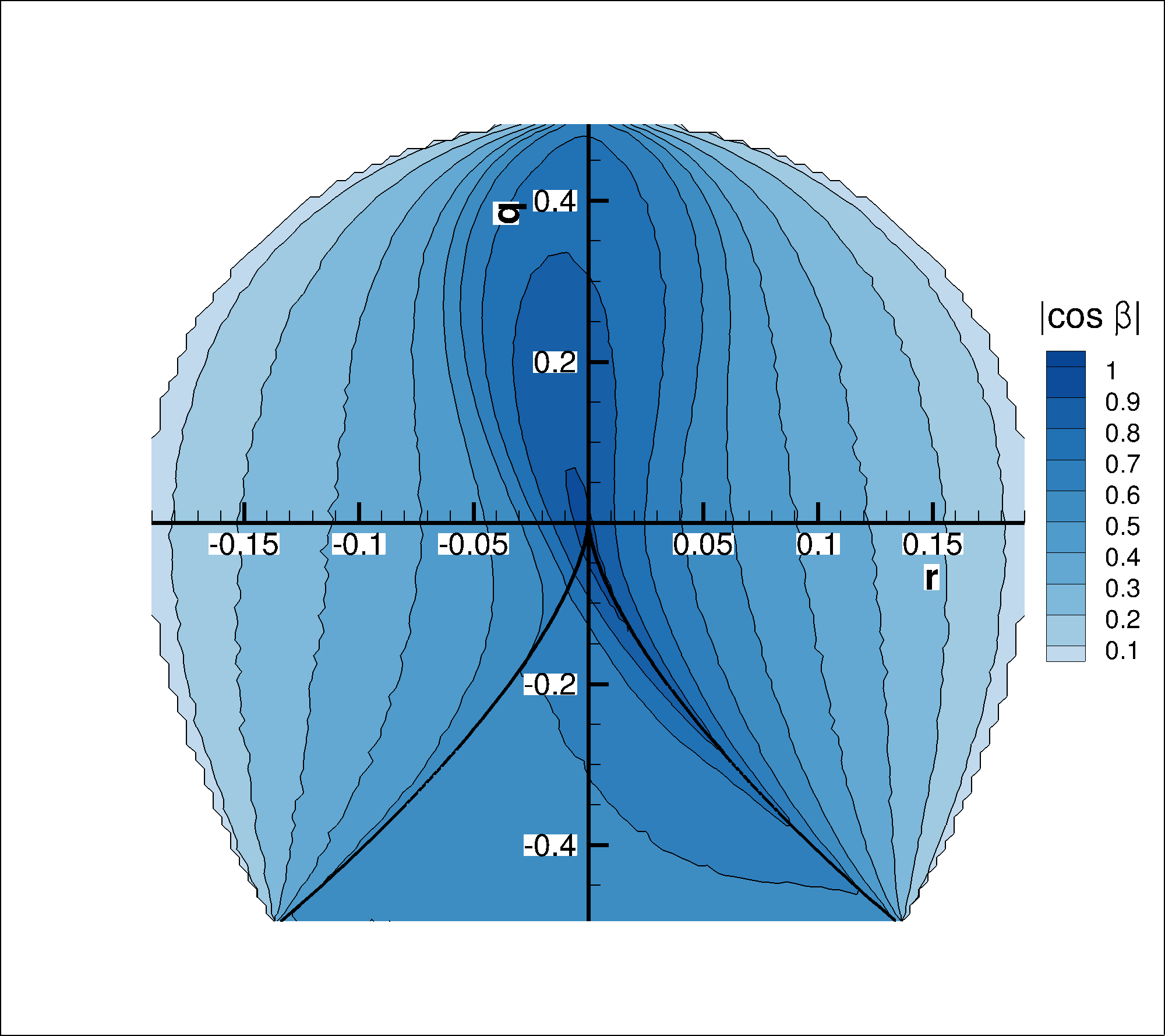}};
\node at (10pt,145pt) {(\textit{b})};
\end{tikzpicture}
\caption{Conditional average of (a) intermediate strain-rate: $\langle \; a_2\;|q,r \rangle$ and (b) angle of alignment between vorticity and intermediate strain-rate eigenvector: $\langle \; |cos \beta|\;|q,r \rangle$}.
\label{fig:a2w2}
\end{figure}

It is of interest to examine the internal alignment properties as a function of the invariants.
The conditional mean intermediate strain-rate eigenvalue ($a_2$) in the $q$-$r$ plane (figure \ref{fig:a2w2} a) shows that $a_2$ is strongly positive along the densely-populated right discriminant line, resulting in a positive global average of $a_2$ \citep{ashurst1987alignment}.
In the nodal streamline region of the $q$-$r$ plane, $a_2$ shows negligible dependence on $q$ and a monotonic increase with $r$.
At any $q$-value, $a_2$ is minimum (most negative) along the left boundary and maximum (most positive) along the right boundary of the plane.
As mentioned before, minimum $a_2$ for a given $q$ implies that $a_1$ is maximum for that $q$. Thus, the left boundary represents SFS streamlines with the maximum possible expansive strain-rate $a_1$. Similarly, the right boundary of the $q$-$r$ plane represents UFC streamlines with the maximum possible compressive strain-rate $a_3$.

As shown in section \ref{ssec:eqns2d}, vorticity vector ($\vec{\omega}$) is aligned with the most expansive strain-rate eigenvector at the left boundary and with the most compressive strain-rate eigenvector at the right boundary (figure \ref{fig:a2w2} b). 
The conditional mean $|\omega_1|$ and $|\omega_3|$ in the $q$-$r$ plane (not presented) for the DNS datasets further reaffirms this result. 
Thus, \textit{the left boundary of the $q$-$r$ plane represents orthogonal stretching of stable spiral and the right boundary represents orthogonal compression of unstable spiral}.

Figure \ref{fig:a2w2} (b) further shows that $\vec{\omega}$ is most likely to be aligned with the intermediate strain-rate direction in the region around $r=0$ line for $q>0$ and along the densely populated right discriminant line for $q<0$.
Assuming perfect alignment along the right $d=0$ line, i.e.
\begin{equation}
    \omega_2 \approx \sqrt{\frac{1}{4}+\frac{q}{2}} \;\;\; \text{when} \;\;\; r^2 + \frac{4}{27}q^3 =0 \;, \;\; q \leq 0 
\end{equation}
and solving equations (\ref{eq:10}) and (\ref{eq:11}), we obtain the following solution for $a_2$ along the right $d=0$ line,
\begin{equation}
    a_2 \approx \sqrt{-q/3}
   \label{eq:a2d0}
\end{equation}
This analytical expression of $a_2$ agrees reasonably well with the conditional mean $a_2$ values along the right $d=0$ line in figure \ref{fig:a2w2} (a).



\section{\label{sec:Res2}VGT evolution in $\bm{q}$-$\bm{r}$ space}

The objective is to characterize turbulence velocity gradient dynamics conditioned upon shape-parameters.
The evolution of streamline geometric-shape is investigated in the phase plane of frame-invariant shape-parameters - $q$ and $r$. 
Evolution in other shape parameter spaces will be considered in future work.
In addition to providing improved insight into turbulence physics, the study is also expected to serve as a foundation for developing Lagrangian velocity-gradient models.

\subsection{\label{sec:Res2a}$Q$-$R$ conditional mean trajectories (CMTs)}

For reference we first present the conditional mean trajectories (CMTs) in the $Q$-$R$ plane for the $Re_\lambda=225$ case in figure \ref{fig:CMT_total} (a).
Following the works of \cite{martin1998dynamics} and \cite{ooi1999study}, the CMTs in $Q$-$R$ space are obtained by time-integration of the conditional mean phase velocity vector field given by $\langle V_Q \equiv dQ/dt \;| Q,R \rangle$ and $\langle V_R \equiv dR/dt \;| Q,R \rangle$.
The contours in the background represent the normalized conditional mean phase velocity magnitude, 
\begin{equation}
\langle \overline{V}|Q,R \rangle = \frac{\sqrt{{\langle V_Q | Q,R \rangle}^2 + {\langle V_R |Q,R \rangle}^2}}{1/\tau_\eta} \;\;\; \text{where} \;\;\; \tau_\eta=\sqrt{\nu/\epsilon} \sim 1/\langle A\rangle 
\end{equation}
to indicate the speed of the trajectories in different parts of the plane normalized by the global Kolmogorov time scale ($\tau_\eta$). 
Consistent with the findings of \cite{martin1998dynamics}, \cite{ooi1999study} and \cite{chevillard2008modeling}, $Q$-$R$ CMTs tend to spiral in a clockwise manner around the origin, which is a stable focus of the phase-space.
Large values of $Q$ and $R$, away from the origin, implies high VG magnitude. 
The figure shows that the velocity of the trajectories is higher in regions of large $Q$ and $R$. The evolution rate slows down significantly near the origin and along the right discriminant line.
It is difficult to infer many other details from $Q$-$R$ CMTs, especially in the regions of large $Q$ and $R$.
We now demonstrate that the CMTs in $q$-$r$ space provide further information about various turbulent processes, not evident in $Q$-$R$ space.

\subsection{\label{sec:Res2a2}$q$-$r$ conditional mean trajectories (CMTs)}

\begin{figure}
\centering
\begin{tikzpicture}
\node[above right] (img) at (0,0) {\includegraphics[width=0.48\textwidth,trim={1.3cm 4cm 3cm 3cm},clip]{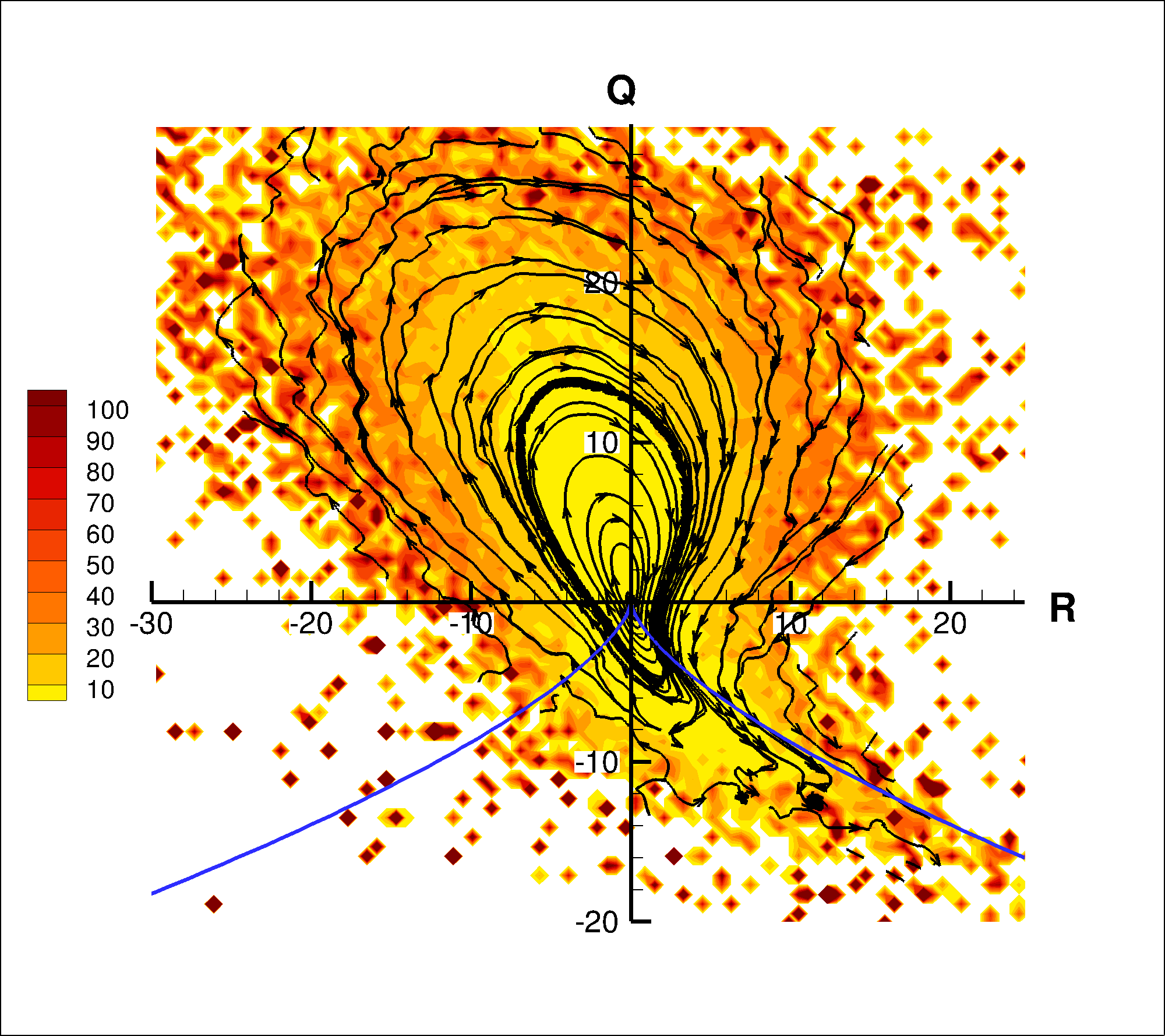}};
\node at (10pt,150pt) {(\textit{a})};
\end{tikzpicture}
\begin{tikzpicture}
\node[above right] (img) at (0,0) {\includegraphics[width=0.48\textwidth,trim={0.3cm 1.3cm 1cm 1cm},clip]{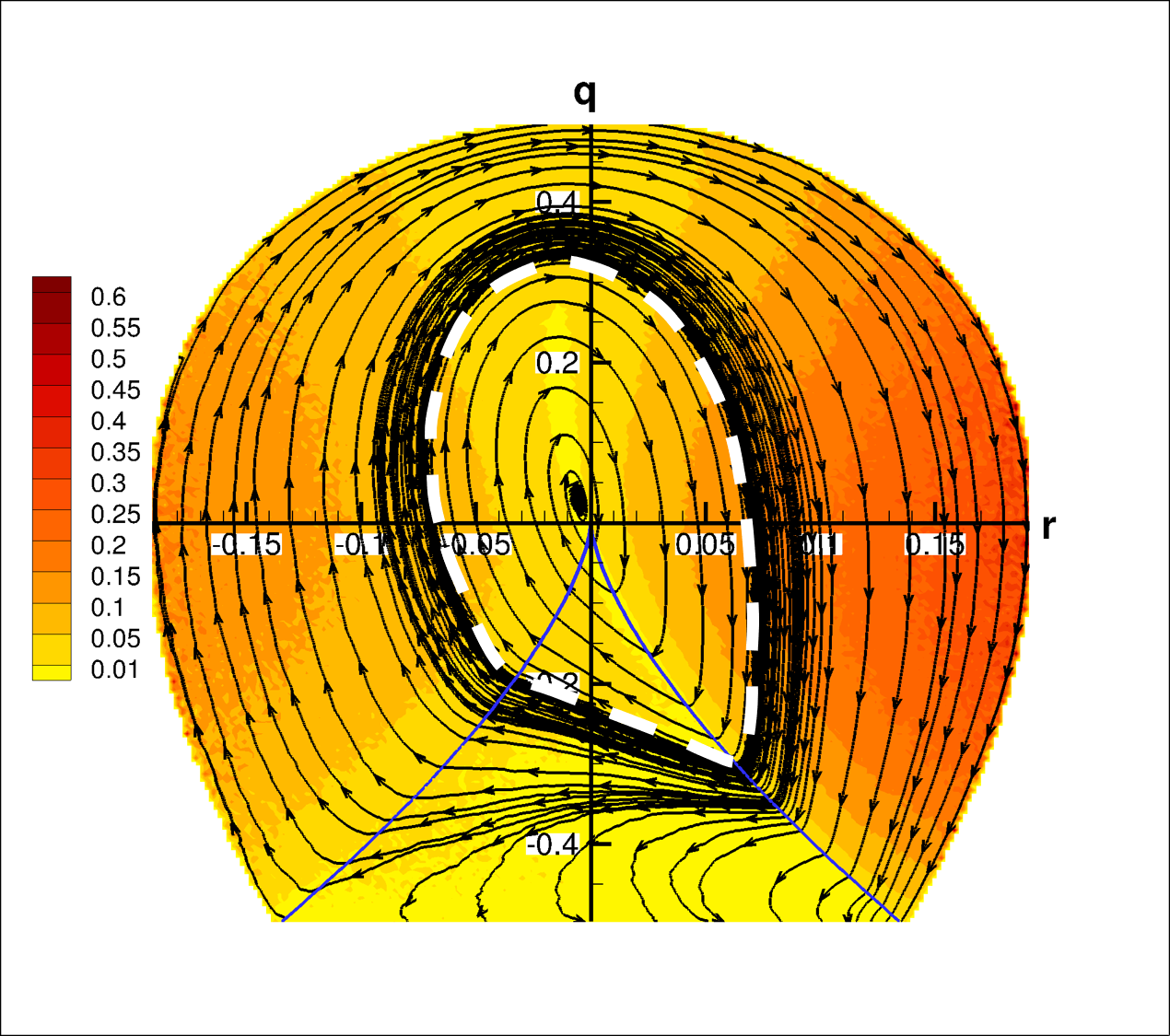}};
\node at (10pt,150pt) {(\textit{b})};
\end{tikzpicture}
\begin{tikzpicture}
\node[above right] (img) at (0,0) {\includegraphics[width=0.48\textwidth,trim={0.3cm 1.3cm 1cm 1cm},clip]{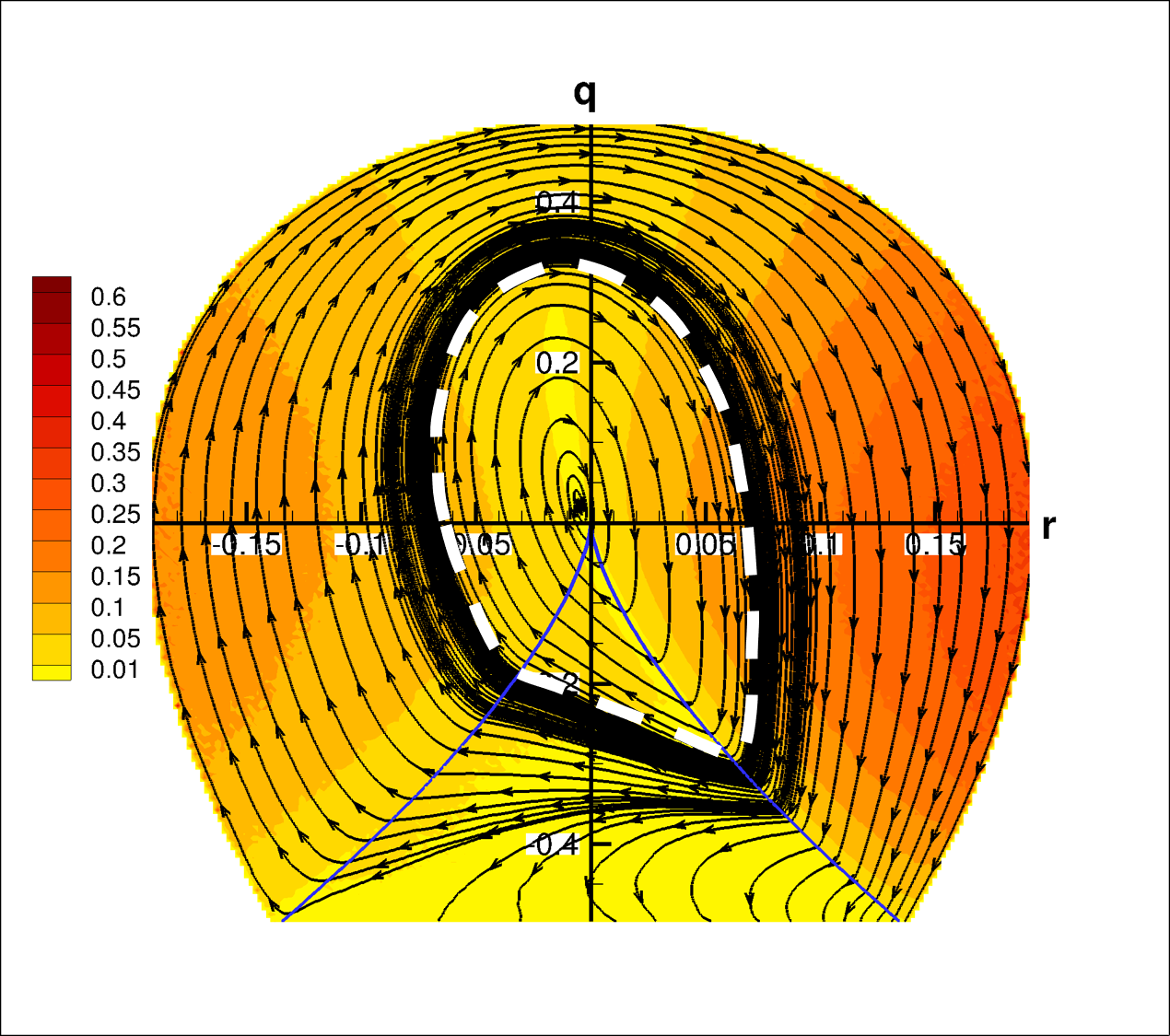}};
\node at (10pt,150pt) {(\textit{c})};
\end{tikzpicture}
\begin{tikzpicture}
\node[above right] (img) at (0,0) {\includegraphics[width=0.48\textwidth,trim={0.3cm 1.3cm 1cm 1cm},clip]{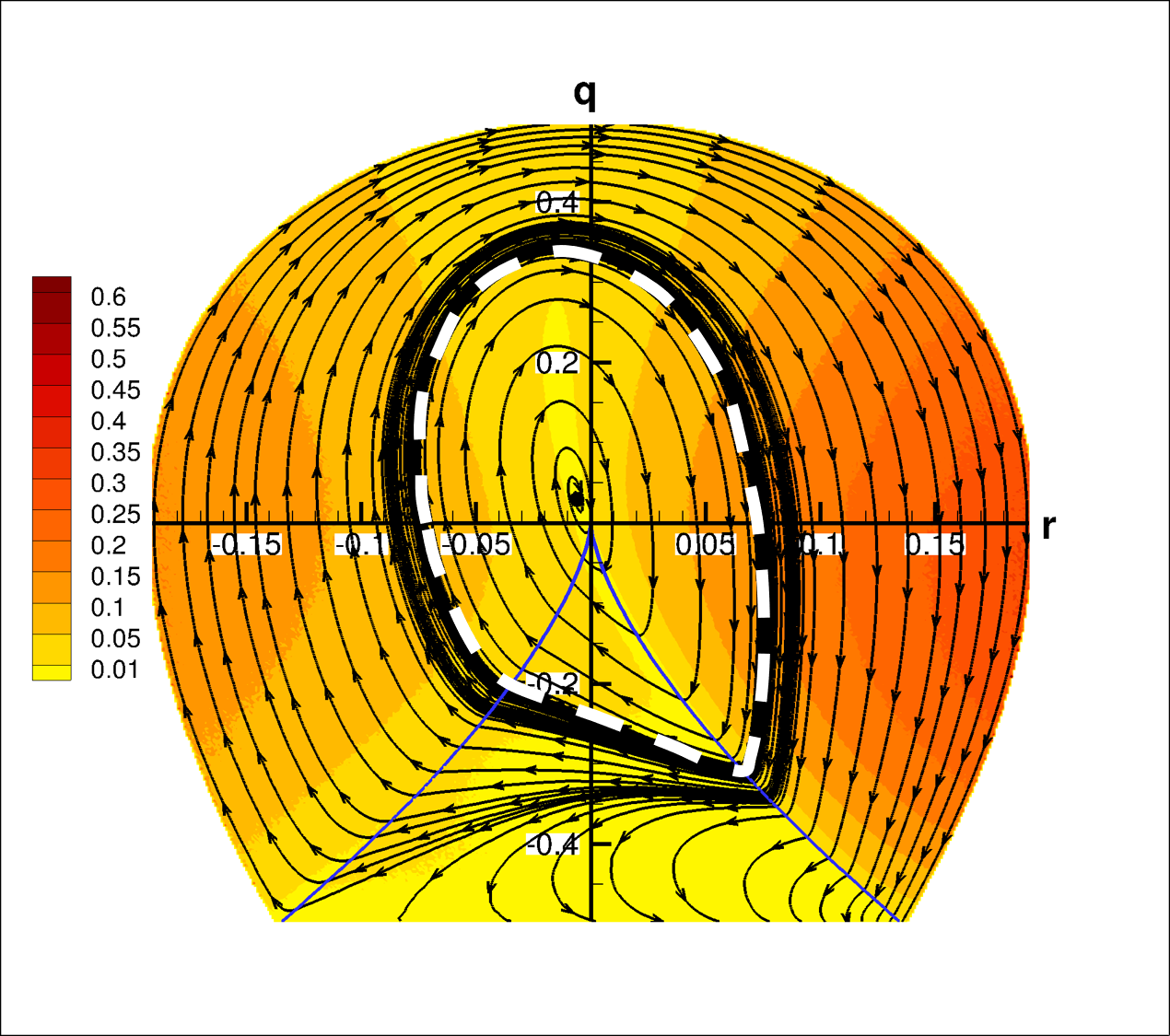}};
\node at (10pt,150pt) {(\textit{d})};
\end{tikzpicture}
\caption{Conditional mean trajectories (CMT) 
in (a) $Q$-$R$ plane for $Re_\lambda=225$ (b) $q$-$r$ plane for $Re_\lambda=225$ (c) $q$-$r$ plane for $Re_\lambda=385$ (d) $q$-$r$ plane for $Re_\lambda=588$ .
Background contours indicate the speed of the trajectory at each point, normalized by Kolmogorov time scale. White dashed lines represent the corresponding separatrices.}
\label{fig:CMT_total}
\end{figure}
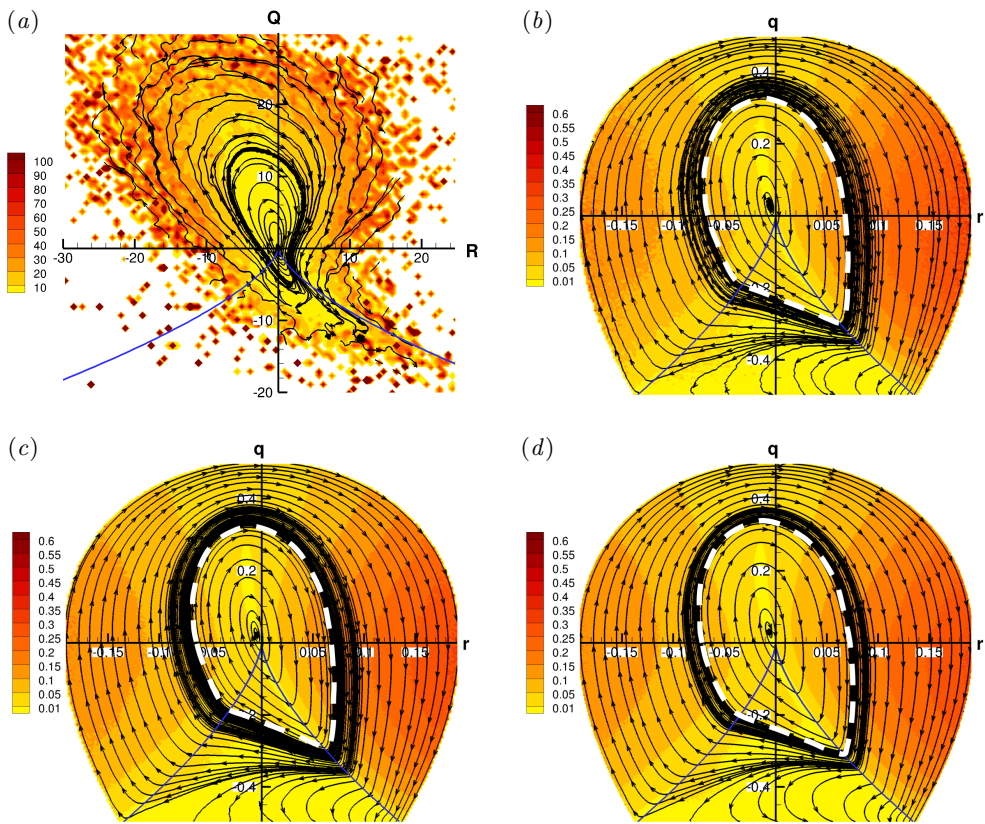

The CMTs in the normalized invariant $q$-$r$ plane are displayed for different Reynolds number cases in figure \ref{fig:CMT_total} (b-d).
The $q$-$r$ CMTs are determined by time-integration of the conditional mean phase velocity vector field given by $\langle v_q \equiv dq/dt \;| q,r \rangle$ and $\langle v_r \equiv dr/dt \;| q,r \rangle$, computed from equations (\ref{eq:b3}) and (\ref{eq:b4}). 
The speed of the trajectories or the normalized conditional mean phase velocity magnitude, 
\begin{equation}
\langle \overline{v}|q,r \rangle = \frac{\sqrt{{\langle v_q | q,r \rangle}^2 + {\langle v_r |q,r \rangle}^2}}{1/\tau_\eta} \;\;\; \text{where} \;\;\; \tau_\eta  \sim 1/\langle A\rangle 
\end{equation}
is indicated by the background contours. 

It is evident from figures \ref{fig:CMT_total} (b-d) that the compact $q$-$r$ CMTs are well-behaved throughtout the domain and highlight many more features than the $Q$-$R$ CMTs. 
The $q$-$r$ CMTs can be divided into two distinct types - inner and outer trajectories, separated by the white dashed loop (separatrix) marked in the figure.  
Inner CMTs spiral clockwise towards the origin while outer CMTs asymptote to the lower boundary of the plane in a clockwise manner. 

The $q$-$r$ equations can be considered a dynamical system in a compact phase space. The description of the behavior of this dynamical system is of much value for understanding and modeling VG dynamics.
The system consists of two attractors - the attracting focus ($q \approx 0, r\approx 0$) represents pure-shear geometry and the attracting manifold ($q=-1/2$ line) represents pure strain shape, as shown in section \ref{ssec:eqns2d}.
The dashed loop is the separatrix - an invariant manifold that separates the domain of attraction of the two attractors. Trajectories originating on the separatrix loop continue to circle along the loop until a small deviation causes it to gradually leave the loop. 

The evolution is generally faster in the focal streamline region of the $q$-$r$ plane, particularly when vorticity is more aligned with the most expansive or most compressive strain-rates.
CMTs become considerably slower after crossing the right discriminant line. 
The nodal streamline shapes with extremely low rate of evolution, therefore, constitute the long-lived structures in turbulent flow.  
The extremely low velocities near the discriminant line in $Q$-$R$ plane prompted many to believe that the discriminant line is an attractor \citep{martin1998dynamics}. However, the $q$-$r$ plane clearly indicates that even though the trajectories slow down near the discriminant line, it is not an attractor of the system.

The $q$-$r$ CMTs for all the investigated Reynolds numbers, $Re_\lambda=225$, $385$ and $588$, are very similar: (i) The two attractors of the system are identically located. (ii) However, the location of the separatrix appears to depend weakly on the Reynolds number. 
More work is required to establish a clear $Re_\lambda$-dependence of the separatrix. As will be shown later, this $Re_\lambda$-dependence of separatrix may be attributed to the viscous contribution in $q$-$r$ evolution.

\subsection{\label{sec:Res2b}Cycle time period of CMT}

\begin{figure}
\centering
\begin{tikzpicture}
\node[above right] (img) at (0,0) {\includegraphics[width=0.3\textwidth,trim={2cm 4cm 3cm 3cm},clip]{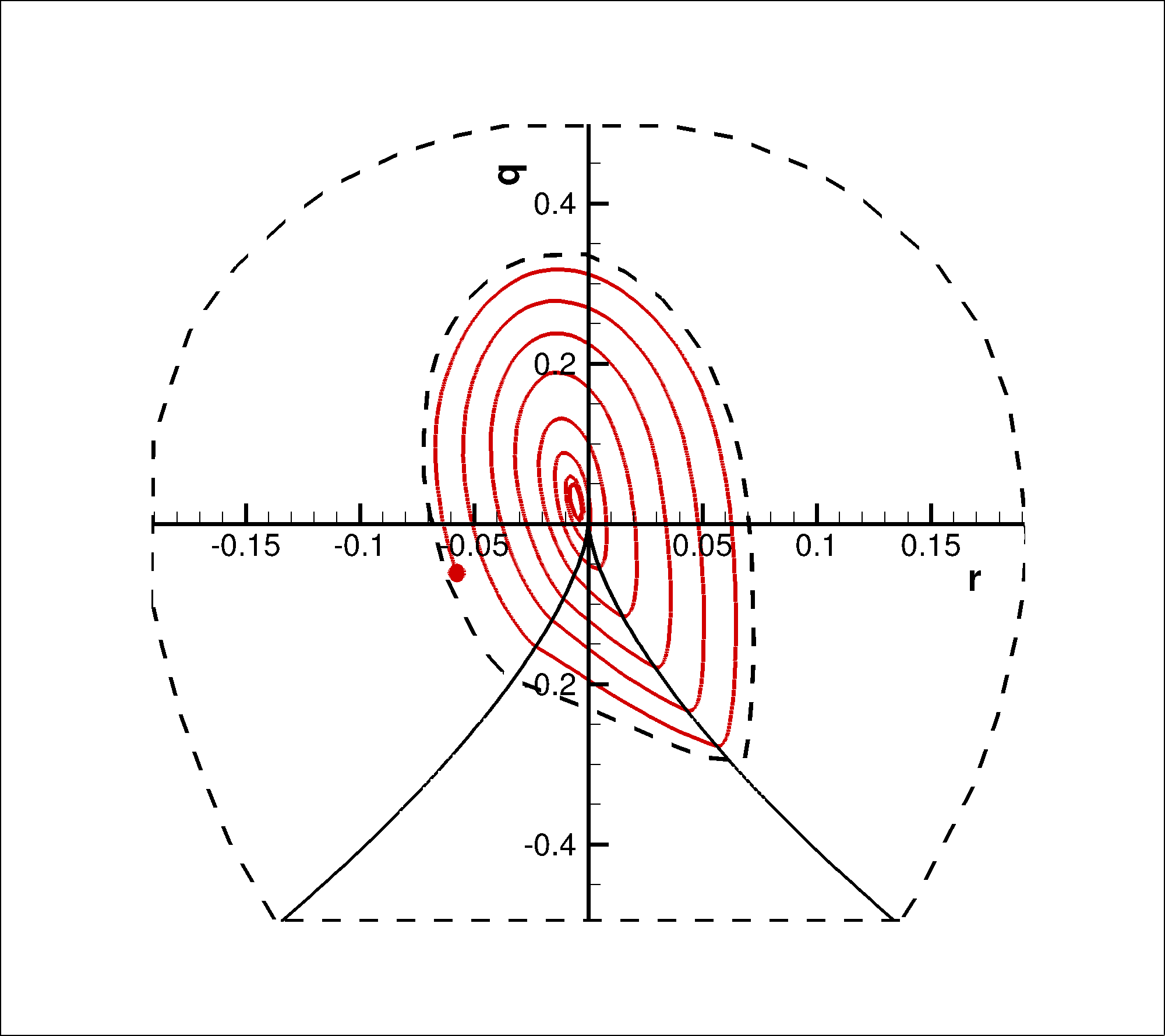}};
\node at (8pt,90pt) {(\textit{a})};
\end{tikzpicture}
\begin{tikzpicture}
\node[above right] (img) at (0,0) {\includegraphics[width=0.3\textwidth,trim={1cm 4cm 3cm 3cm},clip]{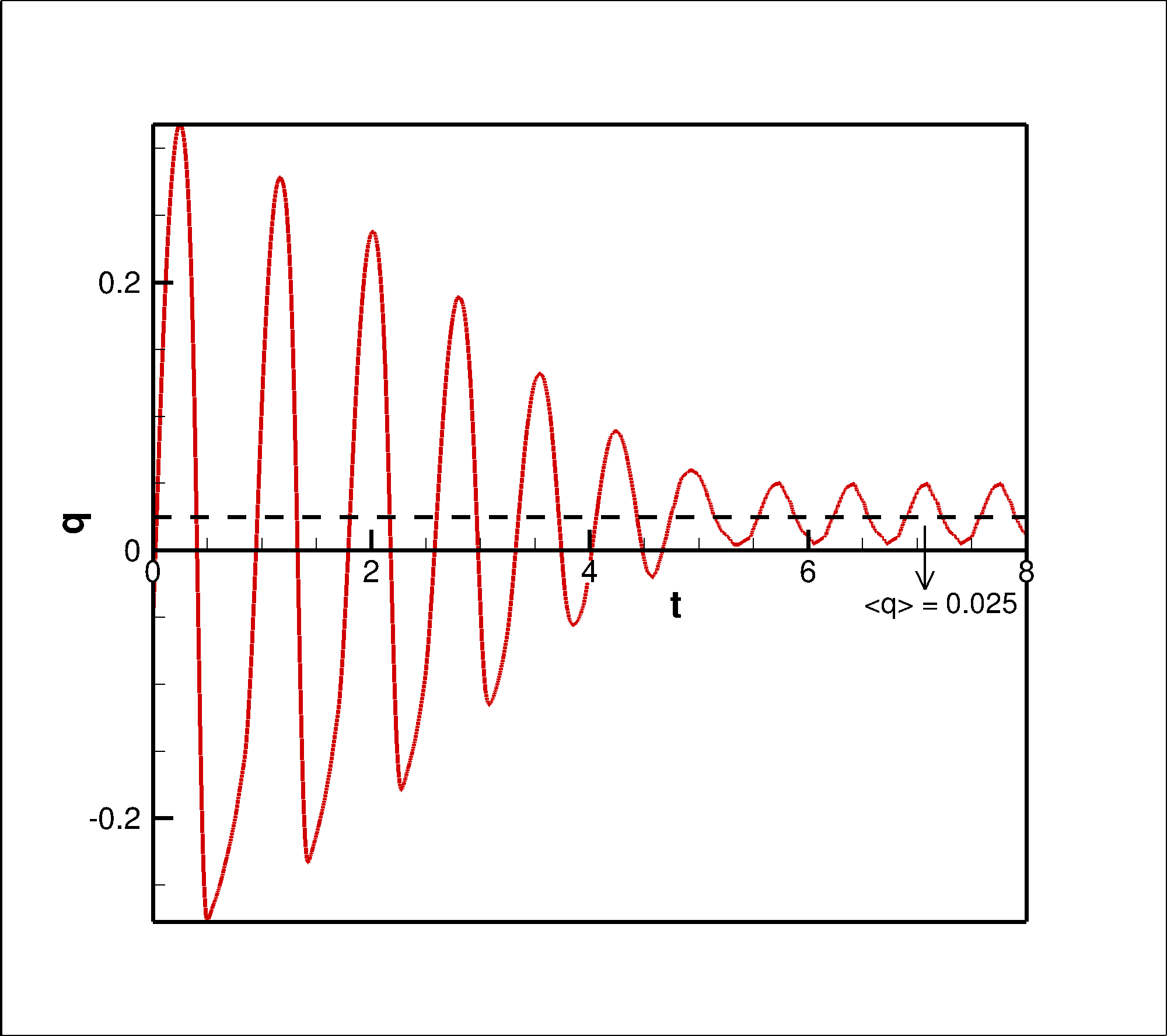}};
\node at (5pt,90pt) {(\textit{b})};
\end{tikzpicture}
\begin{tikzpicture}
\node[above right] (img) at (0,0) {\includegraphics[width=0.3\textwidth,trim={1cm 4cm 3cm 3cm},clip]{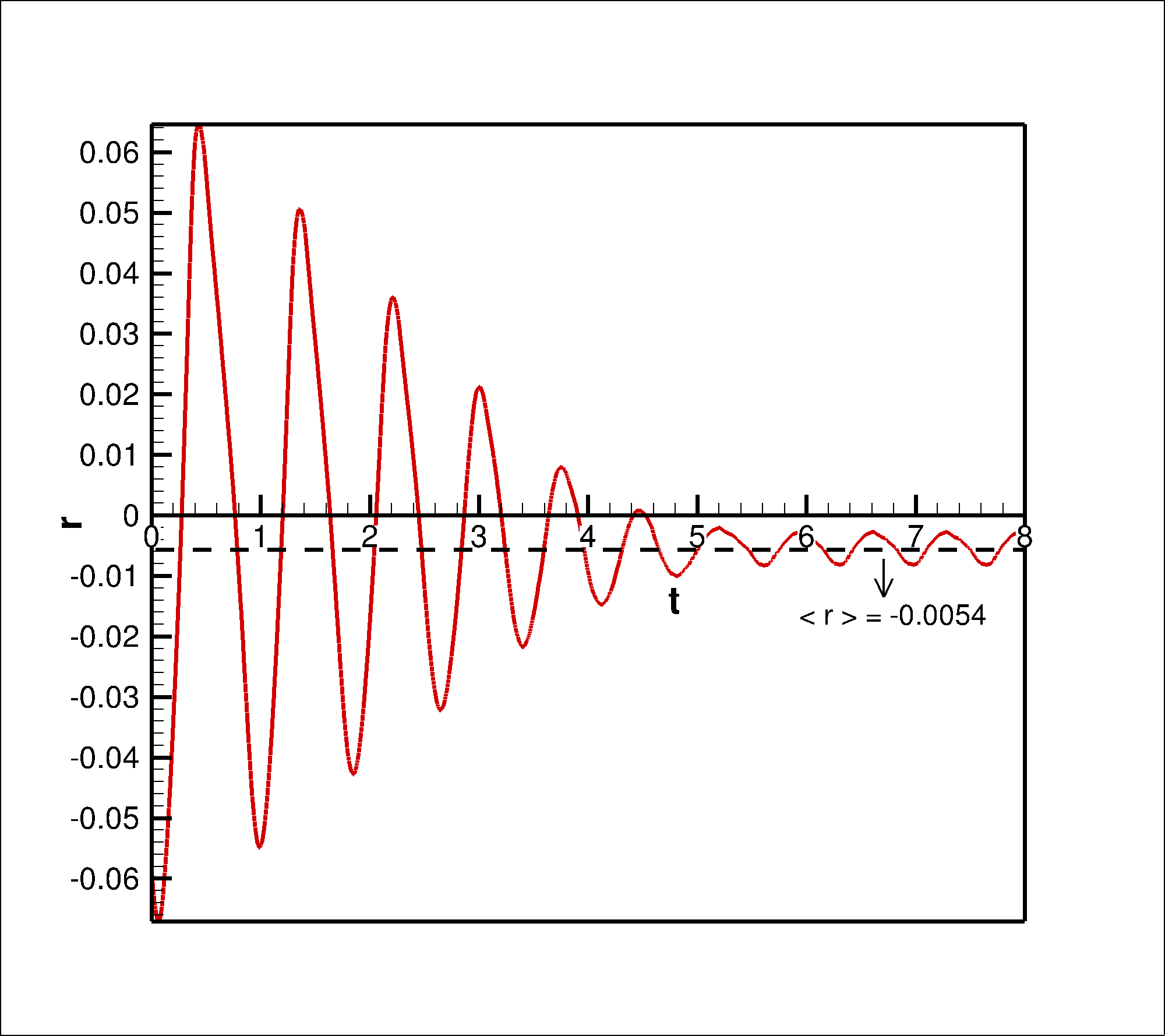}};
\node at (3pt,90pt) {(\textit{c})};
\end{tikzpicture}

\begin{tikzpicture}
\node[above right] (img) at (0,0) {\includegraphics[width=0.3\textwidth,trim={2cm 4cm 3cm 3cm},clip]{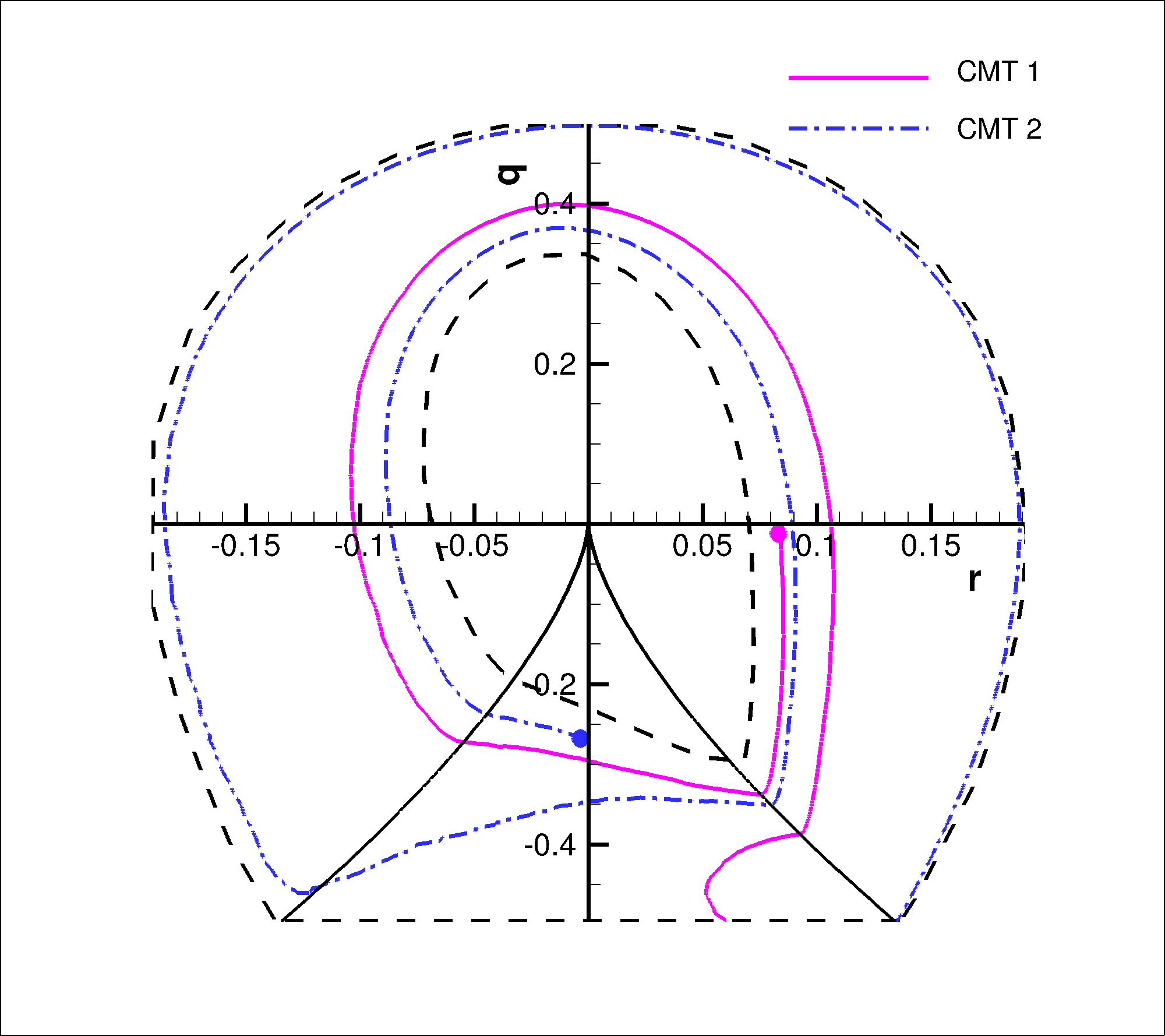}};
\node at (8pt,90pt) {(\textit{d})};
\end{tikzpicture}
\begin{tikzpicture}
\node[above right] (img) at (0,0) {\includegraphics[width=0.3\textwidth,trim={1cm 4cm 3cm 3cm},clip]{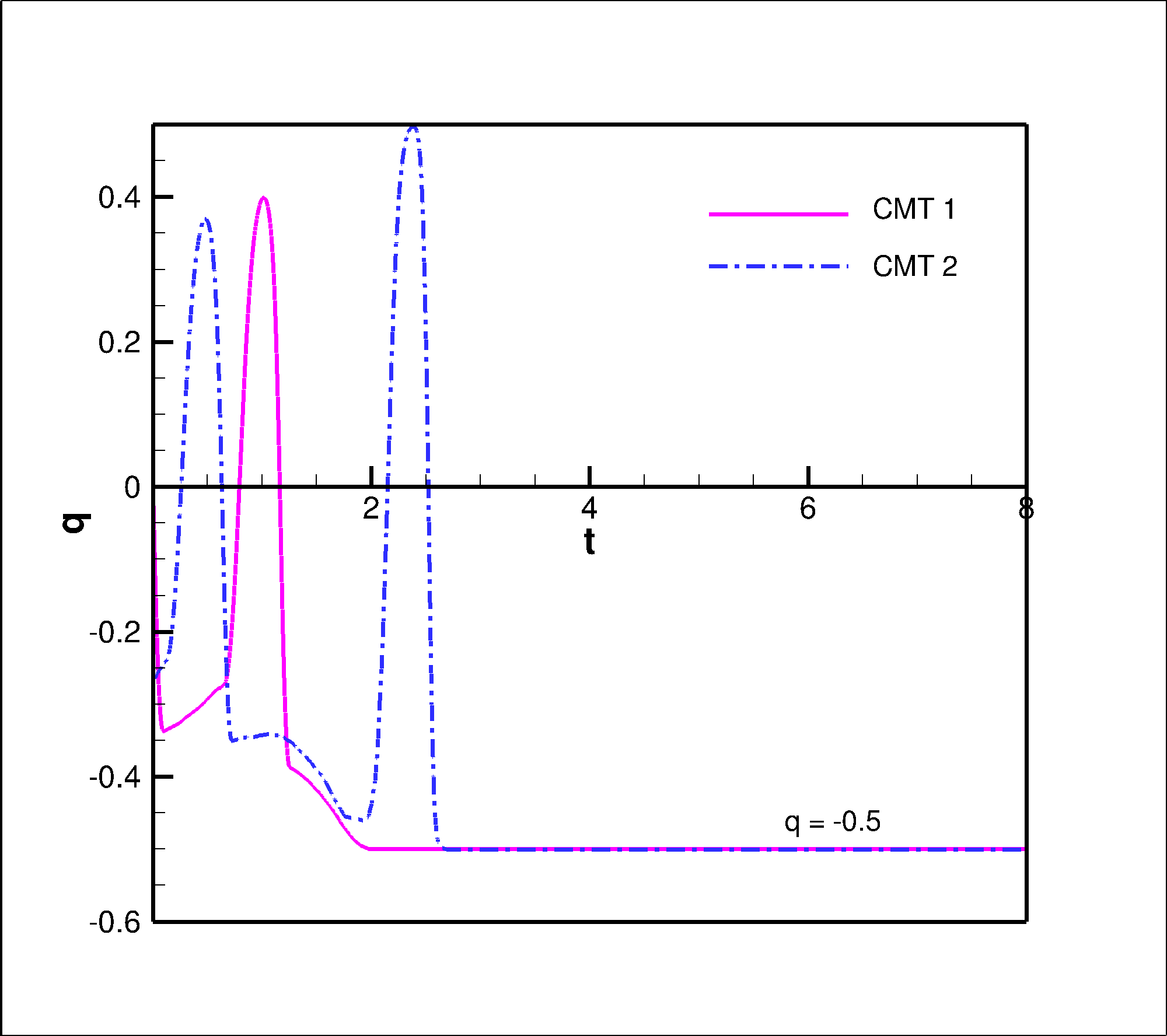}};
\node at (5pt,90pt) {(\textit{e})};
\end{tikzpicture}
\begin{tikzpicture}
\node[above right] (img) at (0,0) {\includegraphics[width=0.3\textwidth,trim={1cm 4cm 3cm 3cm},clip]{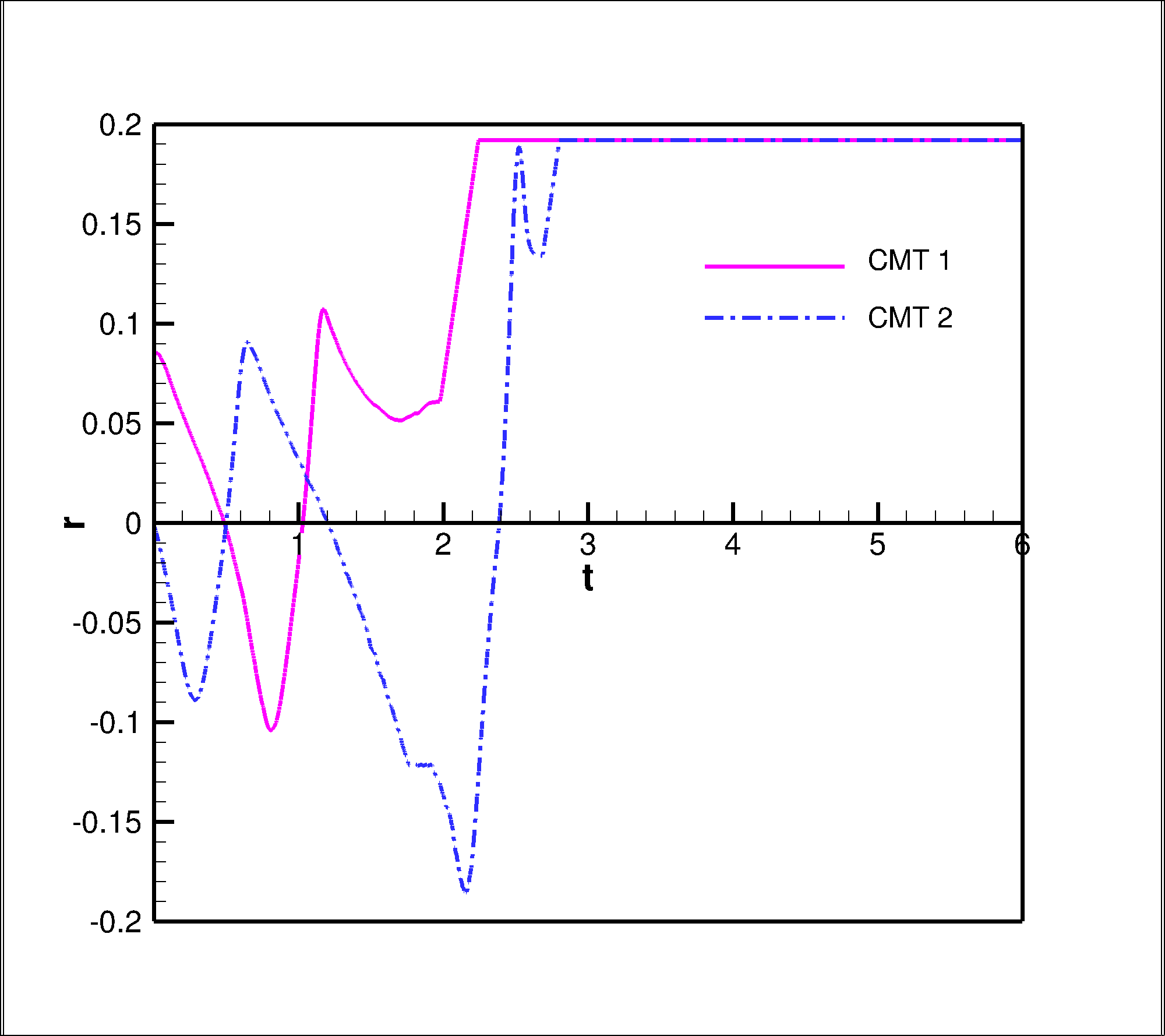}};
\node at (3pt,90pt) {(\textit{f})};
\end{tikzpicture}

\caption{(a) A representative inner CMT (point of origin marked by circle) and corresponding (b) $q$ evolution with time (c) $r$ evolution with time.
(d) Two representative outer CMTs (points of origin marked by circles) and corresponding (e) $q$ evolution with time (f) $r$ evolution with time.   
Dashed lines in (a,d) represent the separatrix loop and boundary of realizable region of $q$-$r$ plane.}
\label{fig:indiv_CMT}
\end{figure}
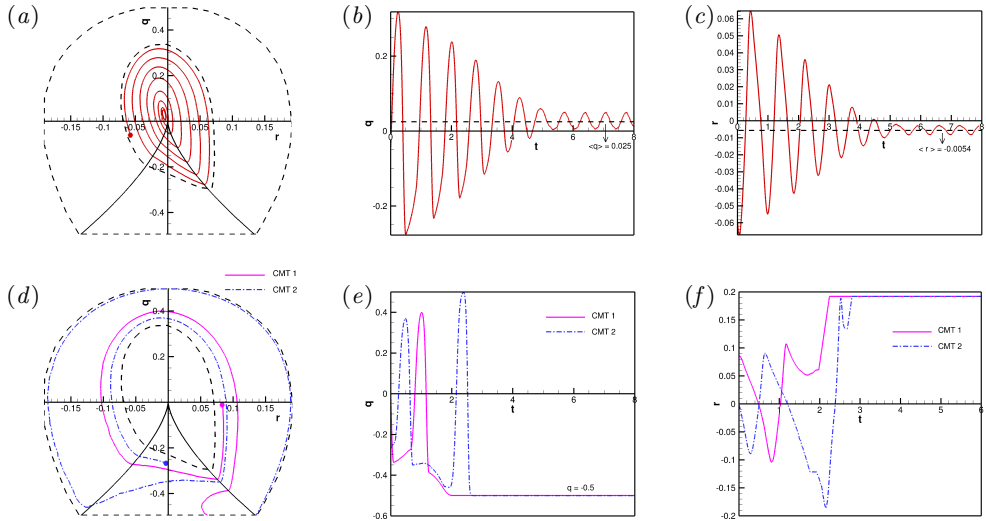

Figure \ref{fig:indiv_CMT} (a-c) illustrates the time evolution of $q$ and $r$ for a representative CMT in inner $q$-$r$ plane.
The inner trajectories spiral towards the attractor in nearly periodic cycles. 
Both $q$ and $r$ oscillate in time with decreasing amplitude and asymptote to 
the close neighborhood of the attractor ($q \approx 0, r\approx 0$). 
The time period of cycle decreases monotonically as the trajectory goes from the separatrix to the attractor:
$T_{sep} \sim 35 \tau_\eta$ at the separatrix and $T_{in} \sim 25 \tau_\eta$ at the innermost oscillations about the attractor.
Similar behavior is observed at other Reynolds numbers ($Re_\lambda=385$, $588$). 
This range, $T \in (25\tau_\eta,35\tau_\eta)$, includes the characteristic cycle time of $30 \tau_\eta$ reported by \cite{martin1998dynamics} for spiralling trajectories in $Q$-$R$ plane in a lower $Re_\lambda$ forced homogeneous isotropic turbulence. 

The outer trajectories exhibit aperiodic behavior and tend toward the pure strain attracting manifold. This is illustrated by two CMTs in figure \ref{fig:indiv_CMT} (d-f) at the two extreme ends of the outer region - one remains close to the separatrix (solid magenta line) and the other travels towards the boundary of the realizable region (dash-dot blue line).
A CMT originating on the separatrix continues to move along the separatrix for a number of cycles with a time period $\sim T_{sep}$ until it slightly drifts out of this loop on either one of the sides and evolves toward the corresponding attractor.

\subsection{\label{sec:Res2c} CMT residence time}

\begin{table}  
  \begin{center}
\def~{\hphantom{0}}
\setlength{\tabcolsep}{10pt}
  \begin{tabular}{lcccc}
        $~~$ & SFS & UFC & UN/S/S  & SN/S/S \\[1pt]
	\hline
	\hline
        Inner region $\%$ composition     & $~~44.7~~$ & $~~27.6~~$ & $~~21.5~~$ & $~~6.2~~$ \\
        Inner CMT residence time ($\%$)	& $~~44.0~~$ & $~~26.8~~$ & $~~22.4~~$ & $~~6.7~~$ \\
	\hline
        Outer region $\%$ composition     & $~~26.0~~$ & $~~20.9~~$ & $~~40.7~~$ & $~~12.3~~$ \\	
	Outer CMT1 residence time ($\%$) 	& $~~31.9~~$ & $~~19.3~~$ & $~~34.0~~$ & $~~14.8~~$ \\	
	Outer CMT2 residence time ($\%$) 	& $~~22.7~~$ & $~~13.8~~$ & $~~29.6~~$ & $~~33.9~~$ \\
	\hline
  \end{tabular}
  \caption{Percentage composition of topologies in inner and outer regions and residence time ($\%$) of representative CMTs (figure \ref{fig:indiv_CMT}) in each topology.}
  \label{tab:resi}
  \end{center}
\end{table}

The percentage of total time spent by a $q$-$r$ CMT in each topology type is the residence time (referred to as mean lifetime by \cite{parashar2019lagrangian}) for that topology. The residence time is a measure of the average lifetime of that topology in turbulence. 
Therefore, it is compared with the percentage composition of each topology in a turbulent flow field.
The residence time of the representative CMTs shown in figure \ref{fig:indiv_CMT}, are listed in table \ref{tab:resi} along with the percentage composition of each topology in inner and outer regions for comparison. 
The residence time percentages of the inner CMTs are very close to the population percentages of the corresponding topologies in the inner region. 
In the outer region, CMT-1 (closer to the separatrix) has residence times fairly close to the population percentages while the residence times of CMT-2 show significant deviation from the population fractions. 
The high-density inner region constitutes majority of the population and the residence time here conforms well with the population percentages.
The results agree well with the work of \cite{parashar2019lagrangian}, who showed that the residence time of Lagrangian trajectories (obtained from particle tracking) in each topology is in the same proportion as the percentage composition of the topology.
In contrast, the residence times based on $Q$-$R$ CMTs investigated by \cite{martin1998dynamics} and \cite{elsinga2010evolution} do not compare well with the Lagrangian results.
This demonstrates yet another advantage of examining the CMTs in the $q$-$r$ space.


\subsection{\label{sec:Res2d} Action of different physical processes}

The evolution of the invariants, $q$ and $r$, depends on three distinct physical processes - inertial, pressure and viscous. The effect of each of these processes is examined in isolation in the following sub-sections.


\subsubsection{\label{sec:Res2d1} Inertial effects}

\begin{figure}
\centering
\begin{tikzpicture}
\node[above right] (img) at (0,0) {\includegraphics[width=0.48\textwidth,trim={1.3cm 4cm 3cm 3cm},clip]{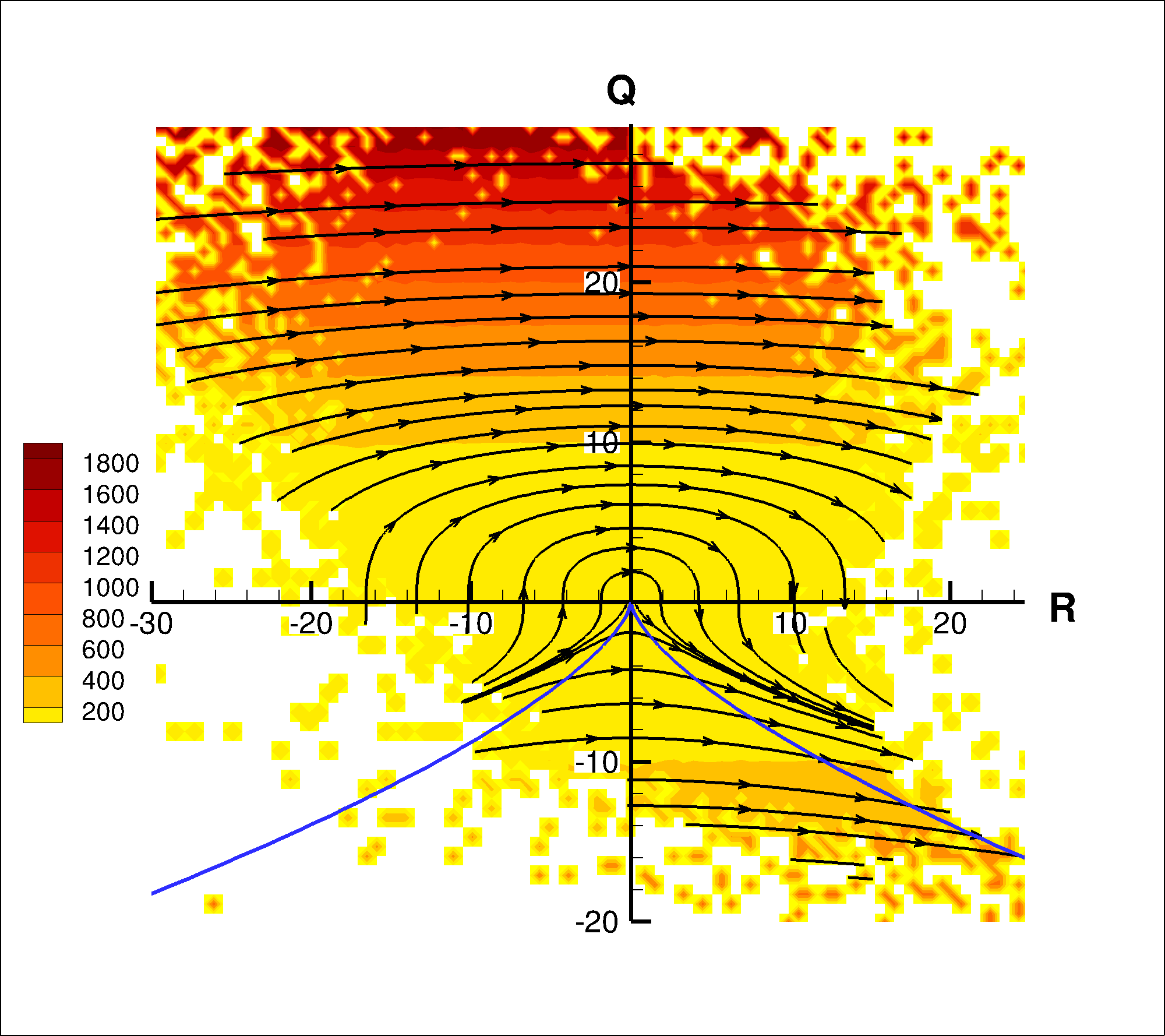}};
\node at (10pt,150pt) {(\textit{a})};
\end{tikzpicture}
\begin{tikzpicture}
\node[above right] (img) at (0,0) {\includegraphics[width=0.48\textwidth,trim={1.3cm 4cm 3cm 3cm},clip]{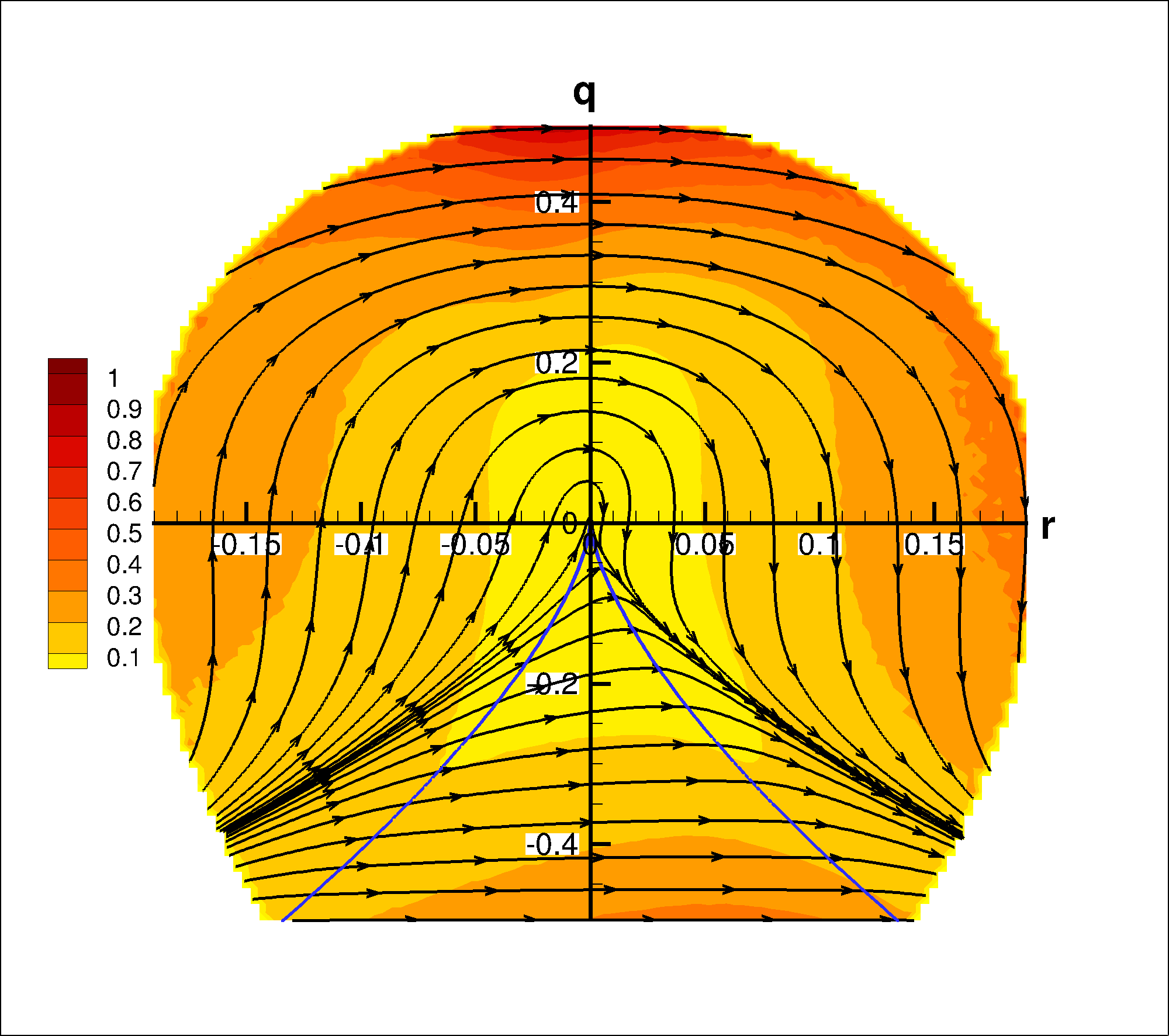}};
\node at (10pt,150pt) {(\textit{b})};
\end{tikzpicture}
\caption{Inertial CMTs in (a) $Q$-$R$ plane and (b) $q$-$r$ plane for $Re_\lambda=225$. Background contours indicate the speed of the trajectory at each point, normalized by Kolmogorov time scale.}
\label{fig:CMT_inertial}
\end{figure}

\textit{$Q$-$R$ plane}: The inertial CMTs in $Q$-$R$ plane, as depicted in figure \ref{fig:CMT_inertial} (a), move from left to right in the plane, rendering stable topologies unstable.
This is in line with the CMTs obtained from Restricted Euler solution \citep{cantwell1992exact} shown by \cite{martin1998dynamics2}, \cite{ooi1999study} and \cite{chevillard2008modeling}. 
Some of the trajectories tend to asymptote to a line above the right $D=0$ line (invariant line of Restricted Euler solution) due to the absence of the isotropic pressure term. This line is also different from the invariant line of Burger's equation, shown in \cite{bikkani2007role}, since the incompressibility condition is not enforced in that case. 

\textit{$q$-$r$ plane}: The $q$,$r$-evolution due to inertial effects, as given by the following terms from equations (\ref{eq:b3}) and (\ref{eq:b4}), 
\begin{equation}
	{\bigg(\frac{dq}{dt} \bigg)}_{I} = A (-3r + 2q b_{ij}b_{ik}b_{kj}) \;\;, \;\;\; {\bigg(\frac{dr}{dt} \bigg)}_{I} = A ({2}q^2 + 3rb_{ij}b_{ik}b_{kj})
\end{equation}
is plotted in figure \ref{fig:CMT_inertial} (b) for $Re_\lambda=225$. Since the inertial terms only depend on $\bm{b}$, the CMTs for other $Re_\lambda$ cases are exactly identical and are not displayed separately.
As observed for the $Q$-$R$ plane \citep{martin1998dynamics2}, the origin of the $q$-$r$ plane also appears to be a degenerate saddle point.
Clearly, in this compact and bounded phase space, the left boundary acts as a repellor and the right boundary acts as an attractor of the system.
Referring back to figure \ref{fig:qrmap}, it is evident that inertia, whilst acting in isolation, causes the intermediate strain-rate to become more positive and the vorticity vector to be more aligned with the most compressive strain-rate eigen-direction.


\subsubsection{\label{sec:Res2d2} Pressure effects}

\begin{figure}
\centering
\begin{tikzpicture}
\node[above right] (img) at (0,0) {\includegraphics[width=0.48\textwidth,trim={1.3cm 4cm 3cm 3cm},clip]{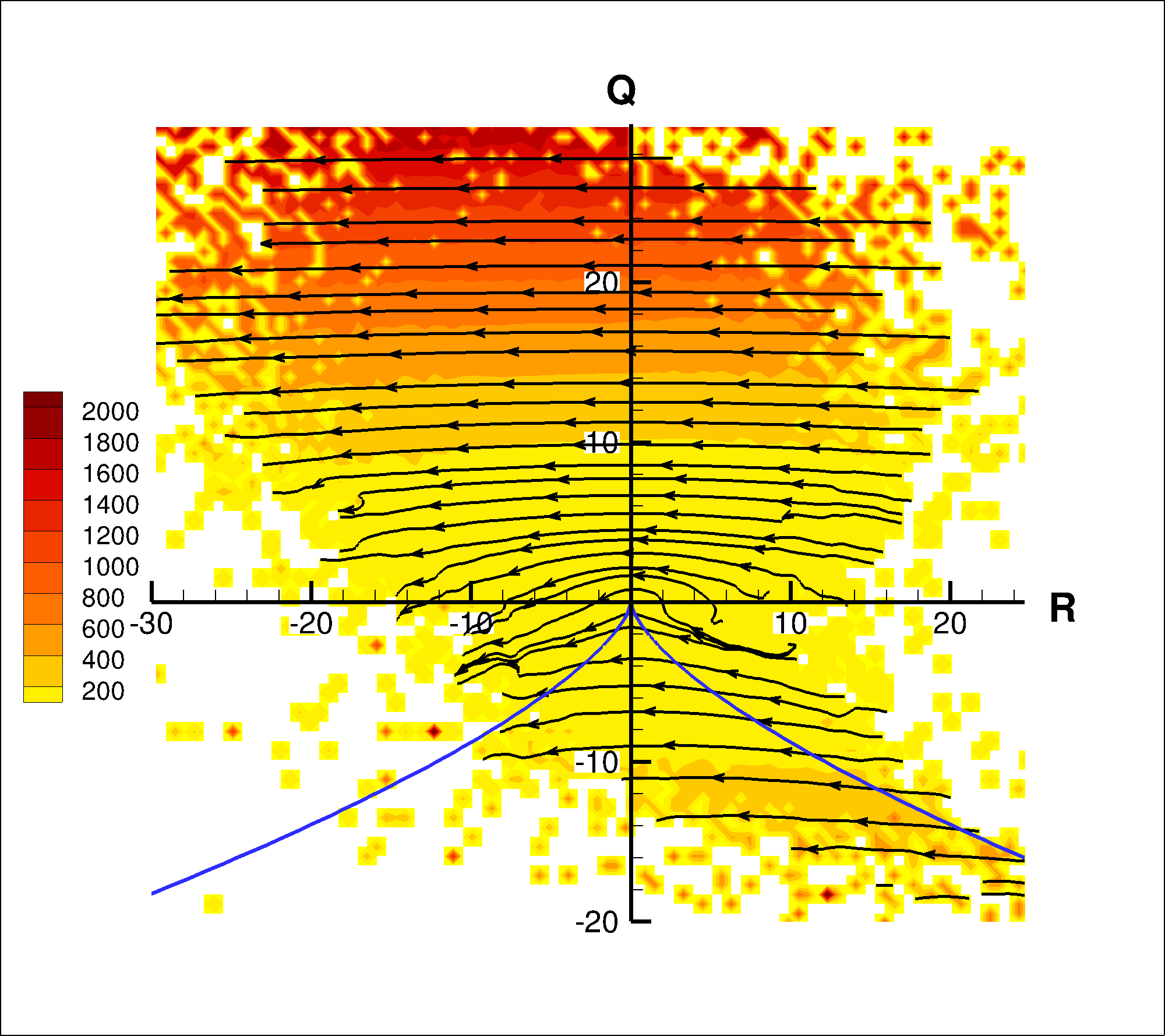}};
\node at (10pt,150pt) {(\textit{a})};
\end{tikzpicture}
\begin{tikzpicture}
\node[above right] (img) at (0,0) {\includegraphics[width=0.48\textwidth,trim={0.3cm 1.3cm 1cm 1cm},clip]{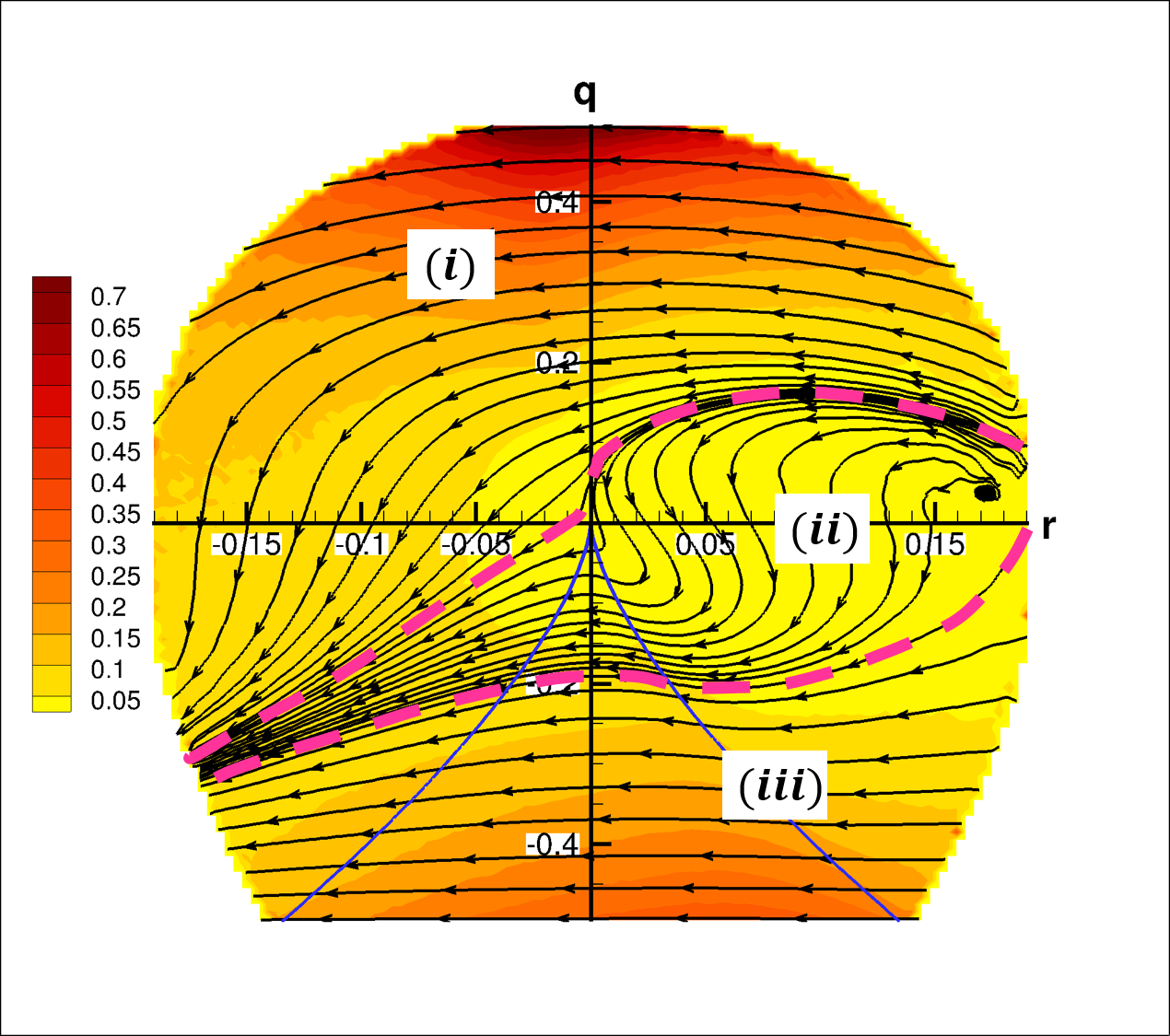}};
\node at (10pt,150pt) {(\textit{b})};
\end{tikzpicture}
\caption{Pressure CMTs in (a) $Q$-$R$ plane and (b) $q$-$r$ plane for $Re_\lambda=225$. Background contours indicate the speed of the trajectory at that point, normalized by Kolmogorov time scale. The dashed lines separate the three types of pressure CMTs in $q$-$r$ space.}
\label{fig:CMT_pressure}
\end{figure}

\textit{$Q$-$R$ plane}: Figure \ref{fig:CMT_pressure} (a) shows the $Q$-$R$ CMTs due to pressure action.
As opposed to the anisotropic pressure Hessian contribution to $Q$-$R$ evolution illustrated in previous studies \citep{chevillard2008modeling,johnson2016closure}, the complete pressure Hessian term has been plotted here.
It shows that pressure action opposes the inertial action. 

\textit{$q$-$r$ plane}: The $q$-$r$ CMTs based on the pressure terms, 
\begin{equation}
    {\bigg(\frac{dq}{dt} \bigg)}_{\mathcal{P}} = A (- h_{ij}(b_{ji} + 2qb_{ij})) \;\;, \;\;\; {\bigg(\frac{dr}{dt} \bigg)}_{\mathcal{P}} = A \Big(- \frac{4}{3}q^2  - h_{ij}(b_{ki}b_{jk} + 3rb_{ij}) \Big)
\end{equation}
considered in isolation from the other terms in equations (\ref{eq:b3}) and (\ref{eq:b4}), are plotted in Figure \ref{fig:CMT_pressure} (b) for the $Re_\lambda=225$ case.
It is clear that there are three types of pressure CMTs - \textit{(i)} CMTs of rotation-dominated focal streamlines near the top of the plane travel directly from the right boundary to the left boundary with very high speeds. 
\textit{(ii)} CMTs in the middle region of the plane are repelled from a small region close to the right boundary and are attracted towards a small region in the left boundary. These relatively slow-moving CMTs tend to traverse along a line in the upper UFC region, get deflected by what appears to be a degenerate saddle point at origin and converges to the attractor in the lower SFS region.
\textit{(iii)} The strain-dominated streamlines near the bottom of the plane evolve reasonably fast, straight from the right to left boundary. 
Overall, pressure action causes UFC streamlines with vorticity along most compressive strain-rate (right boundary of the plane) to change towards SFS streamlines with vorticity aligned along the most expansive strain-rate (left boundary of the plane). 
In general the pressure trajectories oppose the inertial trajectories except the type-\textit{ii} CMTs in the middle region of UFC topology, where pressure contribution aligns with the inertial contribution.
The isotropic pressure term ($dq/dt= 0, dr/dt= -4Aq^2/3$) simply drives the streamline shapes from right to left in straight horizontal lines. All the additional features of the pressure CMTs stem from the anisotropic pressure term.
Pressure $q$,$r$-CMTs at other Reynolds numbers (not displayed) are nearly identical to the present case and therefore effect of pressure on local streamline shape can be deemed nearly independent of Reynolds number at high enough $Re_\lambda$.

\subsubsection{\label{sec:Res2d3} Pressure and inertial effects}

In incompressible flows, the role of pressure is to counter inertial action in a manner that the velocity field is divergence free. As the action of pressure is a response to inertial effect on the velocity field, it may be useful to examine the combined outcome of the two processes.

\begin{figure}
\centering
\begin{tikzpicture}
\node[above right] (img) at (0,0) {\includegraphics[width=0.48\textwidth,trim={1.3cm 4cm 3cm 3cm},clip]{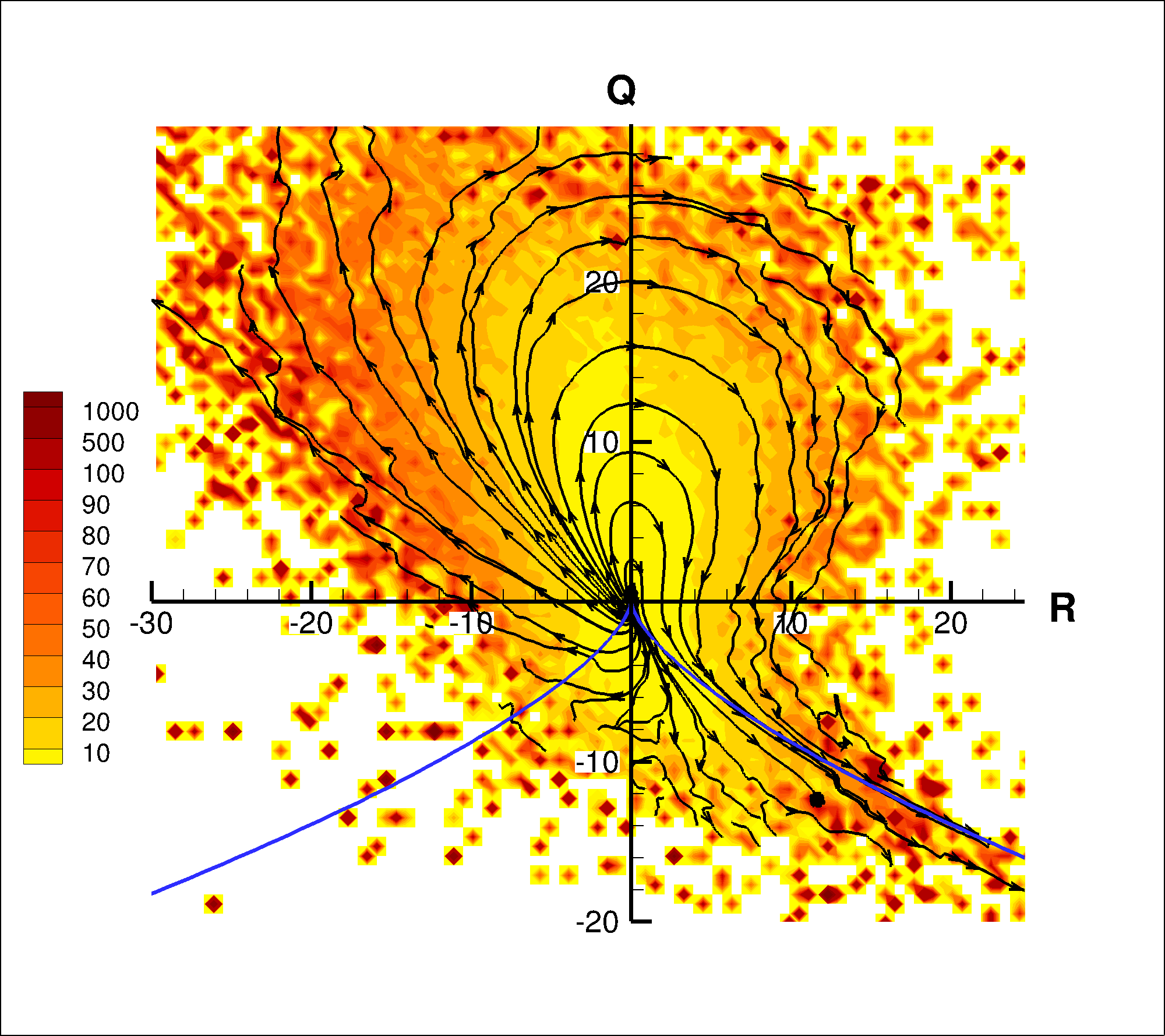}};
\node at (10pt,150pt) {(\textit{a})};
\end{tikzpicture}
\begin{tikzpicture}
\node[above right] (img) at (0,0) {\includegraphics[width=0.48\textwidth,trim={1.3cm 4cm 3cm 3cm},clip]{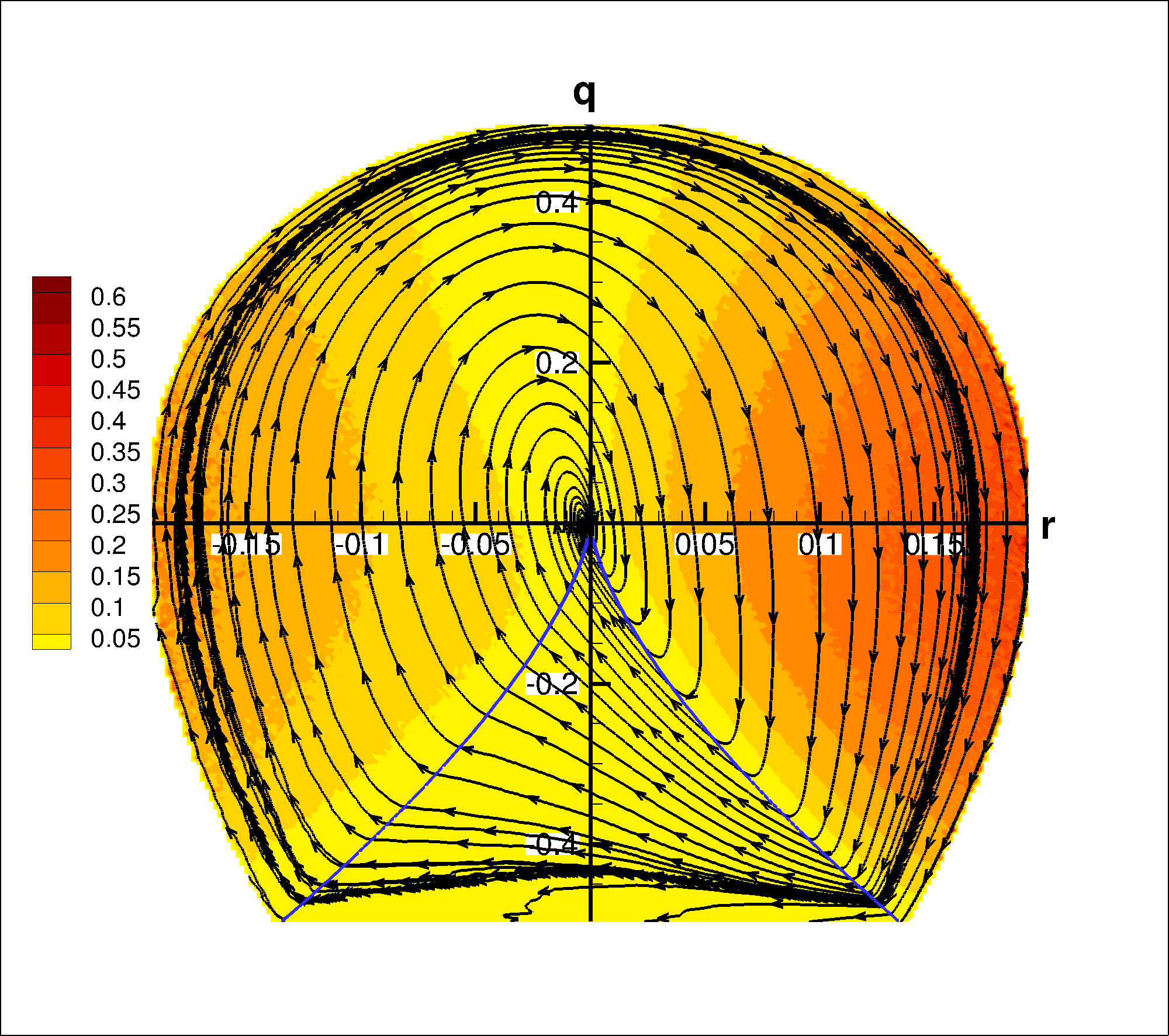}};
\node at (10pt,150pt) {(\textit{b})};
\end{tikzpicture}
\caption{Pressure-inertial CMTs in (a) $Q$-$R$ and (b) $q$-$r$ plane for $Re_\lambda=225$. Background contours indicate the speed of the trajectory at that point, normalized by Kolmogorov time scale.}
\label{fig:CMT_PI}
\end{figure}

\textit{$Q$-$R$ plane}: 
The $Q$-$R$ CMTs due to pressure and inertia are plotted in figure \ref{fig:CMT_PI} (a) for $Re_\lambda=225$. 
All trajectories appear to be repelled from the origin and moving outwards in clockwise direction. This indicates that pressure and inertia together increases the VG magnitude, evolving the streamline topology outward in the $Q$-$R$ plane. 

\textit{$q$-$r$ plane}: The $q$-$r$ CMTs of pressure and inertial processes are plotted in figure \ref{fig:CMT_PI} (b). 
The trajectories suggest a very different picture.
The pressure-inertial $q$-$r$ CMTs are attracted towards the origin, which is in contrast to the pressure-inertial $Q$-$R$ CMTs. The reason for this difference is that the $Q$-$R$ trajectories predominantly reflect the growth in magnitude (altering the scale of streamline structure) and are unable to distinguish the effects on streamline shape. 

There are two attractors in the $q$-$r$ phase space - the origin, representing pure-shear, and the $q=-1/2$ line, representing pure strain.
The CMTs spiral clockwise towards both the attractors.
The basin of attraction of pure-shear attactor spans almost the entire $q$-$r$ plane, with only a few CMTs in the slender outer region of extremely low population density that asymptote to the pure-strain attractor.
The pressure-inertial CMTs are exactly identical in other Reynolds number cases (not displayed).
Therefore, we conclude that \textit{pressure responds to inertial action in a manner that most of the geometric shapes of streamlines evolve towards pure-shear geometry irrespective of the Reynolds number of the flow}.


\subsubsection{\label{sec:Res2d4} Viscous effects}

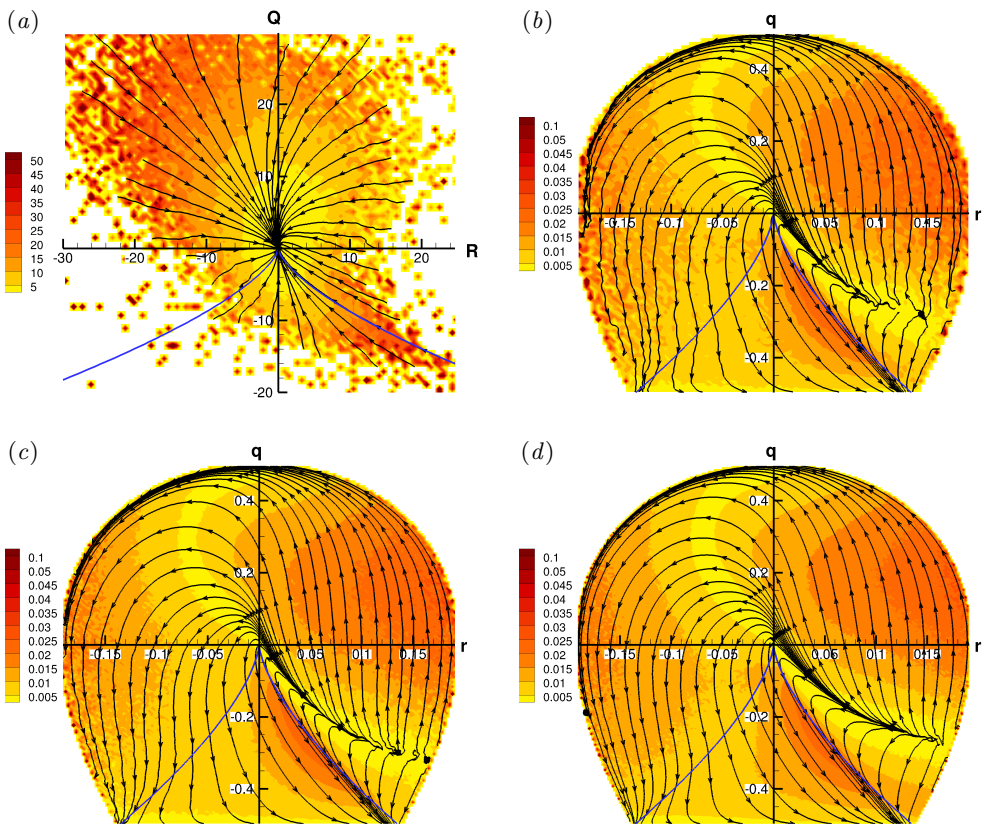
\begin{figure}
\centering
\begin{tikzpicture}
\node[above right] (img) at (0,0) {\includegraphics[width=0.48\textwidth,trim={1.3cm 4cm 3cm 3cm},clip]{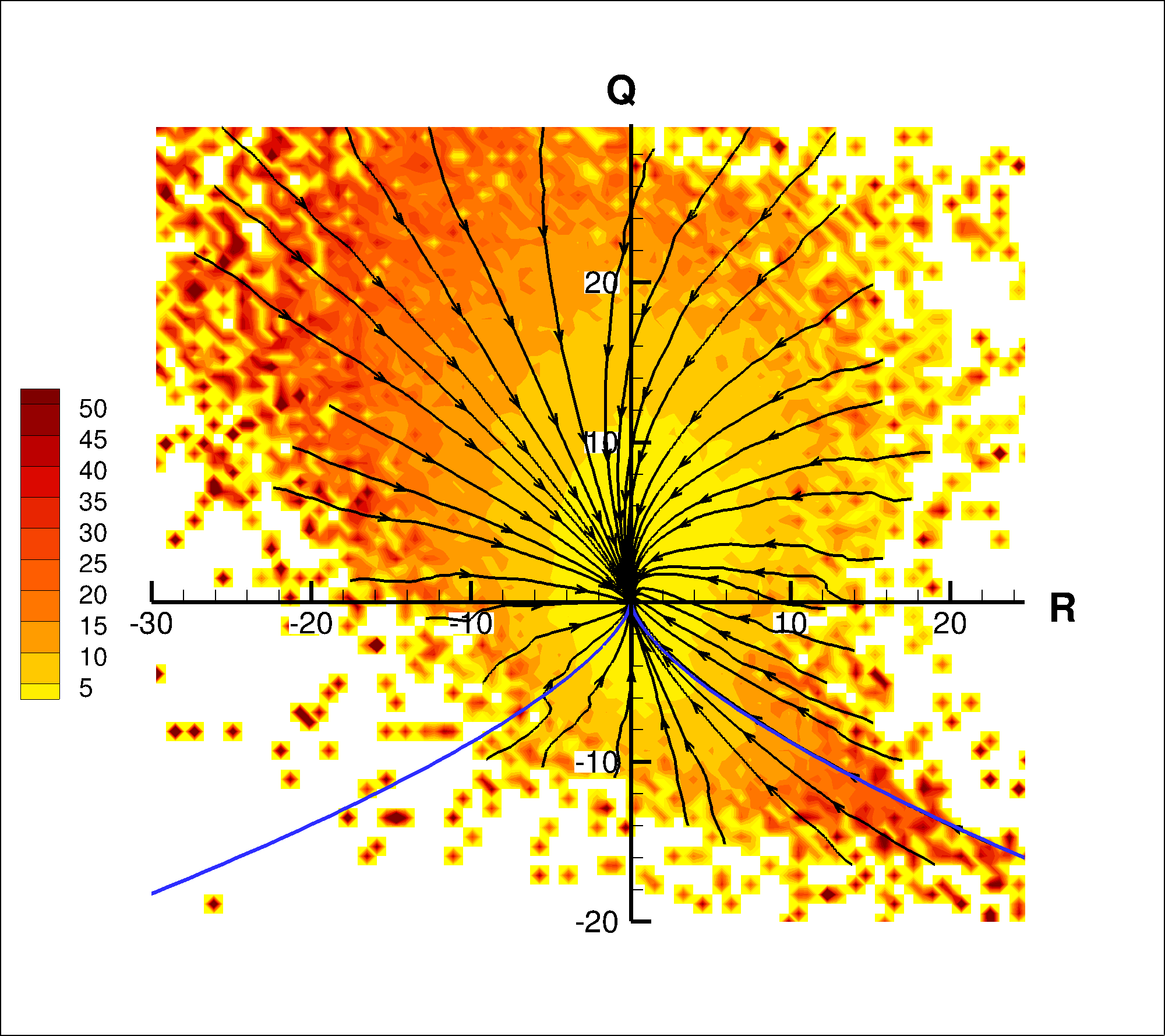}};
\node at (10pt,150pt) {(\textit{a})};
\end{tikzpicture}
\begin{tikzpicture}
\node[above right] (img) at (0,0) {\includegraphics[width=0.48\textwidth,trim={1.3cm 4cm 3cm 3cm},clip]{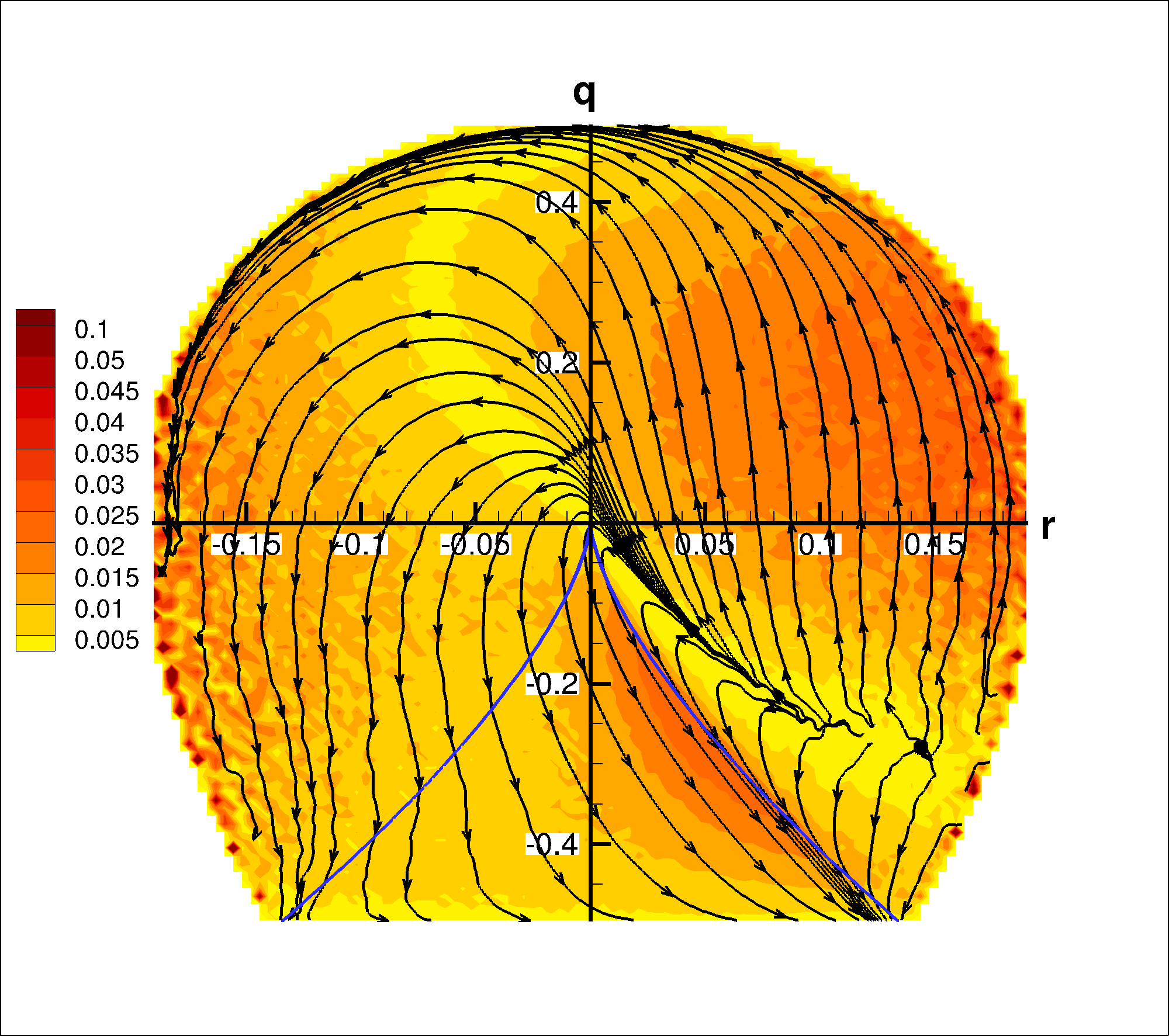}};
\node at (10pt,150pt) {(\textit{b})};
\end{tikzpicture}
\begin{tikzpicture}
\node[above right] (img) at (0,0) {\includegraphics[width=0.48\textwidth,trim={1.3cm 4cm 3cm 3cm},clip]{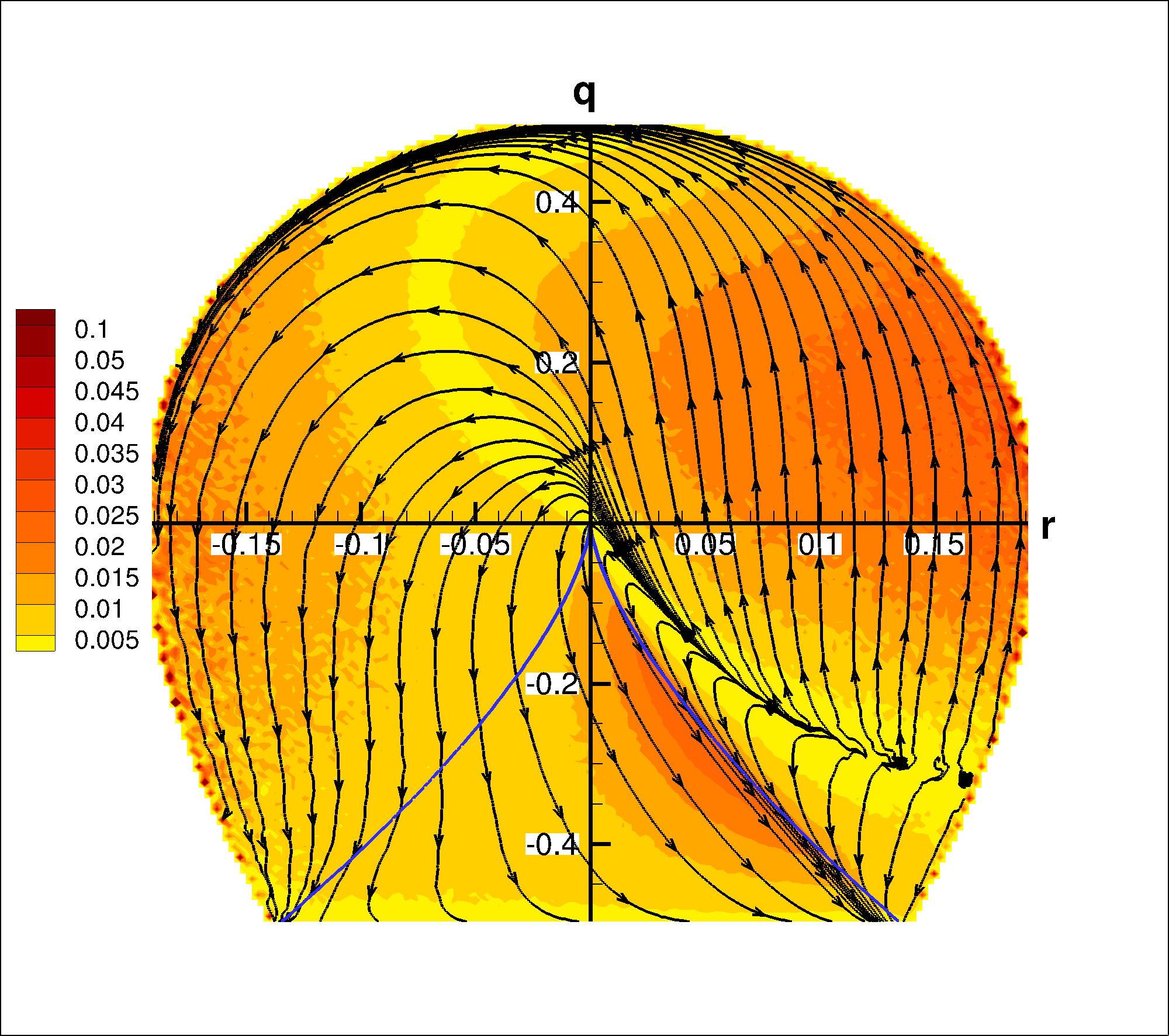}};
\node at (10pt,150pt) {(\textit{c})};
\end{tikzpicture}
\begin{tikzpicture}
\node[above right] (img) at (0,0) {\includegraphics[width=0.48\textwidth,trim={1.3cm 4cm 3cm 3cm},clip]{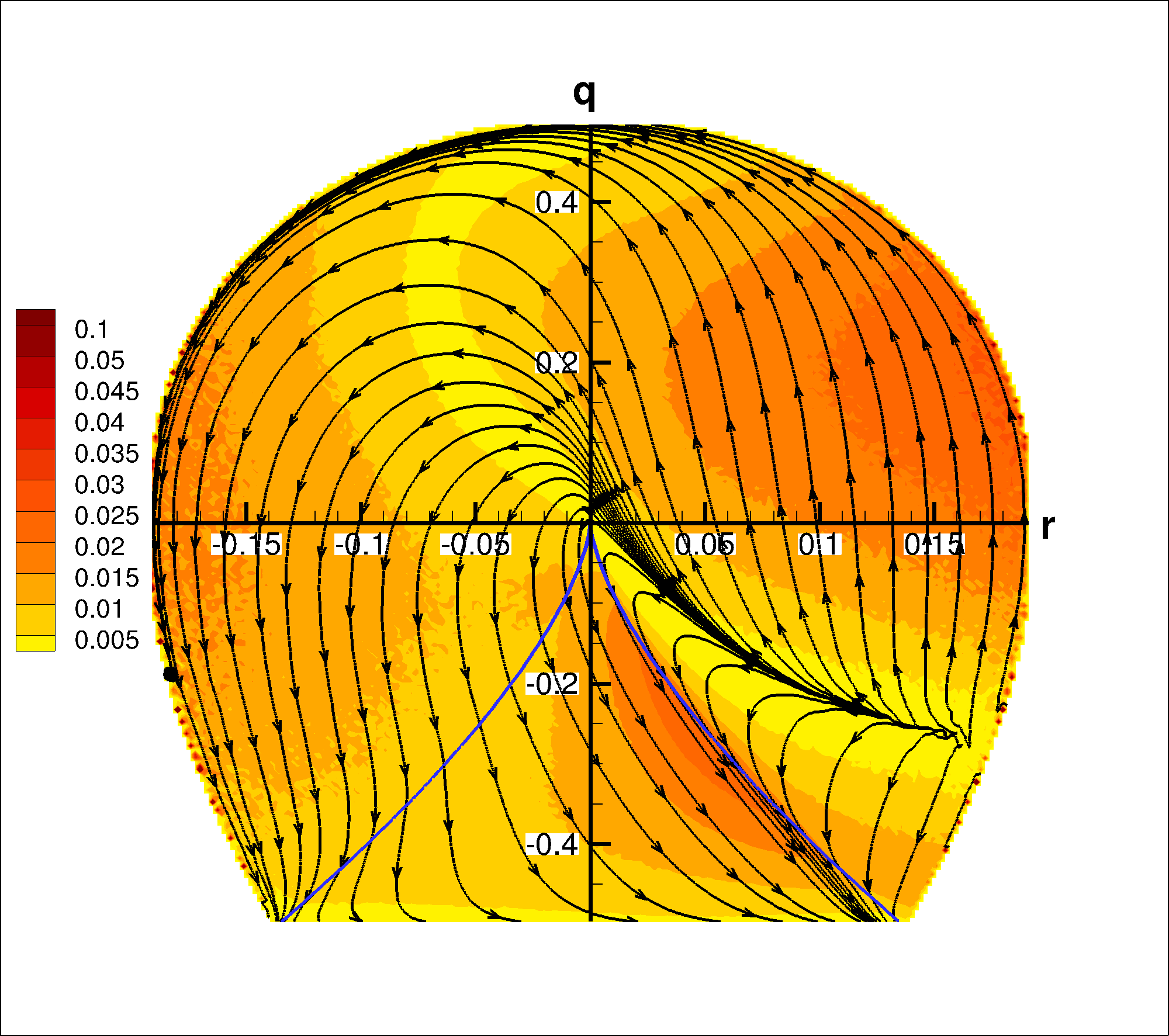}};
\node at (10pt,150pt) {(\textit{d})};
\end{tikzpicture}
\caption{Viscous CMTs in (a) $Q$-$R$ plane for $Re_\lambda=225$ (b) $q$-$r$ plane for $Re_\lambda=225$ (c) $q$-$r$ plane for $Re_\lambda=385$ (d) $q$-$r$ plane for $Re_\lambda=588$. Background contours indicate the speed of the trajectory at that point, normalized by Kolmogorov time scale.}
\label{fig:CMT_viscous}
\end{figure}

\textit{$Q$-$R$ plane}:  The viscous contribution to $Q$-$R$ evolution is plotted in figure \ref{fig:CMT_viscous} (a) and is in accordance with the results of \cite{chevillard2008modeling} and \cite{johnson2016closure}. 
Viscous action directs all trajectories in the $Q$-$R$ plane towards the origin. 
The evolution rate diminishes as the trajectories approach the origin.
This reflects the damping effect of viscous action, which is to reduce the VG magnitudes of all streamline topologies alike.  
However, the effect of viscosity on evolution of geometric-shape can not be surmised from the $Q$-$R$ CMTs. 

\textit{$q$-$r$ plane}: The viscous contribution to $q$-$r$ evolution, based on the following terms from equations (\ref{eq:b3}) and (\ref{eq:b4}),
\begin{equation}
     {\bigg(\frac{dq}{dt} \bigg)}_{V} = - A \tau_{ij}(b_{ji} + 2qb_{ij}) \;\;,\; \;\; {\bigg(\frac{dr}{dt} \bigg)}_{V} = - A\tau_{ij}(b_{ki}b_{jk} + 3rb_{ij})
\end{equation}
are plotted in figure \ref{fig:CMT_viscous} (b-d) for $Re_\lambda=225$, $385$ and $588$.
Again, important features not apparent in the $Q$-$R$ CMTs are evident here.
The viscous $q$-$r$ CMTs demonstrate that aside from reducing VG magnitude, the viscous action has a distinct influence on the geometric shape of the local streamlines. 
There exists a clearly defined repelling manifold in the lower UFC region. 
The only attracting manifold in this system is the $q=-1/2$ line, representing pure-strain streamlines with mutually orthogonal stretching and compression.
Starting at the repelling manifold, viscous CMTs either follow a short path directly towards the attractor through UN/S/S streamline topology or take a longer path through SFS followed by SN/S/S and UN/S/S streamlines. 
Clearly, the viscous evolution of $q$-$r$ depends on its precise point of origin.
Overall, \textit{viscous action directs streamlines away from strain-dominated vortex compression towards pure-strain}. 
Even though the magnitude of viscous contribution is much lower than the pressure-inertial contribution, viscous action alters the pressure-inertial $q$-$r$ evolution by reducing the domain of attraction of pure-shear attractor and expanding that of pure-strain attracting manifold. The viscous action exhibits some degree of $Re_\lambda$-dependence which will be examined in the future using DNS of higher resolution.


\section{\label{sec:conc} Summary and conclusions}

The study provides a description of the local streamline geometry by separating geometric-shape and scale-factor of the streamlines and characterizes velocity gradient dynamics conditioned on key shape-parameters.
This work is composed of three parts.

In the first part, a complete description of local streamline geometry is developed in terms of geometric-shape and scale-factor. It is established that the normalized VGT $b_{ij}$ completely determines the streamline shape while the scale of the streamline structure depends on VG magnitude $A$ (inversely). Geometric-shape is characterized by only four shape-parameters - $q$, $r$, $a_2$ and $|\omega_2|$. 
The study further demonstrates that invariant shape-parameters $q$ and $r$ constitute a bounded phase-space with unique specification of streamline shape and therefore, provide a useful platform to study VG dynamics.

In the second part of this work, DNS data of forced isotropic turbulence is employed to investigate the statistical distribution of streamline shape and scale features in turbulence.
It is demonstrated that the probability distribution of local streamline shape in this four-dimensional shape-parameter space is invariant with Reynolds number for $Re_\lambda > 200$. 
The pdfs reaffirm that the streamlines are most likely to have positive intermediate strain-rate and vorticity aligned with the intermediate strain-rate eigen direction.
Furthermore, DNS data reveals that the VG magnitude is highest (streamline scale-factor is smallest) in stable streamlines undergoing nearly pure rotation with vorticity vector along positive intermediate strain-rate.

The final part of this study characterizes the Lagrangian evolution of geometric shape using conditional mean trajectories (CMTs) in the $q$-$r$ plane. The key findings are summarized below:

\begin{enumerate}
\item The $q$-$r$ phase space can be divided into inner and outer regions. Inner CMTs spiral (clockwise) towards the pure-shear attractor at origin while outer CMTs (also clockwise) asymptote to pure-strain attracting manifold ($q=-1/2$). 
Inner CMTs are periodic in nature with time period, $T \in (25\tau_\eta,35\tau_\eta)$, decreasing progressively inwards while outer CMTs are aperiodic in nature.



\item The residence times of CMTs in different topologies are proportional to their percentage compositions particularly in the inner region. 
Therefore, it is evident that $q$-$r$ CMTs are fairly accurate approximations of mean Lagrangian evolution of local streamline shape.

\item Pressure-inertial effects dominate in the inner region driving all geometric shapes towards pure-shear attractor while in the outer region, pressure-inertial-viscous effects together direct all streamline shapes towards pure-strain.

\item $q$-$r$ CMTs are nearly invariant with $Re_\lambda$ at high enough Reynolds numbers, except for a slight dependence of the separatrix on $Re_\lambda$, possibly due to the viscous effects. 

\end{enumerate}


The study highlights some of the key features of turbulence processes in evolution of internal structure of streamlines, not evident in previously studied frameworks. 
The characterization of pressure and viscous processes in the evolution of geometric-shape indicate the amenability of their closure modeling in the bounded $q$-$r$ phase space, thus leading toward a Lagrangian model for VGT evolution.
In future work, the evolution of the entire four-parameter system - $q$, $r$, $a_2$ and $|\omega_2|$ - will be characterized to determine the complete streamline shape dynamics.

\vspace{0.4cm}
\noindent \textbf{Acknowledgements}
\vspace{0.1cm}

The authors would like to thank Prof. Diego Donzis of Texas A\&M University for providing the DNS data used in this study. The first author would also like to thank Zonta International for awarding the Amelia Earhart Fellowship for the academic year 2019-2020.

\vspace{0.4cm}
\noindent
Declaration of Interests. The authors report no conflict of interest.



\bibliographystyle{jfm}
\bibliography{main}

\end{document}